\documentclass[10pt,journal,compsoc]{IEEEtran}

\ifCLASSOPTIONcompsoc
\usepackage[nocompress]{cite}
\else
\usepackage{cite}
\fi

\usepackage{subfigure}
\usepackage{amsmath}
\usepackage{graphicx} 
\usepackage{caption}
\usepackage{graphicx,subfigure}
\usepackage{multicol} 
\usepackage{graphicx} 
\usepackage{amssymb}
\usepackage{float} 
\usepackage{wrapfig} 
\usepackage{booktabs}
\usepackage{lipsum} 
\usepackage{indentfirst}
\usepackage{bm}
\usepackage{color}
\usepackage{algorithm}
\usepackage{algorithmic}
\usepackage{amsmath}
\usepackage{enumerate}
\usepackage{extarrows}
\usepackage{url}
\usepackage{graphicx}
\usepackage{epstopdf}
\usepackage{multirow}
\usepackage{tablefootnote}
\usepackage[flushleft]{threeparttable}
\usepackage{slashbox}
\usepackage{array}

\newtheorem{assumption}{Assumption}

\newtheorem{problem}{\textbf{Problem}}
\newtheorem{question}{\textbf{Question}}

\newtheorem{lemma}{\textbf{Lemma}}
\newtheorem{corollary}{\textbf{Corollary}}
\newtheorem{proposition}{\textbf{Proposition}}
\newtheorem{definition}{\textbf{Definition}}
\newtheorem{theorem}{\textbf{Theorem}}

\def\eq{\triangleq}

\def\val{\theta}
\def\valmax{\val_\text{max}}
\def\Val{\Theta}
\def\cut{\beta}
\def\dmean{\bar{d}}
\def\dmax{D}

\def\payoff{S}
\def\payoffexp{\bar{S}}

\def\revenue{R}
\def\revenueexp{\bar{R}}
\def\cost{C}
\def\profit{W}
\def\spedata{\tau}

\def\plan{\mathcal{T}}
\def\Flat{\text{f}}
\def\Pure{\text{p}}

\def\dcap{Q}
\def\pcap{\Pi}
\def\adfee{\pi}
\def\mechanism{\kappa}
\def\na{\mathtt{Na}}

\def\QoS{\rho}
\def\load{L}
\def\subscription{\Psi}

\def\flexibility{\mathcal{F}}
\def\tdr{\Omega}

\def\QEDclosed{\mbox{\rule[0pt]{1.3ex}{1.3ex}}} 

\begin{document}

\title{Exploring Time Flexibility in Wireless Data Plans}

\author{Zhiyuan~Wang,
	Lin~Gao,~\IEEEmembership{Senior Member,~IEEE,}
	and~Jianwei~Huang,~\IEEEmembership{Fellow,~IEEE}
	
	\IEEEcompsocitemizethanks{
		\IEEEcompsocthanksitem Part of the results appeared in WiOpt 2017 \cite{wang2017pricing}.
		\IEEEcompsocthanksitem This work is supported by the General Research Funds (Project Number CUHK 14219016) established under the University Grant Committee of the Hong Kong Special Administrative Region, China.
		\IEEEcompsocthanksitem Zhiyuan Wang and Jianwei Huang are with the Network Communications
		and Economics Lab, Department of Information Engineering, The Chinese University of Hong Kong, Shatin, N.T., Hong Kong, China.\protect\\
		E-mail:	$\{$wz016, jwhuang$\}$@ie.cuhk.edu.hk
	\IEEEcompsocthanksitem Lin Gao is with the School of Electronic and Information Engineering, Harbin Institute of Technology (Shenzhen), Shenzhen, China.\protect\\
		E-mail: gaolin@hitsz.edu.cn
}
}

%
%


%

\IEEEtitleabstractindextext{%
	\begin{abstract}
		Recently, the mobile network operators (MNOs) are exploring more time flexibility with the  rollover data plan, which allows  the unused data from the previous month to be used in the current  month.
		Motivated by this industry trend,  we propose a general framework for designing and optimizing the mobile data plan with time flexibility. 
		Such a framework includes the traditional data plan, two existing  rollover data plans, and a new credit data plan as special cases.
		Under this framework, we formulate a monopoly MNO's optimal data plan design as a three-stage Stackelberg game: In Stage I,  the MNO decides the data mechanism; In Stage II, the MNO  further decides the corresponding data cap, subscription fee, and the per-unit fee; Finally in Stage III, users make subscription decisions  based on their own characteristics.
		Through backward induction, we analytically characterize the MNO's profit-maximizing data plan and the corresponding users' subscriptions.
		Furthermore, we conduct a market survey to estimate the distribution of users' two-dimensional characteristics, and  evaluate the performance of different data mechanisms using the real data.
		We find that a more time-flexible data mechanism increases  MNO's profit and users' payoffs, hence improves the social welfare.
	\end{abstract}
	
	}

\maketitle

\IEEEdisplaynontitleabstractindextext

\IEEEpeerreviewmaketitle

\section{Introduction}
\subsection{Background and Motivations}
\IEEEPARstart{D}{ue} to the increasing competition in the telecommunications market, Mobile Network Operators (MNOs) are under an increasing pressure to increase market shares and improve profits \cite{sen2015smart}.
One approach is to adopt various  novel wireless technologies to improve the quality of service (QoS) to attract more subscribers.
However, the technology upgrade is often costly and time-consuming.
A complementary economical approach is to explore various innovative data pricing schemes to better address  heterogeneous user requirements.

Traditionally, most network operators used the \textbf{flat-rate} data plans for wireless data services \cite{sen2015smart}, where users pay a fixed fee for unlimited monthly data usage.
Later in 2010, the Federal Communications Commission (FCC) and Cisco backed usage-based pricing to penalize those heavy users and manage the traffic.
Therefore, MNOs started to adopt the \textbf{usage-based} pricing scheme, where the subscribers are charged based on their actual data consumptions.
A widely used  form of usage-based plan adopted by many MNOs today is the \textbf{three-part tariff} plan, which consists of a monthly subscription fee, a data cap within which there is no additional cost of usage, and a linear unit price for any data consumption exceeding the data cap.

Recent years have witnessed  many MNOs  exploring the \textbf{time flexibility} in their  data plans  to further increase their market competitiveness.
For example, the \textit{rollover data plan}, which allows the unused part of the data cap from the previous month to be used in the current month, has  been implemented by many MNOs, e.g., AT\&T \cite{ATTrollover}, T-Mobile \cite{TMobilerollover}, and China Mobile \cite{CMrollover}.
Such a rollover plan  reduces the users' uncertainty due to stochastic data demands, and offers users more time flexibility.
This motivates us to ask the first key question in this paper: 
\begin{question}\label{Question: who benefit}
	\textit{Who will benefit more from the introduced time flexibility, the MNO or users?}
\end{question}

Although  centering around the same core idea, various rollover data plans in practice can be quite different in terms of the consumption priority and the expiration time.
For example, AT\&T specifies that the rollover data from the previous month  will be used after the current monthly data cap is fully consumed \cite{ATTrollover}, while China Mobile specifies that the rollover data will be used before consuming the current monthly data cap \cite{CMrollover}.
As for the expiration time, both AT\&T and China Mobile require the rollover data to expire after one month, while T-Mobile allows subscribers to accumulate their rollover data over several months \cite{TMobilerollover}.

We observe that a common feature of  various rollover data plans is that a user can only use the ``remaining''  data  cap  from the previous month(s). 
This motivates us to  propose a \emph{credit data plan} that inversely allows users to ``borrow'' their data quota from  future months.\footnote{The telecom market in many countries is based on the real-name registration, hence it is difficult for a user to keep borrowing data and then stop the subscription without paying back. Moreover, the Data Bank platform \cite{DataBank} implemented by China Unicom allows users to borrow data from the MNO, which exhibits a  similar idea to the credit data plan.}
In fact, the rollover and credit data mechanisms represent two different ways of exploring the data dynamics across the time dimension: backward and forward.
This motivates us to ask the second and the third key questions in this paper:

\begin{question}\label{Question: most flexible}
	\textit{Which data mechanism is the most time-flexible?}
\end{question}

\begin{question}\label{Question: which adopt}
	\textit{Which data mechanism should the MNO adopt?}
\end{question}

Furthermore, the MNOs profit from mobile data services through carefully choosing  and optimizing their mobile data plans.
Even though the MNOs have implemented different versions of rollover data plans in practice, there is no systematical understanding on how the time flexibility affects the optimal data plan design.
This motivates us to ask the fourth key question in this paper:
\begin{question}\label{Question: impact}
	\textit{What is the impact of  time flexibility on the MNO's optimal data cap, subscription fee, and per-unit fee?}
\end{question}


In this paper we will study and evaluate these innovative data mechanisms and reveal the impact of time flexibility in a comprehensible way.
We hope that our results in this paper could pave the way for the MNOs to better implement and the public to better understand these  time-flexible mobile data plans.




\subsection{Solutions and Contributions}
In this paper, we study the optimization of the three-part tariff data plan with time flexibility in a monopoly market, and consider four data mechanisms which are different in the \textit{special data} (rollover or credit data) and the \textit{consumption priority}.


We formulate the monopoly MNO's optimal data plan design as a three-stage Stackelberg game, with the MNO as the leader and users as followers.
Specifically, the MNO decides which kind of data mechanism to adopt in Stage I, then further decides its profit-maximizing data cap and the corresponding subscription fee and per-unit fee  in Stage II.
Finally, users make their subscription decisions to maximize their payoffs in Stage III.

To the best of our knowledge, this is the first paper that systematically studies the MNO's optimal three-part tariff plan with time flexibility.
The main results and contributions of this paper are summarized as follows:
\begin{itemize}
	\item \textit{Systematic Study of Data Mechanisms with Time Flexibility:} 
	We propose a general framework that includes the traditional data mechanism and three innovative data mechanisms with rollover or credit data as special cases.
	Based on such a unified framework, we further study the optimal design for mobile data plan with time flexibility.
	\item 	\textit{Three-Stage Decision Model:}
	We model and analyze the interactions between the MNO and users as a three-stage Stackelberg game.
	Despite the complexity of the model, we are able to fully characterize the user subscription in Stage III, the MNO's optimal data cap and pricing strategy in Stage II, and the optimal data mechanism choice in Stage I.
	\item \textit{User Subscription:}
	We consider  users' heterogeneity in the data valuation and the network substitutability.
	We find that under the optimal data plan, the profit-maximizing MNO admits subscribers based only on their data valuations, while treating  users of different network substitutability identically.
	\item \textit{Optimal Data Plan:}
	We study the impact of MNO's quality of service (QoS), operational cost, and capacity cost on its optimal data plan.
	Our analysis reveals a counter-intuitive insight: a better time flexibility dose not necessarily lead to a smaller  data cap; the  data cap can be larger  if the MNO is weak with a  poor QoS and large costs.
	\item \textit{Performance Evaluation:}
	We conduct a market survey to estimate the statistical distribution of users' data valuation and network substitutability.
	The simulations based on the empirical data further reveal that both MNO and users can benefit from the time flexibility. 
	The MNO benefits more than users if the MNO provides good services and experiences  small costs.
	Otherwise, users will benefit more from the time flexibility than the MNO.

\end{itemize}

The remainder of this paper is organized as follows.
In Section \ref{Section: Literature Review}, we review the related works. 
Section \ref{Section: System Model} introduces our system model and the three-stage game. 
Section \ref{Section: Data Mechanisms and Time Flexibility} presents the four data mechanisms with time flexibility in detail.
In Section \ref{Section: Backward Induction of the Three-stage Game}, we analyze the three-stage decision model through backward induction.
Section \ref{Section: Numerical Results} presents the numerical results and Section \ref{Section: Conclusions and Future Works} concludes this paper.


\begin{table*}
	\setlength{\abovecaptionskip}{2pt}
	\renewcommand{\arraystretch}{1.1}		
	\caption{ $\plan\eq\{\dcap,\pcap,\adfee,\mechanism\},\ \mechanism\in\{ 0, 1, 2, 3\}$} 
	\label{table: Various Plans}
	\centering
	\begin{tabular}{c c c c c c}
	\toprule
	Data Mechanism 		& Special Data		& Surplus or Deficit 			& Consumption Priority		& Effective Cap $\dcap_\mechanism^e(\spedata)$		\\
	\midrule
	$\mechanism=0$		& None				& $\ \spedata=0\qquad\quad$		& Cap						& $\dcap\qquad$	\\
	$\mechanism=1$		& Rollover data		& $\spedata\in[0,\dcap]\quad$	& Cap$\Rightarrow$Rollover	& $\dcap+\spedata \in[\dcap,2\dcap]$ 	\\
	$\mechanism=2$		& Rollover data		& $\spedata\in[0,\dcap]\quad$	& Rollover$\Rightarrow$Cap	& $\dcap+\spedata \in[\dcap,2\dcap]$ 	\\
	$\mechanism=3$ 		& Credit data		& $\spedata\in[-\dcap,0]\ $		& Cap$\Rightarrow$Credit	& $2\dcap+\spedata \in[\dcap,2\dcap]\ $ \\
	\bottomrule
	\end{tabular}
\end{table*}

\section{Literature Review\label{Section: Literature Review}}


The optimal design of mobile data plan has been extensively studied in the literature.
The early studies focused on the debate between flat-rate and usage-based schemes \cite{ma2016usage}.
After introducing the data cap, Dai \textit{et al.} in \cite{dai2015effect} demonstrated that heavy users would pay for their usage, while light users would benefit from it.
Then Wang \textit{et al.} in \cite{wang2017role} studied the optimization of the three-part-tariff in the congestion-prone network.
{Xiong \textit{et al.} in \cite{xiong2017economic} focused on the sponsored content and analyzed  a Stackelberg game pricing model on the MNO's data plan optimization.}
{Zheng \textit{et al.} in \cite{zhengoptimizing} studied the dynamics of users' data consumption through  a dynamic programming formulation.}
However, the above studies did not consider various forms of flexibility introduced in recent mobile data plans, including the time dimension \cite{zheng2016understanding,wei2016novel}, user dimension \cite{sen2012economics,zheng2017customized,yu2017mobile}, and location dimension \cite{duan2015pricing,ma2016time}.

\textit{Time flexibility} corresponds to the rollover data plan in practice.
Despite of the increasing popularity of the rollover data plan, the related theoretical study just emerged very recently. 
As far as we know, there existed only two related works  before this work.
Specifically, Zheng \emph{et al.} in \cite{zheng2016understanding} compared the rollover data plan with a traditional three-part tariff, and found that the moderately price-sensitive users can benefit from subscribing to the rollover data plan. 
Wei \emph{et al.} in \cite{wei2016novel} further looked at the choice of expiration time of the rollover data and analyzed the impact of the  rollover period lengths through a contract-theoretic approach. 

\textit{User flexibility} corresponds to the shared data plan and data trading.
Sen \textit{et al.} in \cite{sen2012economics}  introduced an analytical framework for studying the economics of shared data plans.
Zheng \textit{et al.} in \cite{zheng2017customized} examined the ``2CM'' data trading market launched by China Mobile Hong Kong,  and Yu \textit{et al.} in \cite{yu2017mobile} further analyzed users' realistic trading behaviors using prospect theory.

\textit{Location flexibility} corresponds to the global data services, e.g., Skype Wi-Fi and Uroaming.
Duan \textit{et al.} in \cite{duan2015pricing}  studied how the global providers work with many local providers to promote the global mobile data services, and examined the flat-rate and usage-based schemes.
Ma \textit{et al.} in \cite{ma2016time} proposed an optimal design of time and location aware mobile data pricing, incentivizing users to
smoothen their traffic  and reduce network congestion.


Note that each of the above three dimensions deserves  substantial studies and further explorations.
Our focus is the optimization of the three-part tariff mobile data plan with time flexibility (which was not studied in \cite{zheng2016understanding,wei2016novel}).

\section{System Model\label{Section: System Model}}
In this paper, we consider a \textit{monopoly} market with a single MNO who provides mobile data services for heterogeneous users.\footnote{There are multiple competitive MNOs in the practical market. The analysis for the competitive market requires a comprehensive understanding on each MNO's optimal data plan design. Due to space limit, in this paper we focus on the monopoly case. We have reported some preliminary results on the duopoly competition in \cite{Zhiyuan2018Duopoly}.}
The MNO designs a mobile data plan to maximize its profit, and each user decides whether to subscribe to the MNO to maximize his payoff.

\begin{figure} 
	\centering
	\includegraphics[width=0.85\linewidth]{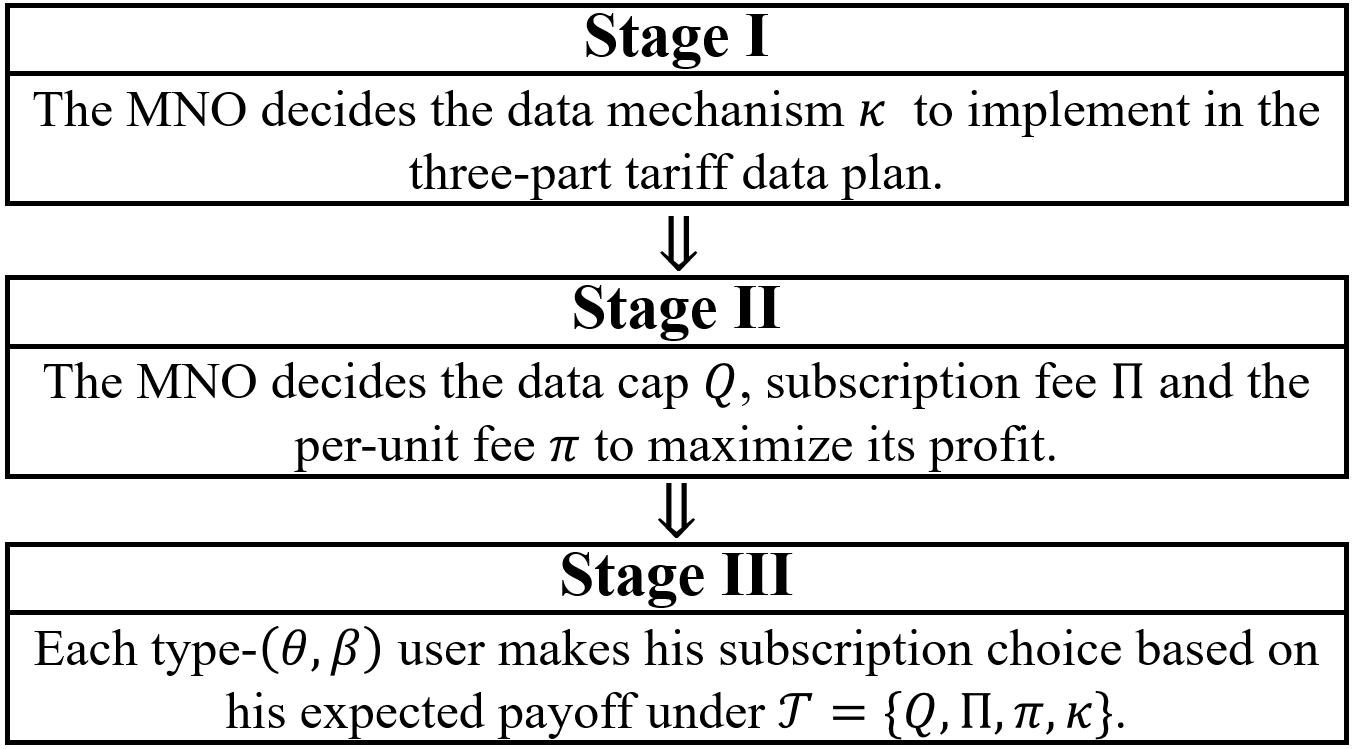}
	\caption{Three-Stage Stackelberg Game.\vspace{-10pt}}
	\label{fig: SystemModel}
\end{figure}

We formulate the economic interactions between the MNO and the mobile users as a three-stage Stackelberg game, as shown in Fig. \ref{fig: SystemModel}. 
The MNO is the Stackelberg leader: it first decides the data mechanism to be implemented within its three-part tariff data plan in Stage I,
then decides the data cap, the subscription fee, and the per-unit fee in Stage II.\footnote{In practice, an MNO adopts a data mechanism (i.e., rollover, credit, or the traditional one) for a relatively long time period (e.g., three or five years), while has the flexibility of updating data mechanism parameters (i.e., the data cap, monthly subscription fee, and the per-unit fee) more often (e.g., on a yearly basis). The three-stage formulation captures the MNO's different decisions at different time scales.}
Finally, the users make their subscription decisions to maximize  their payoffs in Stage III.

Next we first present a unifying framework of  different mobile data plans,  then introduce the model in more details  from the perspectives of users and the MNO, respectively.

\subsection{Mobile Data Plans}
The \textit{three-part tariff data plan with time flexibility} can be characterized by the tuple $\plan=\{\dcap,\pcap,\adfee,\mechanism\}$, where the subscriber pays a fixed lump-sum  subscription fee $\pcap$ for the data usage up to the cap $\dcap$, beyond which the subscriber pays an overage fee $\adfee$ for each unit of additional data consumption.
Here $\mechanism$ represents different \textit{data mechanisms} that gives the subscriber different degrees of time flexibility on their data consumption over time.

Next we first introduce the four data mechanisms, then demonstrate that the pure usage-based data plan and the flat-rate data plan are special cases of the tuple $\plan$.

\subsubsection{Four Data Mechanisms}
We consider four data mechanisms indexed by $\mechanism\in\{0,1,2,3\}$.
The key differences among the different data mechanisms are the \textit{special data} and the \textit{consumption priority}.
To be more specific, the special data could be the rollover data inherited from  the previous month or the credit data that can be borrowed from the next month, both of which can enlarge a subscriber's \textit{effective data cap} (within which no overage fee involved) in the current month.
Moreover, the consumption priority  of the special data and the current monthly data cap further affects how much the effective data cap can be enlarged.

We summarize the key differences of the four data mechanisms in Table \ref{table: Various Plans}.
Here we use $\spedata$ to denote a user's data surplus ($\spedata>0$) or data deficit ($\spedata<0$) at the beginning of a month, and use $\dcap_\mechanism^e(\spedata)$ to denote the corresponding effective data cap. More specifically, 
\begin{itemize}
	\item The case of $\mechanism=0$ denotes the \textbf{traditional data mechanism} without time flexibility. 
		The subscriber has no data surplus or deficit, and the effective cap of each month is $\dcap_0^e(\spedata)=\dcap$; 
	\item The case of $\mechanism=1$ denotes the \textbf{rollover data mechanism} offered by AT\&T. 
		The rollover data $\spedata$ from the previous month is consumed \textit{after} the current monthly data cap {and expires at the end of the current month}. Thus the effective cap of the current month is $\dcap_1^e(\spedata)=\dcap+\spedata$;
	\item The case of $\mechanism=2$ denotes the \textbf{rollover data mechanism} offered by China Mobile.
		The rollover data $\spedata$ from the previous month is consumed \textit{prior} to the current monthly data cap $\dcap$ and {expires at the end of the current month}. Thus the effective cap of the current month is $\dcap_2^e(\spedata)=\dcap+\spedata$; 
	\item The case of $\mechanism=3$ denotes the \textbf{credit data mechanism} proposed in this paper.
		The credit data is from the next month's data cap $\dcap$,{\footnote{Since users can only borrow data from the next one month's data cap, $\dcap$ is the maximal amount that can be borrowed.}} which is consumed \textit{after} the current monthly data cap $\dcap$ (with a data deficit $\spedata$ from the previous month). Thus the effective cap of the current month is $\dcap_3^e(\spedata)=2\dcap+\spedata$;  
\end{itemize}

As  mentioned above, the time flexibility can enlarge the subscriber's effective data cap.
According to Table \ref{table: Various Plans}, the effective data cap of the traditional data mechanism (i.e., $\mechanism=0$) is always $\dcap$; while the potential maximal value of the effective data cap is $2\dcap$ for the rollover and credit data mechanisms (i.e.,  $\mechanism=1,2,3$).
The larger the effective data cap is, the less overage usage is incurred, which will further change the users' subscription decisions.

Table \ref{table: Example} provides a numerical example with five months' data consumptions and the corresponding payments under the four data mechanisms.
Among the four schemes, the last two schemes (i.e., $\mechanism=2,3$) lead to the same least total payment of $\$250$, and the first scheme (i.e., $\mechanism=0$) leads to the maximum total payment of $\$280$.
Is this a coincidence?
We will further discuss it in Section \ref{Section: Data Mechanisms and Time Flexibility}.


\subsubsection{Special Cases of Three-Part Tariff\label{Subsubsection: Special Cases of Three-Part Tariff}}
Next we illustrate that the pure usage-based data plan and the flat-rate data plan are two special cases of the tuple $\plan=\{\dcap,\pcap,\adfee,\mechanism\}$, $\mechanism\in\{0,1,2,3\}$.

Under the \textit{pure usage-based} data plan, the subscriber has a zero monthly data cap and only needs to pay a usage-based fee for each unit of data consumption.
Therefore, we can denote the pure usage-based data plan as $\plan_\Pure=\{0,0,\adfee_\Pure,\na\}$, where $\na$ represents that the data mechanism index $\mechanism$ has no impact under the pure usage-based data plan (as there is no data cap).


Under the \textit{flat-rate} data plan, the subscriber pays a fixed monthly  subscription fee for unlimited data usage without any additional overage fee, thus the per-unit fee $\adfee$ has no impact.
Strictly speaking, the data cap of the flat-rate data plan is  infinity.
However, in this paper we model the user's data demand as a random variable $d\in\{0,1,2,...,\dmax\}$ (to be defined in Section \ref{Subsection: User Model}), where $\dmax$ is the  maximal data demand.
Therefore, in practice it is enough to set  the data cap to be  no less than $\dmax$ to achieve the effect of a flat-rate plan.
Accordingly, we denote the flat-rate data plan as $\plan_\Flat=\{\dcap_\Flat,\pcap_\Flat,\na,\na\}$, where $\dcap_\Flat\ge\dmax$, and $\na$ represents that the per-unit fee $\adfee$ and the data mechanism $\mechanism$ have no impact.

Our later analysis on the MNO's optimal data plan is based on the general tuple $\plan=\{\dcap,\pcap,\adfee,\mechanism\}$. 
Meanwhile,  we will characterize the conditions under which the MNO's optimal three-part tariff with time flexibility would degenerate into 
the pure usage-based data plan $\plan_\Pure$ or the flat-rate data plan $\plan_\Flat$.

\subsection{User Model\label{Subsection: User Model}}
Next we introduce a user's  three characterizations: the data demand $d$, the data valuation $\val$, and the network substitutability $\cut$. 
Based on these, we derive a user's expected payoff. 

First, we model the user's monthly data demand $d$ as a discrete random variable with a probability mass function $f(d)$, a mean value of $\dmean$, and a finite integer support $\{0,1,2,...,\dmax\}$.
Here the data demand is measured in the minimum data unit (e.g, 1KB or 1MB according to the MNO's billing practice).

Second, we follow \cite{ma2016usage} by denoting $\val$ as a user's utility from consuming one unit of data, i.e., the user's data valuation.
According to our market survey conducted in mainland China, $\val$ falls into the range  between 10 RMB/GB and 60 RMB/GB with a large probability.
We will further discuss it in Section \ref{Subsection: data}.

Third, we further explore a user's  behavior change when he reaches the {effective data cap} $\dcap_\mechanism^e(\spedata)$, since further data  consumption leads to an additional payment. 
Although the user will still continue to consume data in this case, he will rely more heavily on other alternative networks (such as Wi-Fi).
As in \cite{sen2012economics}, we use the network substitutability $\cut\in[0,1]$ to denote the fraction of overage usage shrink. 
A larger $\cut$ value represents more overage usage cut (thus a better network substitutability).

A user's mobility pattern can significantly influence the availability of alternative  networks.
Hence, different users usually have heterogeneous network substitutabilities.
For example, a businessman who is always on the road may have a poor network substitutability (hence has a small value of $\cut$); while a student can take the advantage of the school Wi-Fi network (hence has a large value of $\cut$).
According to our market survey, $\cut$ falls into the range  between 0.7 and 1 with a large probability.
We will further discuss it in Section \ref{Subsection: data}.

\begin{table}
\renewcommand{\arraystretch}{1.13}
\setlength{\abovecaptionskip}{2pt}
\setlength{\belowcaptionskip}{1pt}
\caption{Numerical Example for $\mechanism=0,1,2,3$. }
\label{table: Example}
\centering
\begin{threeparttable}
	\begin{tabular}{cc|ccccccc}
		\bottomrule
		\multicolumn{2}{c|}{ Month}								& Jan.		& Feb. 		& Mar.		& Apr. 		& May 		& Total\\
		\multicolumn{2}{c|}{Data Consumption} 					& 2GB 		& 2GB 		& 4GB 		& 4GB		& 1GB		& 13GB\\
		\hline\hline
		$\mechanism=0$ 						& Payment			& $\$$50	& $\$$50	& $\$$65	& $\$$65	& $\$$50	& $\bm{\$280}$ \\
		\hline
		\multirow{2}{*}{$\mechanism=1$} 	& $\tau\ $ 			& 0 		& 1GB		& 1GB		& 0			& 0			& \multirow{2}{*}{$\bm{\$265}$} \\
											& Payment 			& $\$$50	& $\$$50	& $\$$50	& $\$$65	& $\$$50	& \\
		\hline
		\multirow{2}{*}{$\mechanism=2$} 	& $\tau\ $			& 0 		& 1GB		& 2GB		& 1GB		& 0			& \multirow{2}{*}{$\bm{\$250}$} \\
											& Payment 			& $\$$50	& $\$$50	& $\$$50	& $\$$50	& $\$$50	&  \\
		\hline
		\multirow{2}{*}{$\mechanism=3$} 	& $\tau\ $			& 0 		& 0			& 0			& -1GB		& -2GB		& \multirow{2}{*}{$\bm{\$250}$} \\
											& Payment 			& $\$$50	& $\$$50	& $\$$50	& $\$$50	& $\$$50	&\vspace{-2pt}  \\
		\bottomrule
	\end{tabular}
	\begin{tablenotes}
	\item {\scriptsize Here the data cap is 3GB, the subscription fee is $\$$50, the overage fee is $\$$15$/$GB, and $\tau$ denotes the data surplus or deficit of each month.\vspace{-10pt}}
	\end{tablenotes}
\end{threeparttable}
\end{table}

Without loss of generality, we normalize the total user population size to  one in the rest of the paper.
We assume that users are homogeneous in the data demand distribution $f(d)$,
\footnote{
According to the statistical analysis in \cite{lambrecht2007does},\cite{nevo2016usage}, users' monthly demand can be estimated by a log-normal distribution.
For analysis tractability,  we consider a homogeneous demand distribution.
In the future, we will consider the heterogeneous case and collect users' data usage records to estimate the demand distributions as in \cite{zhengoptimizing,zheng2017customized}.} and investigate the heterogeneity in the data valuation $\val$ and network substitutability $\cut$.
Therefore, we model each user by the two-dimensional characteristics $(\val,\cut)$, and define the whole user market as $\mathcal{M}=\{(\val,\cut):0\le\val\le\valmax ,0\le\cut\le 1\}$ with probability density functions $h(\val)$ and $g(\cut)$, since our market survey shows that $\val$ and $\cut$ are independent with the Pearson correlation coefficient less than $0.05$.
Furthermore, we denote $\subscription(\plan)\subseteq\mathcal{M}$ as the subscriber set, i.e., the type-$(\val,\cut)$ user subscribes to the MNO if and only if $(\val,\cut)\in\subscription(\plan)$.

A subscriber's payoff is the difference between his utility and total payment.
More specifically, for a type-$(\val,\cut)$ subscriber with $d$ units of data demand and an effective data cap $\dcap_\mechanism^e(\spedata)$, his actual data usage is $d-\cut[d-\dcap_\mechanism^e(\spedata)]^+$, where $[x]^+=\max\{0,x\}$.
{Moreover, we use $\QoS$ to represent the MNO's \textit{average} quality of service (QoS).\footnote{In practice, an MNO's wireless data service depends on the network congestion, which has been studied before (e.g., \cite{ma2016usage,sen2015smart}). In this work, instead of modeling the detailed congestion-aware control, we are more interested in the \textit{long-term average quality} of the MNO's wireless data service.}}
Mathematically, $\QoS$ is a utility multiplicative coefficient, thus the subscriber's utility is $\QoS\val(d-\cut[d-\dcap_\mechanism^e(\spedata)]^+)$.
In addition, the subscriber's total payment consists of the subscription fee $\pcap$ and the overage payment $\adfee(1-\cut)[d-\dcap_\mechanism^e(\spedata)]^+$.
Therefore, \textit{the payoff of a type-$(\val,\cut)$ subscriber} with $d$ units of data demand and an effective cap $\dcap_\mechanism^e(\spedata)$ is given by
\begin{equation}
\begin{aligned} 
& \payoff(\plan,\val,\cut,d,\spedata)\\
=& \QoS\val\left(d-\cut[d-\dcap_\mechanism^e(\spedata)]^+\right) -\adfee(1-\cut)[d-\dcap_\mechanism^e(\spedata)]^+-\pcap,
\end{aligned}	
\end{equation}
where the data demand $d$ and the data surplus (or deficit) $\spedata$ are two random variables that change in each month. 
After taking the expectation over $d$ and $\spedata$, we obtain  a \textit{type-$(\val,\cut)$ subscriber's expected monthly payoff} as
\begin{equation}\label{Equ: system model payoffexp}
\begin{aligned}
\payoffexp(\plan,\val,\cut)=&\mathbb{E}_{d,\spedata}\left[ \payoff\left( \plan,\val,\cut,d,\spedata \right) \right] \\
=&\QoS\val\left[\dmean-\cut A_\mechanism(\dcap)  \right]  - \adfee(1-\cut) A_\mechanism(\dcap)-\pcap .
\end{aligned}
\end{equation}

Here $A_\mechanism(\dcap)$ is the type-$(\val,0)$ subscriber's \textit{expected monthly overage data consumption} under the data mechanism $\mechanism$, defined as follows:
\begin{equation}\label{Equ: system model A_i(Q_i)}
\begin{aligned}
A_\mechanism(\dcap)=&\mathbb{E}_{d,\spedata} \big\{\left[ d-\dcap_\mechanism^e(\spedata) \right]^+ \big\} \\
=& \textstyle \sum\limits_{\spedata}	\sum\limits_{d}	[d-\dcap_\mechanism^e(\spedata)]^+f(d)p_\mechanism(\spedata),
\end{aligned}
\end{equation}
{where the summation range of $d$ is from $0$ to the maximal demand $\dmax$, while the range of $\spedata$ depends on the data mechanism $\mechanism$, which is given in (\ref{Equ: A 1}), (\ref{Equ: A 2}), and (\ref{Equ: A 3}).}
The $p_\mechanism(\spedata)$  represents the probability mass function of $\spedata$ under the data mechanism $\mechanism$.
Moreover, $p_\mechanism(\cdot)$ is the key difference among the four data mechanisms, since the data mechanism $\mechanism$ affects a subscriber's payoff through  $A_\mechanism(\dcap)$ in (\ref{Equ: system model payoffexp}).
In Section \ref{Section: Data Mechanisms and Time Flexibility}, we will further explain how to compute $p_\mechanism(\spedata)$ and $A_\mechanism(\dcap)$ in detail.

Furthermore, the \textit{expected total payoff of the whole market} under $\plan$ is the integration over all the subscribers in $\subscription(\plan)$, as follows:
\begin{equation}
\begin{aligned}
\tilde{\payoff}(\plan)=& \iint_{\subscription(\plan)} \payoffexp(\plan,\val,\cut)h(\val)g(\cut) d\val d\cut.  \\
\end{aligned}
\end{equation}

So far, we have introduced users' characteristics and derived their payoffs.
Next we move on to model the profit-maximizing  MNO.

\subsection{MNO Model}
In the following we formulate the MNO's revenue, cost, and profit, respectively.
\subsubsection{\textbf{MNO's Revenue}}
The MNO's revenue obtained from a subscriber consists of the subscription fee and the possibly overage fee.
Therefore, the \textit{MNO's revenue from a type-$(\val,\cut)$ subscriber} with $d$ units of data demand and an effective cap $\dcap_\mechanism^e(\spedata)$ is
\begin{equation}
\revenue(\plan,\val,\cut,d,\spedata)=\adfee(1-\cut)\left[d-\dcap_\mechanism^e(\spedata)\right]^+ +\pcap.
\end{equation}

Since $d$ and $\spedata$ are two random variables that change in each month,  we take the expectation and obtain  the \textit{MNO's expected monthly revenue from a type-$(\val,\cut)$ subscriber} as follows:
\begin{equation}
\begin{aligned}
\revenueexp(\plan,\val,\cut)=&\mathbb{E}_{d,\spedata}\left[ \revenue\left( \plan,\val,\cut,d,\spedata \right) \right] \\
=& \adfee(1-\cut)A_\mechanism(\dcap) +\pcap,
\end{aligned}
\end{equation}
where $(1-\cut)A_\mechanism(\dcap)$ is the type-$(\val,\cut)$ subscriber's expected overage usage.
Again we will provide more details on $A_\mechanism(\dcap)$ in Section \ref{Section: Data Mechanisms and Time Flexibility}.
Moreover, the \textit{MNO's expected total revenue} from the entire  market under $\plan$ is
\begin{equation}
\begin{aligned}
\tilde{\revenue}(\plan)=& \iint_{\subscription(\plan)} \revenueexp(\plan,\val,\cut)h(\val)g(\cut) d\val d\cut . \\
\end{aligned}
\end{equation}

\subsubsection{\textbf{MNO's Cost}}
In reality, the MNO's cost is quite a complicated function that is related to many factors \cite{sen2015smart}. 
In this paper, we consider two kinds of costs experienced  by the MNO, i.e.,  the capacity cost and the operational cost.

The MNO's capacity cost mainly arises from its capital expenditure (CapEx), the investment on its network capacity.
In reality, the data cap helps the MNO manage the network congestion and ration the scarce network capacity \cite{rogerson2016economics}, and most MNOs imposed the data cap to alleviate the network congestion \cite{lyons2012impact}.
Therefore, once the MNO decides a data cap to be offered in the market, it should make sure a corresponding network capacity is in place to support the traffic. 
Motivated by this phenomenon, we model the \textit{MNO's capacity cost} as an increasing function $J(\dcap)$ on the data cap $\dcap$.
Intuitively, a larger data cap corresponds to a more severe  network congestion, which  requires more investment on the network capacity in advance.
{The MNO's capacity investment affects the network congestion, which will change its QoS and eventually affect the user utility of consuming data. 
Here we will not incorporate the congestion-aware formulation in this work, but refer interested readers to the related studies in \cite{ma2016usage,sen2015smart}.
}

Furthermore, the MNO's operational expense (OpEx) mainly arises from the system management.
After the MNO decides the mobile data plan to implement in the market, the subscribers' total data consumption will affect the MNO's operational expense.
Specifically, the \textit{expected total data consumption} from the whole market is
\begin{equation}
{\load}(\plan)= \iint_{\subscription(\plan)} \left[ \dmean-\cut A_\mechanism(\dcap) \right]h(\val)g(\cut) d\val d\cut.
\end{equation} 

For analysis tractability, we follow \cite{nabipay2011flat} by considering a linear operational cost, and denote $c$ as the marginal operational cost from unit data consumption.\footnote{Such a linear-form cost has been widely used to model an operator's operational cost (e.g.,  \cite{nabipay2011flat,duan2012duopoly}).}
Accordingly, the \textit{MNO's operational cost} is $\load(\plan)\cdot c$.

Putting  the capacity cost and the operational cost together, we compute the \textit{MNO's expected total cost} as follows: 
\begin{equation}
	\tilde{\cost}(\plan)= {\load}(\plan)\cdot c + J(\dcap).
\end{equation}

\subsubsection{\textbf{MNO's Profit}}
The MNO's profit is the difference between its revenue and cost.
Hence the \textit{MNO's expected total profit} under $\plan$ is 
\begin{equation}
\begin{aligned}
\tilde{\profit}(\plan)	
&=\tilde{\revenue}(\plan)-\tilde{\cost}(\plan) .
\end{aligned}
\end{equation}

So far we have introduced the four data mechanisms, users' payoffs, and MNO's profit.
In the following, we first compare the degrees of time flexibility offered by the four data mechanisms in Section \ref{Section: Data Mechanisms and Time Flexibility}, then study the three-stage game in Section \ref{Section: Backward Induction of the Three-stage Game}.

\begin{figure*}
	\centering
	\setlength{\abovecaptionskip}{2pt}
	\subfigure[$\mechanism=1$]{\label{fig:Usage-T1}\includegraphics[width=0.32\linewidth]{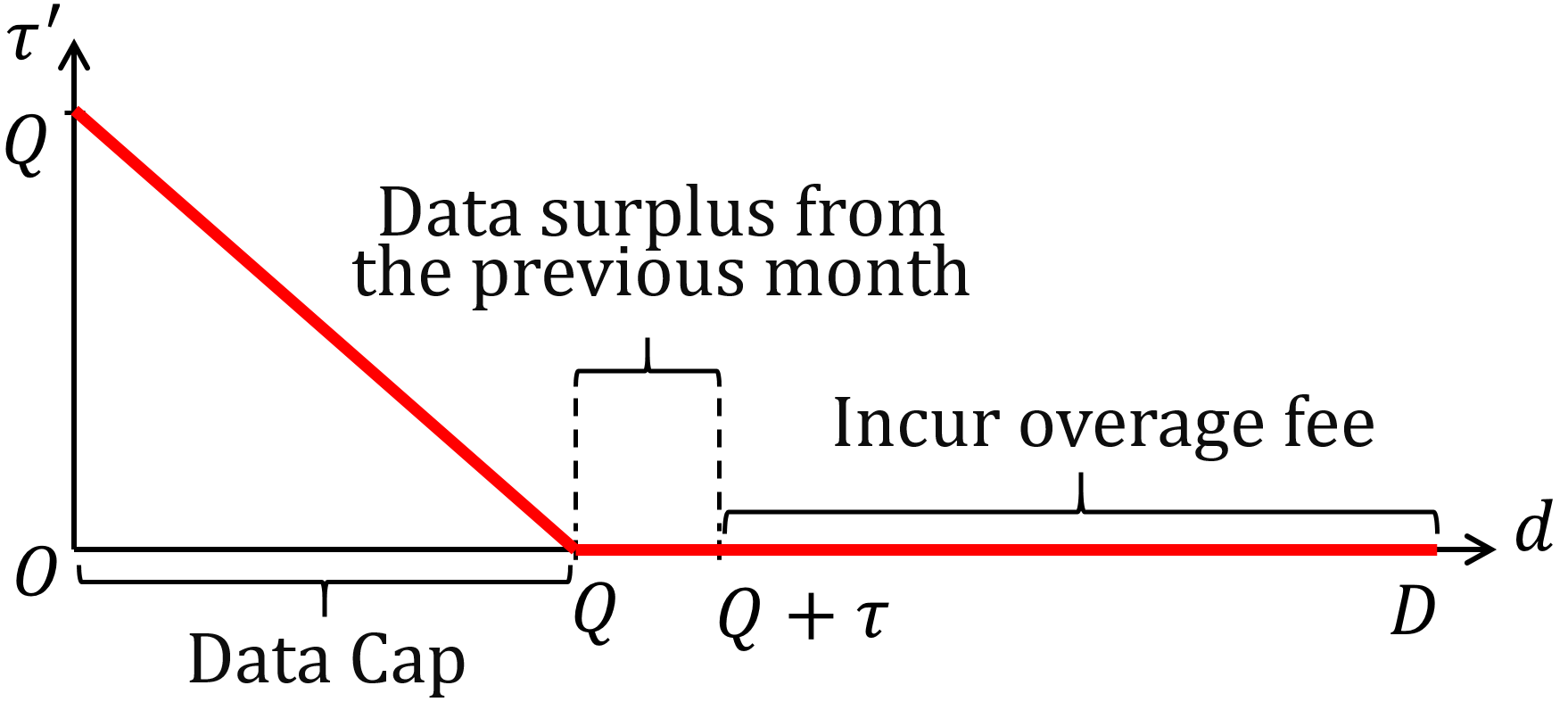}} \quad
	\subfigure[$\mechanism=2$]{\label{fig:Usage-T2}\includegraphics[width=0.32\linewidth]{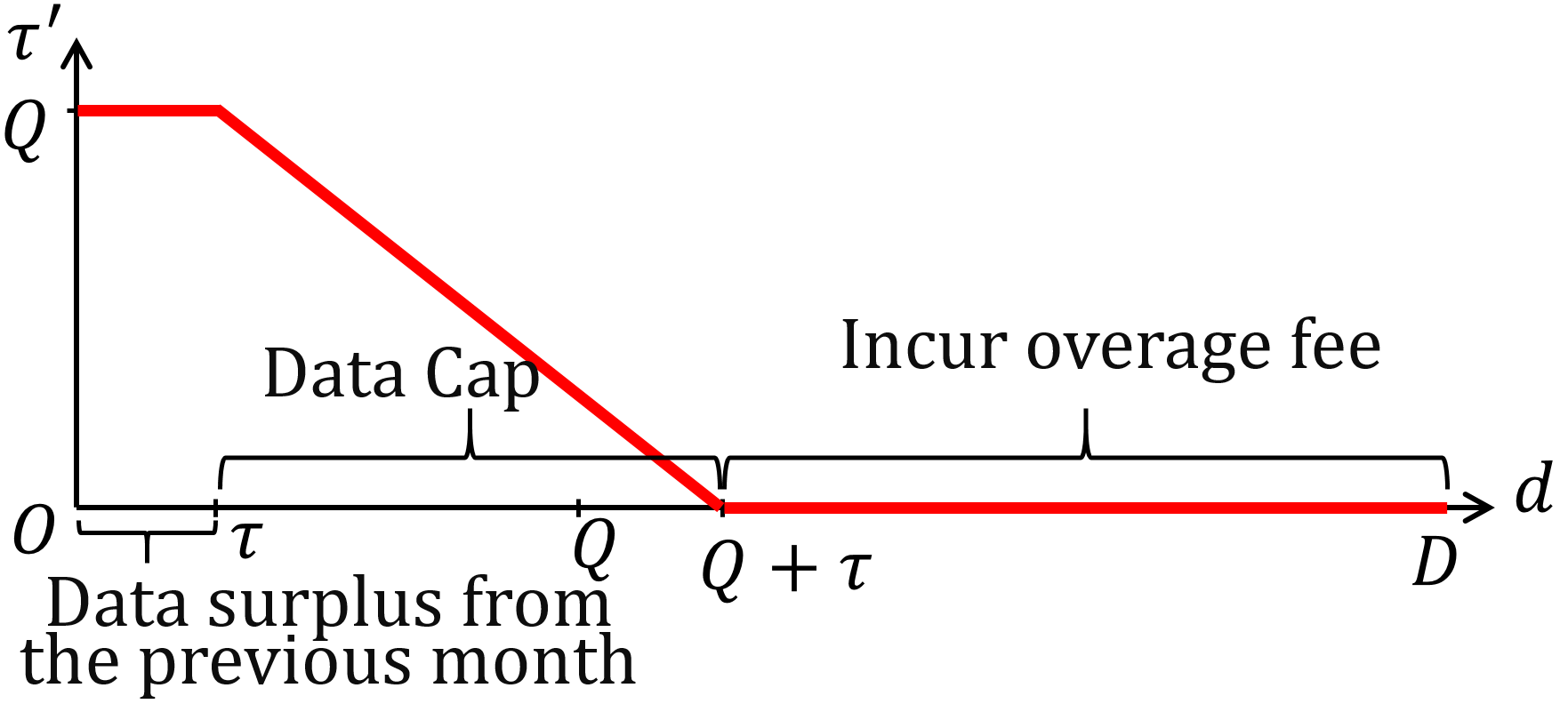}} \quad
	\subfigure[$\mechanism=3$]{\label{fig:Usage-Tc}\includegraphics[width=0.312\linewidth]{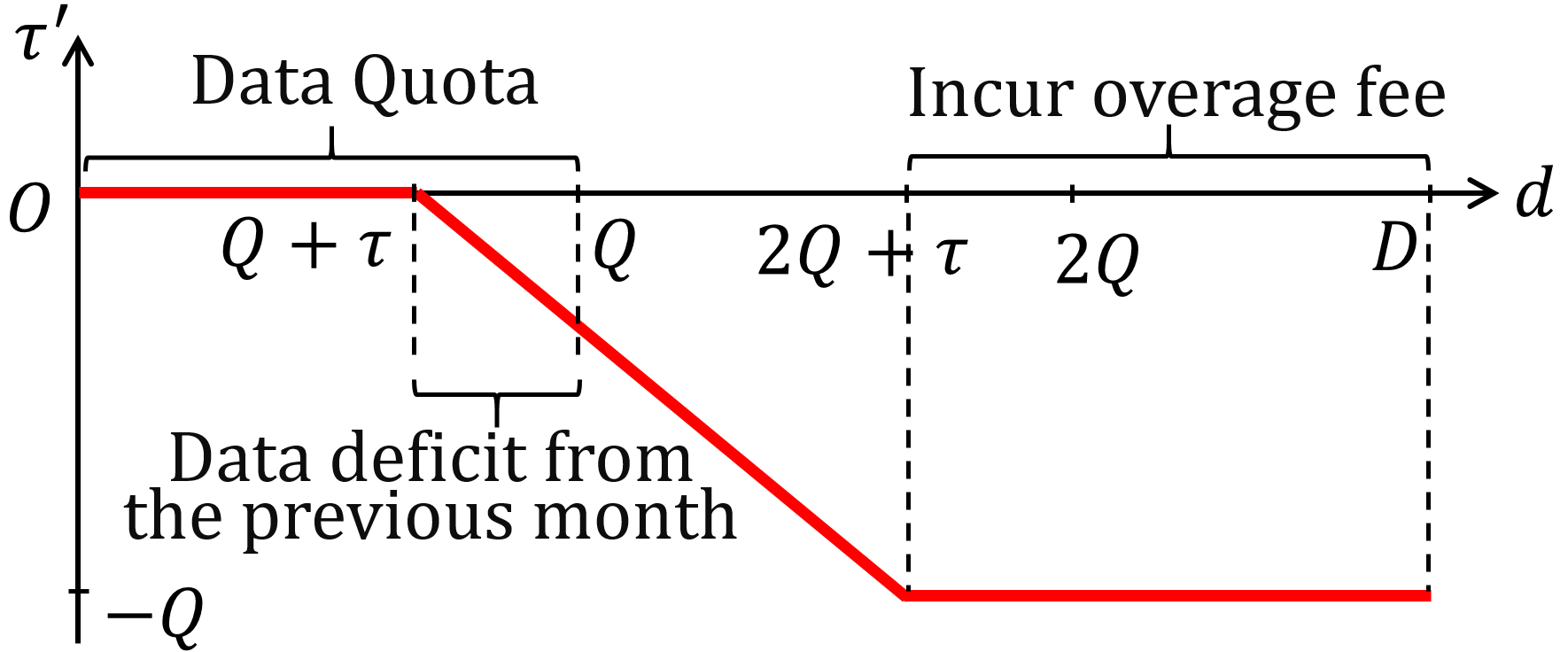}}
	\caption{Transition from $\spedata$ (data surplus or deficit from the previous month) to $\spedata'$ (data surplus or deficit of the next month).\vspace{-10pt}}
	\label{fig: Usage}
\end{figure*}

\section{Data Mechanisms and Time Flexibility\label{Section: Data Mechanisms and Time Flexibility}}
Recall that the numerical example in Table \ref{table: Example} shows that the total payments for $\mechanism=2,3$ are the same and the least, while the payment for $\mechanism=0$ is the most.
In this section we will demonstrate that \textit{it is not a coincidence, but a general conclusion reflecting  the data mechanisms' degrees of time flexibility}.

In the following, we introduce how to compute $A_\mechanism(\dcap)$ in Section \ref{Subsection: Data Mechanisms}, then answer \textit{\textbf{Question \ref{Question: most flexible}}} (i.e., which data mechanism is the most time-flexible) in Section \ref{Subsection: Time Flexibility}.

\subsection{Data Mechanisms\label{Subsection: Data Mechanisms}}
As mentioned in Section \ref{Subsection: User Model}, the key difference among the four data mechanisms is the distribution of the subscriber's data surplus or deficit $p_\mechanism(\tau)$, which further determines $A_\mechanism(\dcap)$ according to (\ref{Equ: system model A_i(Q_i)}).
Particularly, for $\mechanism=0$, the data surplus or deficit is always zero, i.e., $\tau=0$, since it does not offer subscribers any special data.
However, for $\mechanism\in\{1,2,3\}$, we need to consider the data demand dynamic between successive months.

We illustrate the transition of users' data surplus or deficit $\tau$ between two successive months in Fig. \ref{fig: Usage}.
Specifically, the horizontal axis corresponds to users' random data demand $d\in[0,\dmax]$, and the vertical axis represents users' data surplus or deficit $\tau'$ for the next month (given his data surplus or deficit $\tau$ in the current month and the data demand $d$).
The differences among the three red curves in Fig. \ref{fig: Usage} indicate the differences among the three data mechanisms $\mechanism\in\{1,2,3\}$.

In the following, we analyze the distribution $p_\mechanism(\spedata)$ and compute $A_\mechanism(\dcap)$ under the four data mechanisms.

\subsubsection{Traditional Data Mechanism $\mechanism=0$}
For a $\plan=\{\dcap,\pcap,\adfee,0\}$ subscriber, there is no special data to use, i.e., $\spedata=0$ and $\dcap_0^e(\spedata)=\dcap$. Thus we only need to take the expectation over the data demand $d$.
Thus $A_0(\dcap)$ is  
\begin{equation}\label{Equ: A 0}
	A_0(\dcap)=\textstyle\sum\limits_{d=0}^{\dmax}\left[d-\dcap\right]^+f(d).
\end{equation}

\subsubsection{Rollover Data Mechanism $\mechanism=1$}
For a $\plan=\{\dcap,\pcap,\adfee,1\}$ subscriber, the special data is the rollover data from the previous month, which is consumed \textit{after} the current monthly data cap.
Therefore, the effective data cap consists of the monthly data cap and the rollover data surplus $\spedata\in[0,\dcap]$, i.e., $\dcap_1^e(\spedata)=\dcap+\spedata$. 
Fig. \ref{fig:Usage-T1} plots the rollover data to the next month, denoted by $\spedata'$, versus the subscriber's data demand $d$ in the current month.
Thus we know
\begin{equation}
\spedata'=\left\{
\begin{aligned}
& \dcap-d,	& \text{if }d & <\dcap, \\
& 0	,			& \text{if }d & \ge \dcap.
\end{aligned}
\right.
\end{equation}

Note that the rollover data to the next month $\spedata'$ only depends on the subscriber's monthly data cap $\dcap$ and the data demand $d$, and  is independent of the data surplus $\spedata$ from the previous month. 
Therefore, the probability mass function $p_1(\spedata)$ is
\begin{equation}
p_1(\spedata)=\left\{
\begin{aligned}
& f(\dcap-\spedata),									& \text{if }\spedata & \in(0,\dcap], \\
& \textstyle\sum_{d=\dcap}^{\dmax}f(d),		& \text{if }\spedata & =0.
\end{aligned}
\right.
\end{equation}

Then we need to take the expectation over the data demand $d$ and the rollover data surplus $\spedata$ to compute the expected overage usage $A_1(\dcap)$, as follows:
\begin{equation}\label{Equ: A 1}
	A_1(\dcap)=\textstyle \sum\limits_{\spedata=0}^{\dcap}	\sum\limits_{d=0}^{\dmax}	[d-\dcap_1^e(\spedata)]^+f(d)p_1(\spedata).
\end{equation}

\subsubsection{Rollover Data Mechanism $\mechanism=2$}
For a $\plan=\{\dcap,\pcap,\adfee,2\}$ subscriber, the special data is the rollover data from the previous month, which is consumed \textit{prior} to the current monthly data cap.
Therefore, the effective data cap is the same as that for $\mechanism=1$, i.e., $\dcap_2^e(\spedata)=\dcap+\spedata$. 
However, the rollover data is consumed \textit{prior} to the monthly cap. As showed in Fig. \ref{fig:Usage-T2}, we know
\begin{equation}
\spedata'=\left\{
\begin{aligned}
& \dcap,			& \text{if }d\in & [0,\spedata], \\
& \dcap+\spedata-d,	& \text{if }d\in & (\spedata,\dcap+\spedata), \\
& 0,				& \text{if }d\in & [\dcap+\spedata,\dmax].
\end{aligned}
\right.
\end{equation}

It is notable that  the rollover data to the next month $\spedata'$ depends on the monthly data cap $\dcap$, data demand $d$, and the rollover data surplus $\spedata$ from the previous month, resulting in a Markov property on the rollover data surplus $\spedata$.
The one-step transition probability of the  rollover data surplus $\spedata$ is given by
\begin{equation}
p_2(\spedata,\spedata')=\left\{
\begin{aligned}
& \textstyle\sum_{d=0}^{\spedata}f(d),				& \text{if }\spedata' &=\dcap, \\
& f(\dcap+\spedata-\spedata'),						& \text{if }\spedata' &\in(0,\dcap), \\
& \textstyle\sum_{d=\dcap+\spedata}^{\dmax}f(d),	& \text{if }\spedata' &=0.
\end{aligned}
\right.
\end{equation}

Then we can derive the stationary distribution of the rollover data surplus $\spedata$, denoted by $p_2(\spedata)$, according to the above transition probability \cite{bass2011stochastic}. 
Thus $A_2(\dcap)$ is given by
\begin{equation} \label{Equ: A 2}
	A_2(\dcap)=\textstyle\sum\limits_{\spedata=0}^{\dcap}\sum\limits_{d=0}^{\dmax}\left[d-\dcap_2^e(\spedata)\right]^+f(d)p_2(\spedata).
\end{equation}

\subsubsection{Credit Data Mechanism $\mechanism=3$}
For a $\plan=\{\dcap,\pcap,\adfee,3\}$ subscriber, the special data is the credit data borrowed from the next month, which is used after the current monthly data cap.
Therefore, the effective data cap of a subscriber with a data deficit $\spedata\in[-\dcap_3,0]$ consists of the remaining current monthly data cap (with a deficit $\spedata$) and the maximum credit data that he can borrow from the next month (which is $\dcap$), i.e., $\dcap_3^e(\spedata)=2\dcap+\spedata$. 
According to Fig. \ref{fig:Usage-Tc}, the data deficit in the next month, denoted by $\spedata'$, is given by
\begin{equation}
\spedata'=\left\{
\begin{aligned}
& 0,				& \text{if }d\in & [0,\dcap+\spedata], \\
& \dcap+\spedata-d,	& \text{if }d\in & (\dcap+\spedata,2\dcap+\spedata), \\
& -\dcap,			& \text{if }d\in & [2\dcap+\spedata,\dmax].
\end{aligned}
\right.
\end{equation}

Similar to the case when $\mechanism=2$, we note that the data deficit $\spedata'$ in the next month depends on the monthly data cap $\dcap$, the data demand $d$, and the data deficit $\spedata$ in the current month, which indicates a Markov property on the data deficit $\spedata$ for $\mechanism=3$.
The corresponding one-step transition probability of the data deficit $\spedata$ is
\begin{equation}
p_3(\spedata,\spedata')=\left\{
\begin{aligned}
& \textstyle\sum_{d=0}^{\dcap+\spedata}f(d),			& \text{if }\spedata' &=0, \\
& f(\dcap+\spedata-\spedata'),							& \text{if }\spedata' &\in(-\dcap,0), \\
& \textstyle\sum_{d=2\dcap+\spedata}^{\dmax}f(d),		& \text{if }\spedata' &=-\dcap.
\end{aligned}
\right.
\end{equation}

Similarly we can derive the stationary distribution $p_3(\spedata)$ of the data deficit $\spedata$ and compute $A_3(\dcap)$ as follows
\begin{equation}\label{Equ: A 3}
	A_3(\dcap)=\textstyle\sum\limits_{\spedata=-\dcap}^{0}\sum\limits_{d=0}^{\dmax}[d-\dcap_3^e(\spedata) ]^+f(d)p_3(\spedata).
\end{equation}

Now that we have demonstrated how to compute $A_\mechanism(\dcap)$ under the four data mechanisms, next we will compare the degree of time flexibility based on $A_\mechanism(\dcap)$.

\subsection{Time Flexibility\label{Subsection: Time Flexibility}}
In the following we compare the degrees of time flexibility among the four data mechanisms and answer \textit{\textbf{Question \ref{Question: most flexible}}}.
\begin{definition}[Time Flexibility] \label{Defination: Time Flexibility}
	Consider two data mechanisms $i,j\in\{0,1,2,3\}$.
	The data mechanism $i$ has a better time flexibility than the data mechanism $j$, denoted by
	$\flexibility_i>\flexibility_j$, if and only if for an arbitrary data demand distribution $f(d)$, we have $A_i(\dcap)<A_j(\dcap)$ for all $\dcap\in(0,\dmax)$. 
\end{definition}

\textit{Definition \ref{Defination: Time Flexibility}} uses the type-$(\val,0)$ subscriber's expected overage data consumption $A_\mechanism(\dcap)$ to indicate the time flexibility of the data mechanism $\mechanism$.
Intuitively, the better time flexibility the data mechanism offers, the less overage usage is incurred by its subscribers under the same data cap $\dcap$ for an arbitrary data demand distribution $f(d)$.
Note that we require $\dcap\in(0,\dmax)$ in the definition, this is because that the three-part tariff with time flexibility degenerates into the pure usage-based data plan if $\dcap=0$ or the flat-rate data plan if $\dcap\ge\dmax$. 
In these two extreme cases, any data mechanism $\mechanism$ has no impact.

Lemma \ref{Lemma: time flexibility comparison} summarizes the time flexibility of the four data mechanisms.
The proof is given in Appendix \ref{Proof: comparison}.
\begin{lemma}\label{Lemma: time flexibility comparison}
	For an arbitrary data demand distribution $f(d)$, we have $A_0(\dcap)>A_1(\dcap)>A_2(\dcap)=A_3(\dcap)$ for all $\dcap\in(0,\dmax)$.
	Therefore, the four data mechanisms' degrees of time flexibility satisfy   $\flexibility_0<\flexibility_1<\flexibility_2=\flexibility_3$.
\end{lemma}

Lemma \ref{Lemma: time flexibility comparison} provides  the answer to \textbf{\textit{Question \ref{Question: most flexible}}} that we mentioned  in Section 1.
The rollover data mechanism offered by China Mobile (i.e., $\mechanism=2$) and our proposed credit data mechanism (i.e., $\mechanism=3$) are both  the most time-flexible.
The traditional data mechanism (i.e., $\mechanism=0$) is the least time-flexible.

The reason why $\flexibility_2=\flexibility_3$ is twofolds.
First, the two data mechanisms can expand the effective data cap with the same intensity, i.e., $\dcap_\mechanism^e(\spedata)\in[\dcap,2\dcap]$ for $\mechanism\in\{2,3\}$.
Second, the consumption priorities of the two data mechanisms specify that subscribers should first consume the earlier data, i.e., the rollover data prior to the current monthly data cap for $\mechanism=2$, and the current monthly data cap prior to the credit data for $\mechanism=3$.

The reason why $\flexibility_1<\flexibility_2$ is because of the {irregular} consumption priority for $\mechanism=1$.
Recall that the rollover data mechanism offered by AT\&T (i.e., $\mechanism=1$) requires that the current monthly data cap is consumed prior to the rollover data from the previous month, which means the later data (i.e., current monthly data cap) would be consumed prior to the earlier data (i.e., rollover data from the previous month).
Such an irregular consumption priority prevents subscribers from fully utilizing their data quota in the long run, hence reduces the degree of time flexibility.
Nevertheless, it is still time flexible than the traditional data mechanism, i.e., $\flexibility_0<\flexibility_1$.

So far we have compared the degree of time flexibility among the four data mechanisms.
Next in Section \ref{Section: Backward Induction of the Three-stage Game}, we analyze the three-stage game.

\begin{figure*} 
	\centering
	\setlength{\abovecaptionskip}{2pt}
	\subfigure[{Case 1: $\adfee \left[\dmean-A_\mechanism(\dcap)\right] > \pcap $}]{\label{fig: subscription_down}\includegraphics[width=0.25\linewidth]{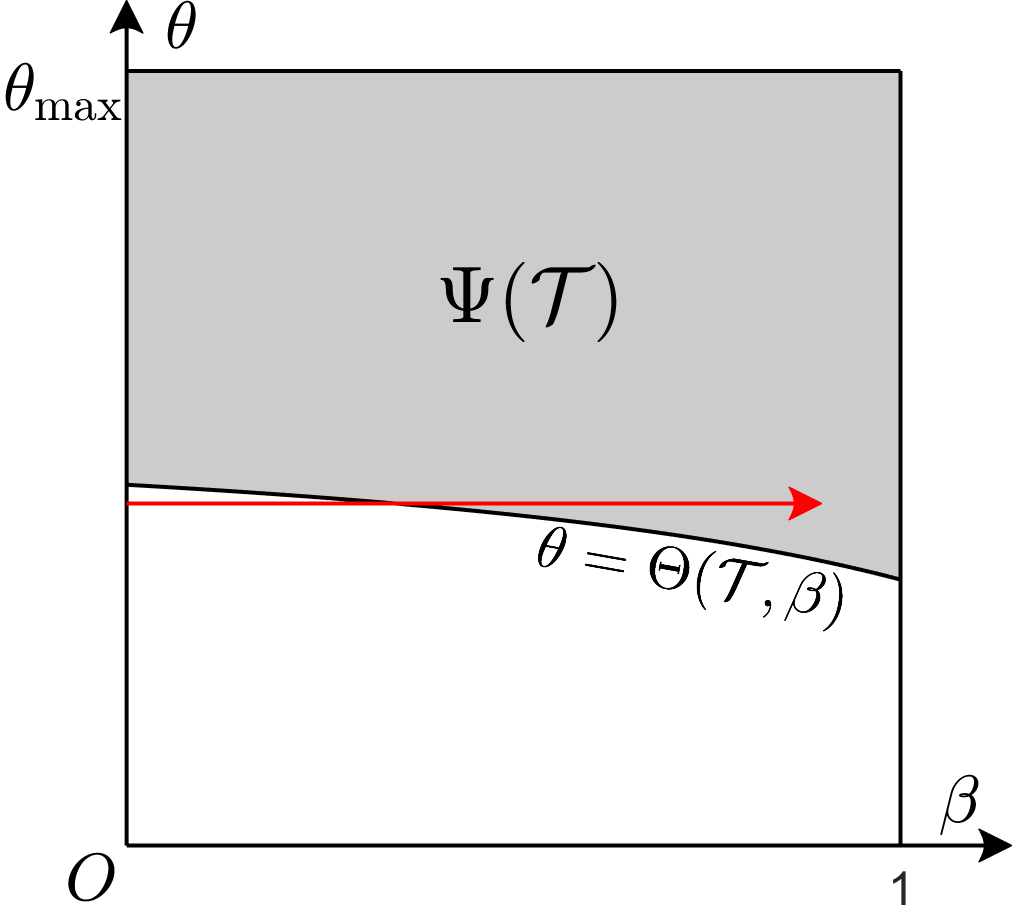}}\qquad\quad
	\subfigure[{Case 2: $\adfee \left[\dmean-A_\mechanism(\dcap)\right] < \pcap $}]{\label{fig: subscription_up}\includegraphics[width=0.25\linewidth]{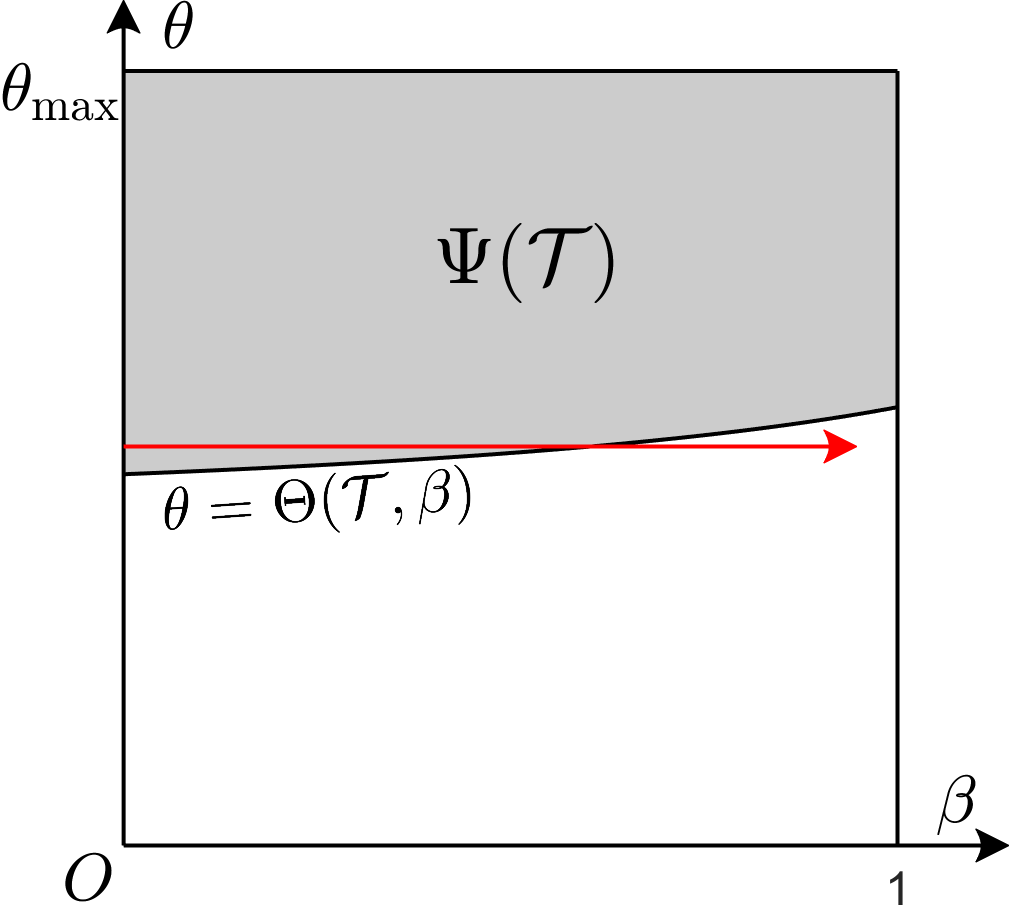}}\qquad\quad
	\subfigure[{Case 3: $\adfee \left[\dmean-A_\mechanism(\dcap)\right] = \pcap $}]{\label{fig: subscription_middle}\includegraphics[width=0.25\linewidth]{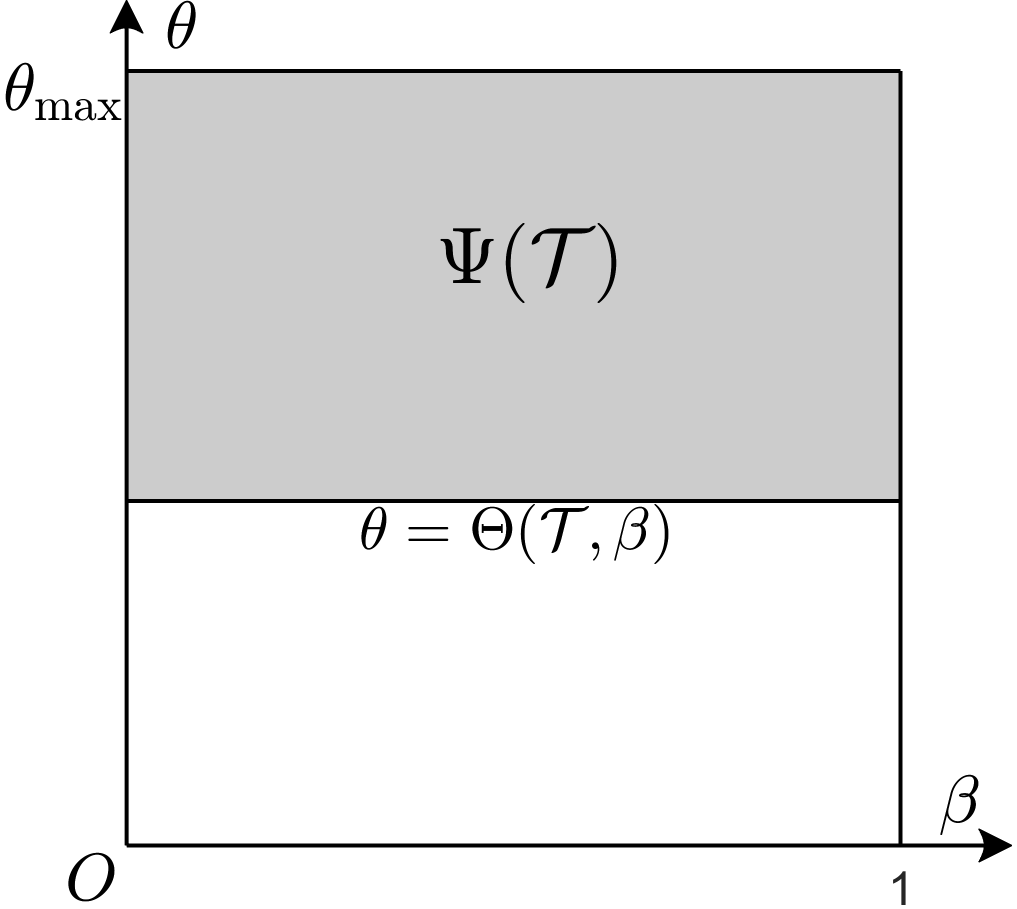}} 
	\caption{Illustration of different market partitions. Gray region: the subscribers $\subscription(\plan)$.\vspace{-10pt}}
	\label{fig: Users' subscription}
\end{figure*}


\section{Backward Induction of the Three-stage Stackelberg Game\label{Section: Backward Induction of the Three-stage Game}}
In this section, we study the Subgame Perfect Equilibrium (SPE, or simply referred to as  equilibrium in this paper) of the three-stage Stackelberg game by backward induction.

\subsection{User's Subscription in Stage III\label{subSection: Stage III}}
In Stage III, each user makes his subscription choice given the data plan $\plan=\{\dcap,\pcap,\adfee,\mechanism\}$ offered by the MNO.


The type-$(\val,\cut)$ user has two choices. 
If he does not subscribe, his payoff will be zero.\footnote{Here we normalize the non-subscription payoff to be zero. If the user has other options, for example, relying purely on Wi-Fi networks, the non-subscription payoff can be positive. Our analysis will still go through in that case with a simple constant shift.} 
Hence the user  will subscribe to the MNO if and only if $\plan=\{\dcap,\pcap,\adfee,\mechanism\}$ brings him a non-negative expected monthly payoff, i.e.,  $\payoffexp(\plan,\val,\cut)\ge0$.
Theorem \ref{Theorem: Market Share} presents the MNO's market share under the data plan $\plan$.
The proof is given in Appendix \ref{Proof: Market Share}.
\begin{theorem}[Market Share]\label{Theorem: Market Share}
	The MNO's market share under data plan $\plan=\{\dcap,\pcap,\adfee,\mechanism\}$ is 
	\begin{equation}
	\subscription(\plan)=\{ (\val,\cut): \Val(\plan,\cut)\le \val \le \valmax, 0\le\cut\le1 \},
	\end{equation}
	where $\Val(\plan,\cut)$ is referred to as the \textit{threshold valuation} under $\plan$ and $\cut$, which is given by 
	\begin{equation}\label{Equ: threshold valuation}
	\Val(\plan,\cut)  \eq  \frac{1}{\QoS}\left[\adfee  + \frac{\adfee\left[\dmean-A_\mechanism(\dcap)\right]-\pcap}{\cut A_\mechanism(\dcap)-\dmean} \right] . 
	\end{equation}	
\end{theorem}

Based on Theorem \ref{Theorem: Market Share}, we further summarize the impacts of users' characteristics $\val$ and $\cut$ on their subscription decisions in \textit{Corollary \ref{Corollary: characteristics val}} and \textit{Corollary \ref{Corollary: characteristics cut}}, respectively.
\begin{corollary}\label{Corollary: characteristics val}[Impact of Data Valuation]
	Given the network substitutability $\cut$, it is more likely for a user to subscribe to the MNO as his data valuation $\val$ increases.
\end{corollary}

\textit{Corollary \ref{Corollary: characteristics val}} indicates that higher valuation users are more likely to subscribe to the MNO.
However, \textit{Corollary \ref{Corollary: characteristics cut}} shows that the impact of network substitutability is more complicated.

\begin{corollary}\label{Corollary: characteristics cut}[Impact of Network Substitutability]
	Given the data valuation $\val$, there are three possibilities for the impact of network substitutability $\cut$:
	\begin{itemize}
		\item Case 1 ($\adfee[\dmean-A_\mechanism(\dcap)]>\pcap$): As the network substitutability improves, users are \textit{more likely} to subscribe to the MNO. 
		\item Case 2 ($\adfee[\dmean-A_\mechanism(\dcap)]<\pcap$): As the network substitutability improves, users are \textit{less likely} to subscribe to the MNO.
		\item Case 3 ($\adfee[\dmean-A_\mechanism(\dcap)]=\pcap$): The network substitutability \textit{does not affect} the subscription decision.	 
	\end{itemize} 
\end{corollary}

In \textit{Corollary \ref{Corollary: characteristics cut}}, $\dmean-A_\mechanism(\dcap)$ represents the total data consumption of a type-$(\val,1)$ subscriber.
This type of subscriber has so good a network substitutability that he stops using mobile data after his effective data cap is used up.
Thus he only needs to pay the subscription fee $\pcap$ in each month.
Accordingly, $\pcap/\left[\dmean-A_\mechanism(\dcap)\right]$ represents the average payment per unit data that he uses in each month.

Therefore, Case 1 in \textit{Corollary \ref{Corollary: characteristics cut}} represents that the per-unit fee $\adfee$ is more expensive compared with the subscription fee $\pcap$ in terms of the average rate of a type-$(\val,1)$ subscriber; 
Case 2 is the opposite of Case 1; 
Case 3 represents that the per-unit fee $\adfee$, and the subscription fee $\pcap$ are comparable for type-$(\val,1)$ subscribers.

Now we illustrate \textit{Corollary \ref{Corollary: characteristics cut}} in Fig. \ref{fig: Users' subscription}, where the two axises correspond to the user's network substitutability $\cut$ and data valuation $\val$, respectively.
The gray region denotes the subscriber set $\subscription(\plan)$.
The red arrow represents the direction where the network substitutability $\cut$ increases.
\begin{itemize}
	\item Fig. \ref{fig: subscription_down}: If the per-unit fee $\adfee$ is more expensive than the average payment per unit data, i.e., $\adfee[\dmean-A_\mechanism(\dcap)]>\pcap$, then the red arrow shows that the users with a better network substitutability are more likely to become subscribers, since they incur less overage usage and thus  less additional payment.
	\item Fig. \ref{fig: subscription_up}: If the average payment for unit data is more expensive than the per-unit fee $\adfee$, i.e., $\adfee[\dmean-A_\mechanism(\dcap)]<\pcap$, then the red arrow shows that the users with a better network substitutability are less likely to become subscribers.
	This is because that they are not willing to pay for an expensive subscription fee, considering the cheap per-unit fee and their good alternative networks.
	\item Fig. \ref{fig: subscription_middle}: If the average payment for unit data and the per-unit fee $\adfee$ are comparable and satisfy $\adfee[\dmean-A_\mechanism(\dcap)]=\pcap$, then the network substitutability does not change users' subscription decisions.
\end{itemize}


Later on we will show that under the MNO's optimal pricing strategy, the market partition is Fig. \ref{fig: subscription_middle}.

Considering the subscription choice derived from  Theorem \ref{Theorem: Market Share}, the MNO's expected total profit is given by 
\begin{equation}\label{Equ: Stage II expected profit}
\begin{aligned}
\tilde{\profit}(\plan)=
& \int_{0}^{1}  \int_{\Val(\plan,\cut)}^{\valmax} \Big\{ 
\underbrace{\adfee(1-\cut)A_\mechanism(\dcap) +\pcap}_\text{Revenue} \\
&  - \underbrace{c  \left[ \dmean-\cut A_\mechanism(\dcap) \right] }_\text{Operational cost}
\Big\} h(\val)g(\cut) d\val d\cut  - \underbrace{J(\dcap).}_\text{Capacity cost}
\end{aligned}
\end{equation}

Next we will further analyze the MNO's optimal data cap and pricing strategy in Stage II.


\subsection{Optimal Data Cap and Pricing Strategy in Stage II\label{subSection: Stage II}}
In Stage II, the MNO determines the profit-maximizing data cap $\dcap^*$ and the pricing strategy $\{\pcap^*,\adfee^*\}$ considering users' subscription decisions from Stage III, given the data mechanism $\mechanism$ obtained in Stage I.



{To make the presentation clear and reveal the key insights,  we first present the MNO's optimal pricing strategy $\{\pcap^*,\adfee^*\}$ given the data cap $\dcap$ in Section \ref{Subsubsection: Optimal Pricing Strategy}, and then we introduce the MNO's optimal data cap $\dcap^*$ in Section \ref{Subsubsection: Optimal Data Cap}.
\footnote{The two-step presentation enables us to illustrate the key insights of the optimal pricing strategy for a particularly data cap. This is practically important, since the MNO usually adopt integral data caps for simplicity, e.g., 1GB, 2GB, and 3GB.}
}

\subsubsection{Optimal Pricing Strategy\label{Subsubsection: Optimal Pricing Strategy}}
Given the data cap $\dcap$, the objective of MNO is to find the optimal subscription fee $\pcap^*$ and per-unit fee $\adfee^*$ that maximize its expected total profit, that is,
\begin{problem}[Optimal Pricing Strategy] \label{Problem: Optimal Pricing Policy}
	\begin{equation}
	\begin{aligned}
	\big\{\pcap^*,\adfee^*\big\} =& \arg\max\limits_{\pcap,\adfee\ge0} \tilde{\profit}(\dcap,\pcap,\adfee,\mechanism).
	\end{aligned}
	\end{equation}
\end{problem}



Before analyzing \textit{Problem \ref{Problem: Optimal Pricing Policy}}, we need to introduce the following assumption on the MNO's QoS $\QoS$ and marginal operational cost $c$.
\begin{assumption}\label{Assumption: benefit}
	The MNO's QoS $\QoS$ and marginal operational cost $c$ satisfy $c<\QoS\cdot\valmax$.
\end{assumption}

{Assumption \ref{Assumption: benefit} is made to avoid a trivial case, where the MNO offers very poor wireless service such that it cannot benefit from the market. 
It is not a technical assumption that limits our analysis and results.
}

Now we characterize the MNO's profit-maximizing subscription fee $\pcap^*$ and per-unit fee $\adfee^*$ in Theorem \ref{Theorem: Optimal Pricing}.
\begin{theorem}[Optimal Pricing Strategy]\label{Theorem: Optimal Pricing}
	Given the data cap $\dcap\ge0$ and the data mechanism $\mechanism$, the MNO's profit-maximizing subscription fee $\pcap^*$ and per-unit fee $\adfee^*$ satisfying the following conditions:
	\begin{equation}
	\left\{
	\begin{aligned}
	&   H\left( \frac{\adfee^*}{\QoS} \right) + \frac{\adfee^*-c}{\QoS}\cdot h\left( \frac{\adfee^*}{\QoS} \right) =1, \\ 
	& \pcap^*=\adfee^*\left[ \dmean-A_\mechanism(\dcap) \right] ,
	\end{aligned}
	\right.
	\end{equation}
	where $h(\cdot)$ and $H(\cdot)$ are the PDF and CDF of the data valuation $\val$. Furthermore,  $\adfee^*$ is unique for an arbitrary $\val$ distribution with  an increasing failure rate (IFR).\footnote{The increasing failure rate condition refers to  $h(\val)/\left[1-H(\val)\right]$ increasing in $\val$. Many commonly used distributions, such as uniform distribution, gamma distribution, and normal distribution, satisfy this condition \cite{brusset2009properties}.} 
%

\end{theorem}

The proof of Theorem \ref{Theorem: Optimal Pricing} is given in Appendix \ref{Proof: Optimal Pricing}.

Based on Theorem \ref{Theorem: Optimal Pricing}, we summarize the impact of several parameters on the optimal per-unit fee $\adfee^*$ and the optimal subscription fee $\pcap^*$ in Proposition \ref{Proposition: optimal pi} and {Proposition \ref{Proposition: optimal PI}, respectively.
\begin{proposition}\label{Proposition: optimal pi}
The optimal per-unit fee $\adfee^*$ increases in  the MNO's {QoS} $\QoS$ and {marginal operational cost} $c$.
It dose not depend on the data mechanism $\mechanism$  or how large the data cap $\dcap$ is.
\end{proposition}


\begin{proposition}\label{Proposition: optimal PI}
The optimal subscription fee $\pcap^*(\dcap,\mechanism)$ is related to the {data cap} $\dcap$ and the {data mechanism} $\mechanism$ in the following ways,
\begin{itemize}
	\item $\pcap^*(\dcap,\mechanism)$ increases in the data cap $\dcap$,
	\item a better time flexibility corresponds to a higher subscription fee, i.e., $\pcap^*(\dcap,0)<\pcap^*(\dcap,1)<\pcap^*(\dcap,2)=\pcap^*(\dcap,3)$ for all $\dcap\in(0,\dmax)$.
\end{itemize}
\end{proposition}

Recall that Corollary \ref{Corollary: characteristics cut} summarizes three possibilities for the impact of network substitutability $\cut$.
Moreover, the optimal subscription fee $\pcap^*$ and the per-unit fee $\adfee^*$ derived in Theorem \ref{Theorem: Optimal Pricing} satisfy $\pcap^*=\adfee^*\left[ \dmean-A_\mechanism(\dcap) \right]$, which is the same as Case 3  discussed in Corollary \ref{Corollary: characteristics cut}.
In this case, the threshold valuation (defined in (\ref{Equ: threshold valuation})) is
\begin{equation}
\textstyle \Val(\dcap,\pcap^*,\adfee^*,\mechanism,\cut)=\frac{\adfee^*}{\QoS},
\end{equation}
which is independent of $\cut$.
This naturally leads to the following corollary on the market partition under the optimal pricing strategy.
\begin{corollary}\label{Corollary: market of optimal pricing}
	Under the optimal pricing strategy specified in Theorem \ref{Theorem: Optimal Pricing}, the network substitutability $\cut$ does not  affect the subscription decision. 
	That is, the type-$(\val,\cut)$ user will subscribe to the MNO if and only if $\val>{\adfee^*}/{\QoS}$.
	The MNO obtains the  market of 
	\begin{equation}
	\subscription\left( \dcap,\pcap^*,\adfee^*,\mechanism \right)=\Big\{(\val,\cut):\textstyle\frac{\adfee^*}{\QoS}\le \val \le \valmax, 0\le\cut\le1 \Big\}.
	\end{equation}
\end{corollary}



Corollary \ref{Corollary: market of optimal pricing} reveals that the MNO tends to select its subscribers based only on their data valuations, while ignoring  the network substitutability.
The intuition behind such a pricing strategy is that the MNO can benefit from good network substitutability users' subscription fee and poor network substitutability users' overage fee.
Therefore, there is no incentive for the MNO to exclude either type of users.

Furthermore, according to Corollary \ref{Corollary: market of optimal pricing}, the MNO's {market share} under the optimal pricing strategy is fixed for any data cap $\dcap$ and data mechanism $\mechanism$.
However, different data caps and data mechanisms bring the MNO different {profit}.
Substitute $\pcap^*(\dcap,\mechanism)$ and $\adfee^*$ into the MNO's expected total profit, then we obtain 
\begin{equation}\label{Equ: Profit Q mechanism}
\begin{aligned}
& \tilde{\profit}(\dcap,\pcap^*(\dcap,\mechanism),\adfee^*,\mechanism ) \\
=& \frac{ \left[\dmean-\bar{\cut}A_\mechanism(\dcap)\right] \left(\adfee^*-c\right)^2 }{\QoS} h\left(\frac{\adfee^*}{\QoS}\right)-J(\dcap) ,
\end{aligned}
\end{equation}
where $\bar{\cut}$ is the mean of the network substitutability among the whole user market, which is given by
\begin{equation}
\bar{\cut}=\int_{0}^{1}\cut g(\cut)d\cut.
\end{equation}

Next we move on to analyze the MNO's optimal data cap, considering the pricing strategy derived in Theorem \ref{Theorem: Optimal Pricing}.

\subsubsection{Optimal Data Cap\label{Subsubsection: Optimal Data Cap}}
The MNO needs to select a data cap $\dcap$ to maximize its expected total profit $\tilde{\profit}(\dcap,\pcap^*(\dcap,\mechanism),\adfee^*,\mechanism )$.
That is, the MNO needs to solve the following problem
\begin{problem}[Optimal Data Cap] \label{Problem: Optimal Data Cap}
	\begin{eqnarray}
		\dcap^*=\arg\max\limits_{\dcap\ge0} \tilde{\profit}(\dcap,\pcap^*(\dcap,\mechanism),\adfee^*,\mechanism ),
	\end{eqnarray}
\end{problem}
where $\pcap^*(\dcap,\mechanism)$ and $\adfee^*$ are the MNO's profit-maximizing subscription fee and per-unit fee obtained from Theorem \ref{Theorem: Optimal Pricing}, respectively.

\textit{Problem \ref{Problem: Optimal Data Cap}} is not difficult to solve, since it is a single variable problem and is convex if $J(\dcap)$ is convex.
To illustrate the key insights of the optimal data cap, hereafter, we follow \cite{dai2015effect} by making the following assumption on the MNO's capacity cost $J(\dcap)$  throughout the rest of the paper.
\begin{assumption}\label{Assumption: capacity cost}
	The MNO's capacity cost takes a linear form, i.e., $J(\dcap) = z\cdot\dcap$ where $z$ is the marginal capacity cost.
\end{assumption}

Before analyzing the MNO's optimal data cap, recall that the differences among the four data mechanisms $\mechanism\in\{0,1,2,3\}$ is entirely captured in $A_\mechanism(\dcap)$, the expected overage data consumption.
To facilitate our later analysis, we refer to  $|A_\mechanism'(\dcap)|$ as the \textit{\textbf{marginal overage data consumption}}, which is the absolute value of the derivative of $A_\mechanism(\dcap)$ with respect to $\dcap$.
Specifically, the marginal overage data consumption $|A_\mechanism'(\dcap)|$ measures the overage data usage decrement for a unit data cap increment on the data cap $\dcap$ under the data mechanism $\mechanism$.

Now we characterize the MNO's optimal data cap in Theorem \ref{Theorem: Optimal Data Cap}.
The proof of Theorem \ref{Theorem: Optimal Data Cap} is given in Appendix \ref{Proof: Optimal Data Cap}.
\begin{theorem}[Optimal Data Cap]\label{Theorem: Optimal Data Cap}
	Given the data mechanism $\mechanism$, the MNO's optimal data cap  $\dcap^*(\mechanism)$ satisfies 
	\begin{equation}\label{Equ: optimal data cap condition}
		 |A_\mechanism'\left( \dcap^*(\mechanism) \right)| =  \tdr(\QoS,c,z),
	\end{equation}
	where $\tdr(\QoS,c,z)$ is given by
	\begin{equation}
		\tdr(\QoS,c,z)=\frac{ z \cdot h\left(\frac{\adfee^*}{\QoS}\right)  }{ \bar{\cut}\QoS \left[ 1-H\left(\frac{\adfee^*}{\QoS}\right) \right]^2 }.
	\end{equation}
	
\end{theorem}

Theorem \ref{Theorem: Optimal Data Cap} indicates that no matter which data mechanism $\mechanism$ that the MNO adopts, the MNO should always choose a data cap  such that the corresponding \textit{marginal overage data consumption} equals to $\tdr(\QoS,c,z)$.
Therefore, we refer to $\tdr(\QoS,c,z)$ as the \textit{\textbf{target marginal overage data consumption}} that the MNO must achieve to maximize its profit.
Note that the target marginal overage data consumption $\tdr(\QoS,c,z)$ is related to the MNO's QoS $\QoS$, marginal operational cost $c$, and the marginal capacity cost $z$.
Therefore, we further summarize how the three parameters affect the optimal data cap in Proposition \ref{Proposition: optimal Q}.
\begin{proposition}\label{Proposition: optimal Q}
	The MNO's optimal data cap $\dcap^*$ increases in the QoS $\QoS$, meanwhile decreases in  the MNO's marginal operational cost $c$ and marginal capacity cost $z$.
\end{proposition}

Now we have characterized the optimal data cap in Theorem \ref{Theorem: Optimal Data Cap}, and revealed how the MNO's QoS (i.e., $\QoS$) and marginal costs (i.e., $c$ and $z$) affect it.
Next we analyze the impact of the data mechanism $\mechanism$ on the optimal data cap, which is related to \textit{Question \ref{Question: impact}} mentioned  in Section 1.

Intuitively, we would think  that the MNO can set a smaller cap under a data mechanism with a better time flexibility (to reduce the capacity cost), since it is  more time-flexible for subscribers.
In the following, however, we will reveal a counter-intuitive insight, i.e., \textit{a better time flexibility does not necessarily correspond to a smaller data cap}.

To fully reveal this counter-intuitive insight, we need to know the detail mathematical expression of  $A_\mechanism(\cdot)$ according to (\ref{Equ: optimal data cap condition}) in Theorem \ref{Theorem: Optimal Data Cap}.
Even though we cannot analytically compute $A_\mechanism(\cdot)$ due to the complexity of the Markov transition matrix, we are able to characterize some properties of $A_\mechanism(\cdot)$ in Lemma \ref{Lemma: Property A(Q)}.
The proof is given in Appendix \ref{Proof: Property A(Q)}.
\begin{lemma}\label{Lemma: Property A(Q)}
	For an arbitrary data demand distribution $f(d)$, $A_\mechanism(\dcap)$ is decreasing and convex in $\dcap$.
	Moreover, $A_\mechanism(0)=\dmean$, $A_\mechanism(\dmax)=0$, $A_\mechanism'(0)=-1$, and $A_\mechanism'(\dmax)=0$ for all $\mechanism\in\{0,1,2,3\}$.
\end{lemma}


To illustrate the counter-intuitive insight, let us consider two data mechanisms $i,j\in\{0,1,2,3\}$, where $j$ offers a better time flexibility, i.e., $\flexibility_i<\flexibility_j$.

Based on Lemma \ref{Lemma: Property A(Q)} and \textit{Definition \ref{Defination: Time Flexibility}}, we can  plot  $A_i(\dcap)$ and $A_j(\dcap)$ versus the data cap $\dcap$ in Fig. \ref{fig: A}, and the corresponding marginal overage data consumptions $|A'_i(\dcap)|$ and $|A'_j(\dcap)|$ in Fig. \ref{fig: A_prime1}.\footnote{
Note that in Fig. \ref{fig: A_prime1} there is only one crossing point for $|A'_i(\dcap)|$ and $|A'_j(\dcap)|$ when $\dcap\in(0,\dmax)$. 
Basically, Lemma \ref{Lemma: Property A(Q)} cannot imply the uniqueness of the crossing point for an arbitrary demand distribution $f(d)$.
Nevertheless, it does not affect the counter-intuitive insight, i.e., \textit{a better time flexibility does not necessarily correspond to a smaller data cap}.
}
\begin{figure}
	\centering
	\setlength{\abovecaptionskip}{2pt}
	\subfigure[$A_\mechanism(\dcap)$ vs $\dcap$.]{\label{fig: A}\includegraphics[width=0.45\linewidth]{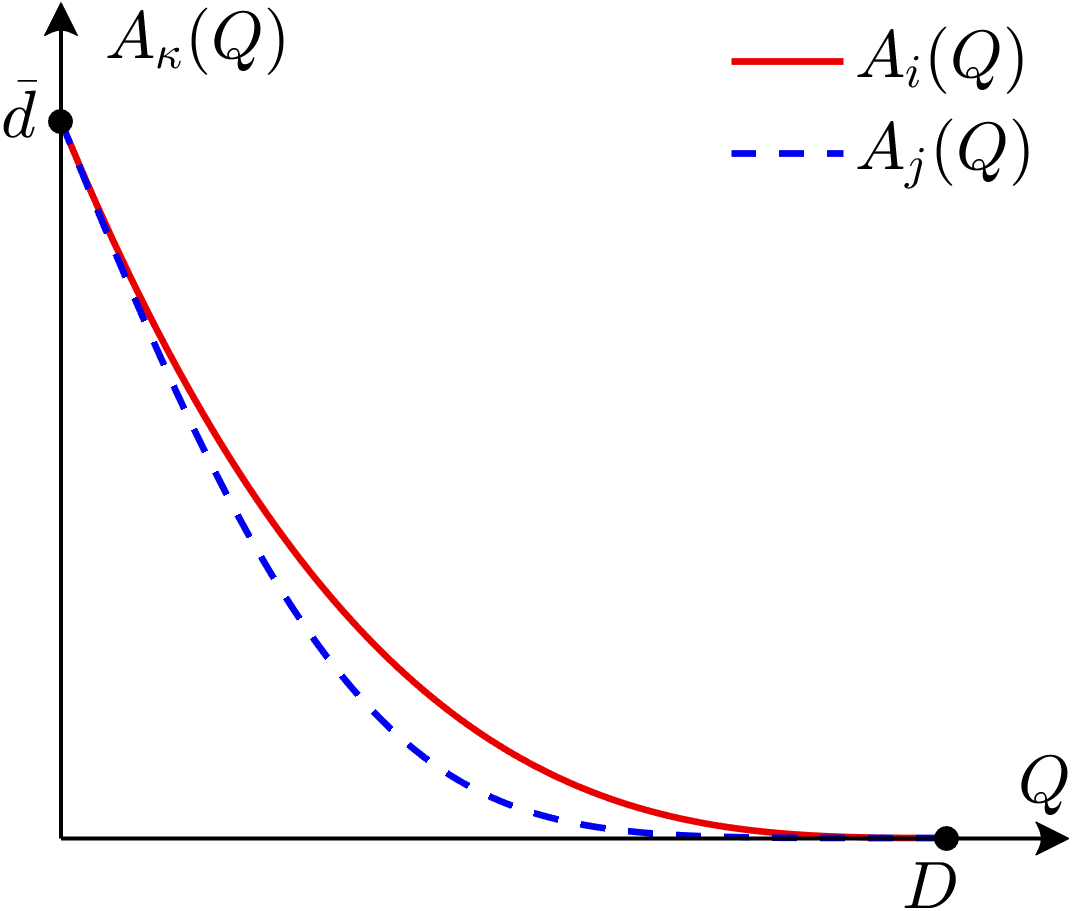}} \quad 
	\subfigure[$|A'_\mechanism(\dcap)|$ vs $\dcap$.]{\label{fig: A_prime1}\includegraphics[width=0.45\linewidth]{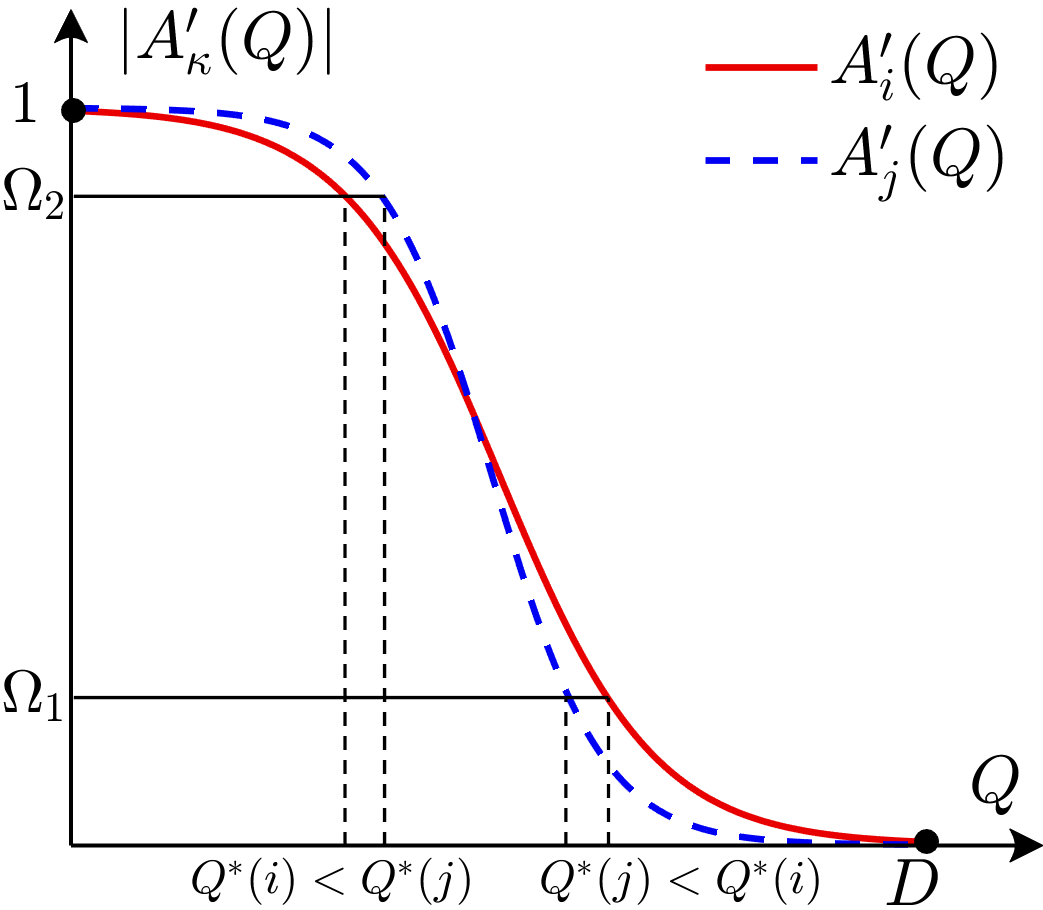}}
	\caption{Illustration for $A_\mechanism(\dcap)$ and $|A'_\mechanism(\dcap)|$.\vspace{-10pt}}
	\label{fig: A A'}
\end{figure}

According to Theorem \ref{Theorem: Optimal Data Cap}, the optimal data cap must make the corresponding marginal overage data consumption equal to the target marginal overage data consumption, i.e., $|A_\mechanism'\left( \dcap^*(\mechanism) \right)| =  \tdr(\QoS,c,z)$.
In Fig. \ref{fig: A_prime1}, we consider two different target marginal overage data consumptions $\tdr_1$ and $\tdr_2$, which correspond to different values of $\QoS$, $c$, and $z$.
\begin{itemize}
	\item To achieve a small target marginal overage data consumption $\tdr_1$, the corresponding optimal data caps satisfy $\dcap^*(j)<\dcap^*(i)$, indicating that the data mechanism $j$ (with a better time flexibility) leads to a smaller data cap;
	\item To achieve a large target marginal overage data consumption $\tdr_2$, the corresponding optimal data caps satisfy $\dcap^*(j)>\dcap^*(i)$, indicating that the data mechanism $j$ (with a better time flexibility) leads to a larger data cap.
\end{itemize}

Later we will further illustrate this counter-intuitive insight in Section \ref{Subsection: simulation}.

So far we have fully characterized the optimal data cap $\dcap^*$ together with the impact of  the data mechanism as well as  the MNO's QoS and costs.
Next we will move on to study the MNO's optimal data mechanism in Stage I.


\begin{figure*}
	\centering
	\begin{minipage}{0.52\textwidth}
		\centering
		\setlength{\abovecaptionskip}{2pt}
		\subfigure[$\val\sim {\rm Gamma}(4.5,0.11)$]{\label{fig: Theta_pdf}{\includegraphics[width=0.49\linewidth]{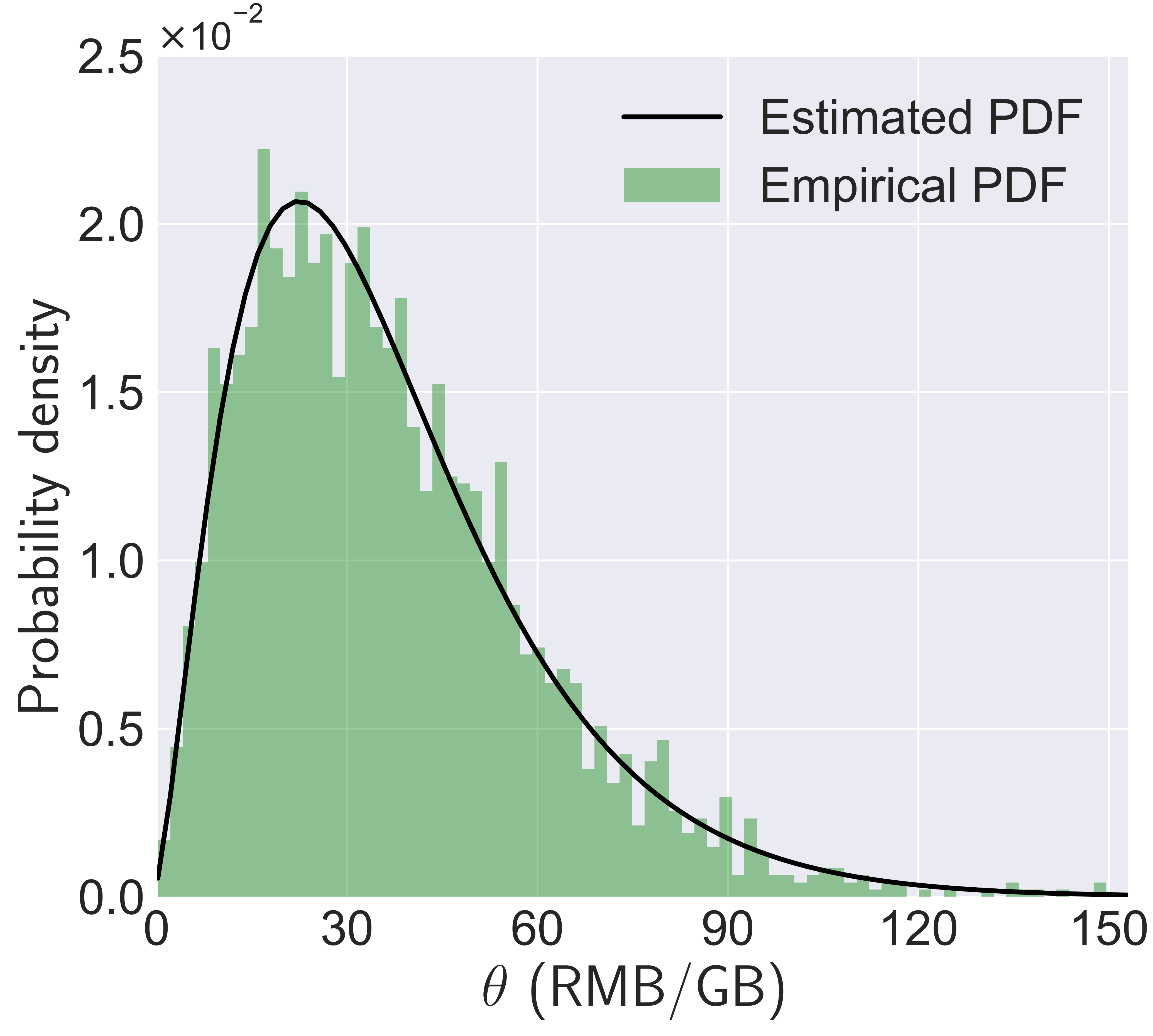}}}   
		\subfigure[$\cut\sim {\rm TN}(0.91,0.22,0,1)$]{\label{fig: Beta_pdf}{\includegraphics[width=0.49\linewidth]{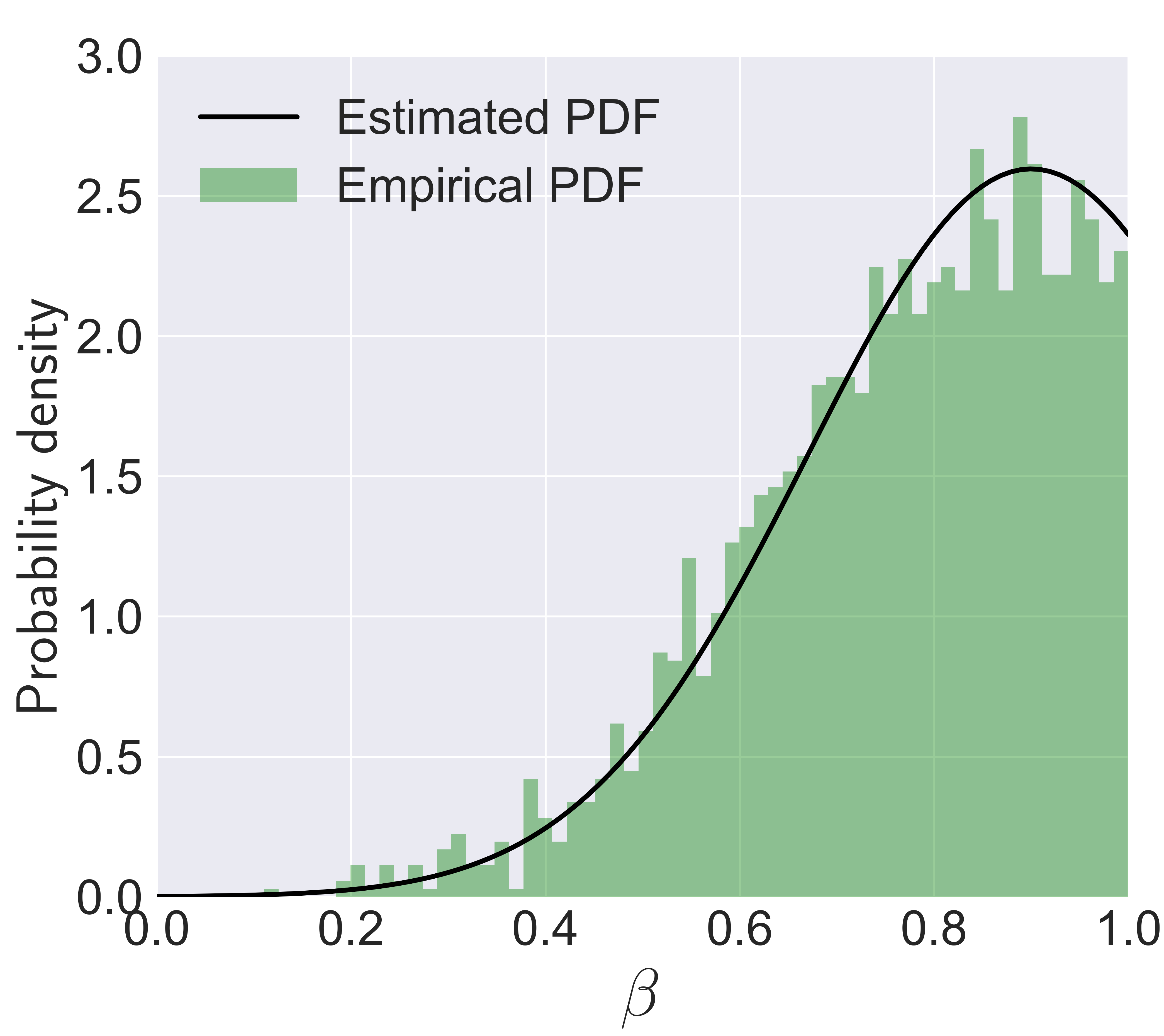}}}  
		\caption{Fitting the PDF of $\val$ and $\cut$. }
		\label{fig: fit}
	\end{minipage}	\quad
	\begin{minipage}{0.45\textwidth}
		\centering
		\includegraphics[width=0.7\linewidth]{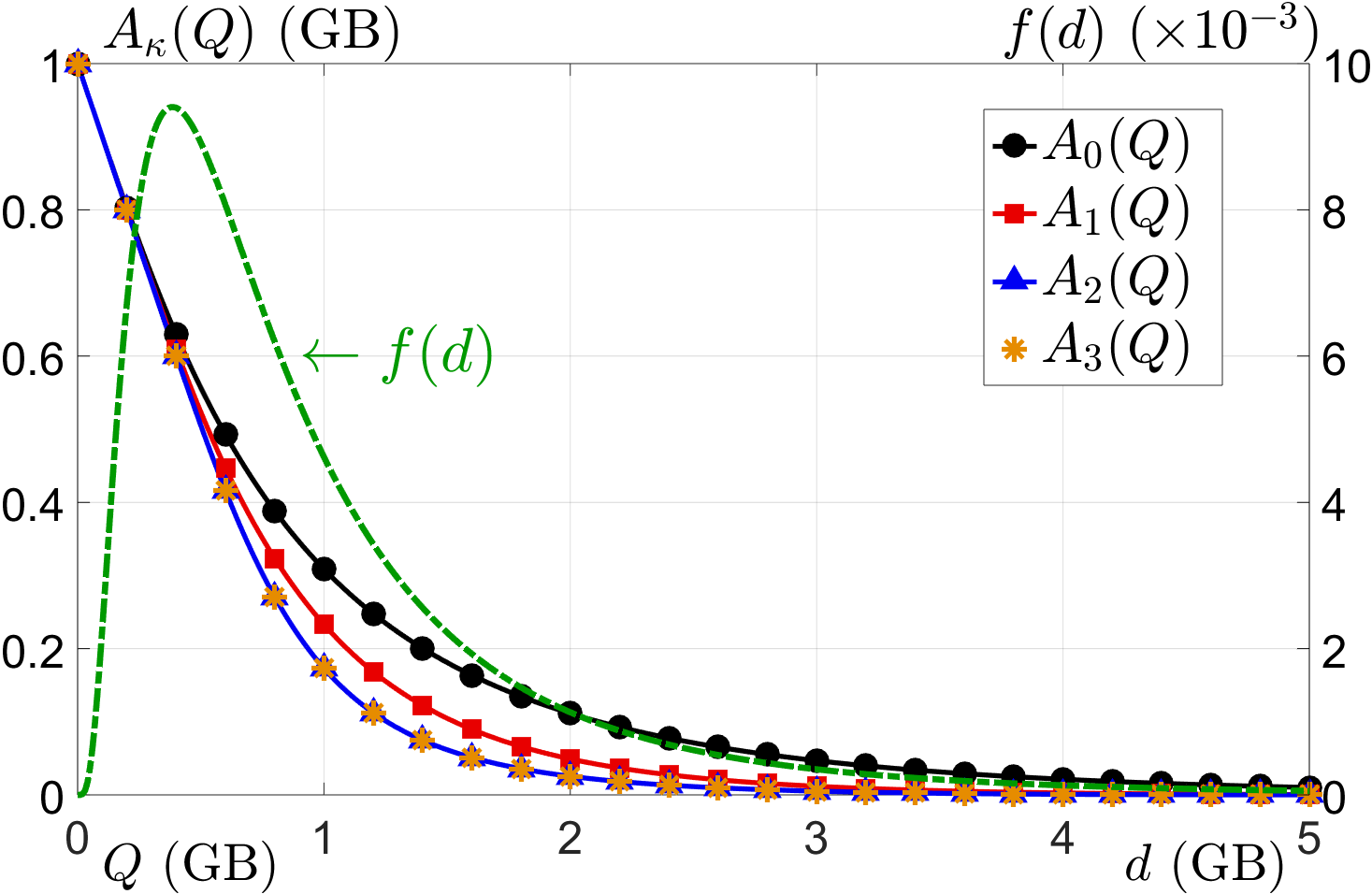}
		\caption{$A_\mechanism(\dcap)$ vs $\dcap$  and $f(d)$ vs $d$. Due to space limit, we only show the part on the interval $[0,{\dmax}/{2}]$.}
		\label{fig: demand distribution}
	\end{minipage} \vspace{-10pt}
\end{figure*}

\subsection{Optimal Data Mechanism in Stage I\label{subSection: Stage I}}	
In Stage I, the MNO determines the optimal data mechanism $\mechanism^*$ to maximize its expected total profit, which answers \textit{Question \ref{Question: which adopt}} mentioned  in Section 1.
\begin{problem}[Optimal Data Mechanism] \label{Problem: Optimal Data Mechanism}
	\begin{equation}
		\begin{aligned}
			\mechanism^* =& \arg\max\limits_{\mechanism\in\{0,1,2,3\}} \tilde{\profit}(\dcap^*(\mechanism),\pcap^*(\dcap^*(\mechanism)),\adfee^*).
		\end{aligned}
	\end{equation}
\end{problem}

Lemma \ref{Lemma: better time flexibility -> higher profit} reveals that a better time flexibility can always bring the MNO a higher profit under the optimal pricing strategy specified in Theorem \ref{Theorem: Optimal Pricing} and the optimal data cap specified in Theorem \ref{Theorem: Optimal Data Cap}, which naturally leads to Theorem \ref{Theorem: Optimal Data Mechanism}.

\begin{lemma}\label{Lemma: better time flexibility -> higher profit}
	Consider two data mechanism $i,j\in\{0,1,2,3\}$, and the data mechanism $i$ has a better time flexibility than $j$, i.e., $\flexibility_i>\flexibility_j$.
	Then the following is true
	\begin{equation}
		\tilde{\profit}(\dcap^*(i),\pcap^*(\dcap^*(i)),\adfee^*) \ge \tilde{\profit}(\dcap^*(j),\pcap^*(\dcap^*(j)),\adfee^*),
	\end{equation}
	where the equality holds if and only if $\dcap^*(i)=\dcap^*(j)=0$ or $\dcap^*(i)=\dcap^*(j)=\dmax$.
\end{lemma}

\begin{theorem}[Optimal Data Mechanism]\label{Theorem: Optimal Data Mechanism}
	Among the four data mechanisms $\{0,1,2,3\}$, the MNO's optimal data mechanism  is $\mechanism^*=2$ and $3$.
\end{theorem}

Theorem \ref{Theorem: Optimal Data Mechanism} shows that the MNO should adopt the rollover mechanism offered by China Mobile or the credit mechanism proposed in this paper to maximize its profit.

Next, we examine the conditions of the system parameters under which  the optimal three-part tariff data plan  degenerates into the pure usage-based plan or the flat-rate plan, in which case the choice of  data mechanism $\mechanism\in\{0,1,2,3\}$ has no effect on the subscribers.

\begin{corollary}[Pure Usage-Based Plan]\label{Corollary: Pure Usage-Based Plan}
	If the target marginal overage data consumption $\tdr(\QoS,c,z) \ge 1$, then the MNO maximizes its expected total profit by offering the pure usage-based data plan $\plan_\Pure=\{0,0,\adfee^*,\na\}$.
\end{corollary}

The condition in Corollary \ref{Corollary: Pure Usage-Based Plan} is satisfied when the MNO 
\begin{itemize}
	\item  provides poor services, i.e., $\QoS$ is small, or
	\item  experiences a large  cost, i.e., $c$ or $z$ is large.
\end{itemize}

The above insight  is consistent  with the reality.
From the users' perspective, they are not willing to pay for any cap if the MNO's QoS is poor.
From the MNO's perspective, it is not beneficial for it to incentivize  more data consumption through a large data cap, if it experiences a large cost from network investment or system management.

As we mentioned in Section 1,  the flat-rate data plan appears earliest in the telecommunication market.
However, most MNOs do not offer the flat-rate data plan anymore.
The following corollary can provide an explanation for this phenomenon.

\begin{corollary}[Flat-Rate Plan]\label{Corollary: Flat-Rate Plan}
	If the marginal capacity cost $z=0$, then the MNO maximizes its expected total profit by offering the flat-rate data plan $\plan_\Flat=\{\dmax,\dmean\adfee^*,\na,\na\}$.
\end{corollary}

Obviously, the condition in Corollary \ref{Corollary: Flat-Rate Plan} corresponds to  an extreme case that does not approximate the current reality well \cite{oughton2017cost}, which explains why the flat-rate data plan has ended  in the past.


\begin{figure*}
	\centering
	\setlength{\abovecaptionskip}{2pt}
	\setlength{\belowcaptionskip}{0pt}
	\subfigure[Optimal cap $\dcap^*(\mechanism)$ vs $\QoS$]{\label{fig: QoS_Optimal_cap}\includegraphics[width=0.28\linewidth]{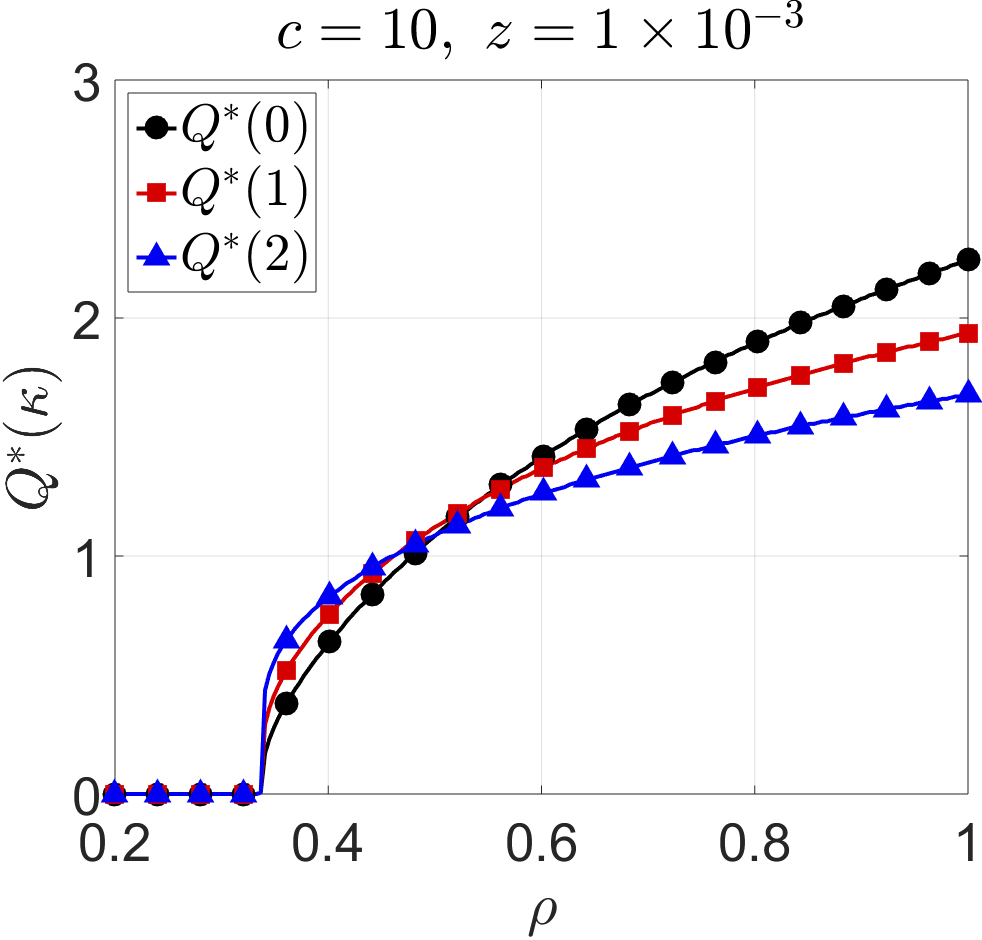}}\qquad
	\subfigure[Optimal cap $\dcap^*(\mechanism)$ vs $c$]{\label{fig: c_Optimal_cap}\includegraphics[width=0.28\linewidth]{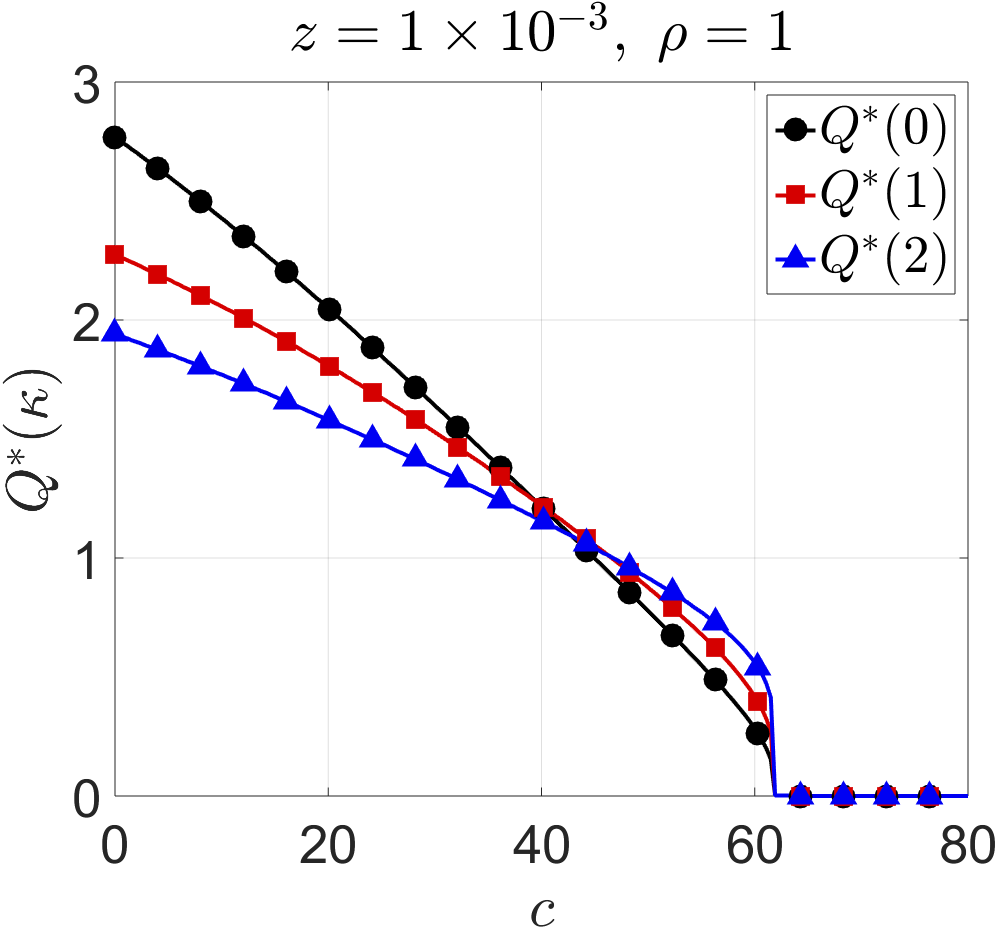}}\qquad
	\subfigure[Optimal cap $\dcap^*(\mechanism)$ vs $z$]{\label{fig: z_Optimal_cap}\includegraphics[width=0.28\linewidth]{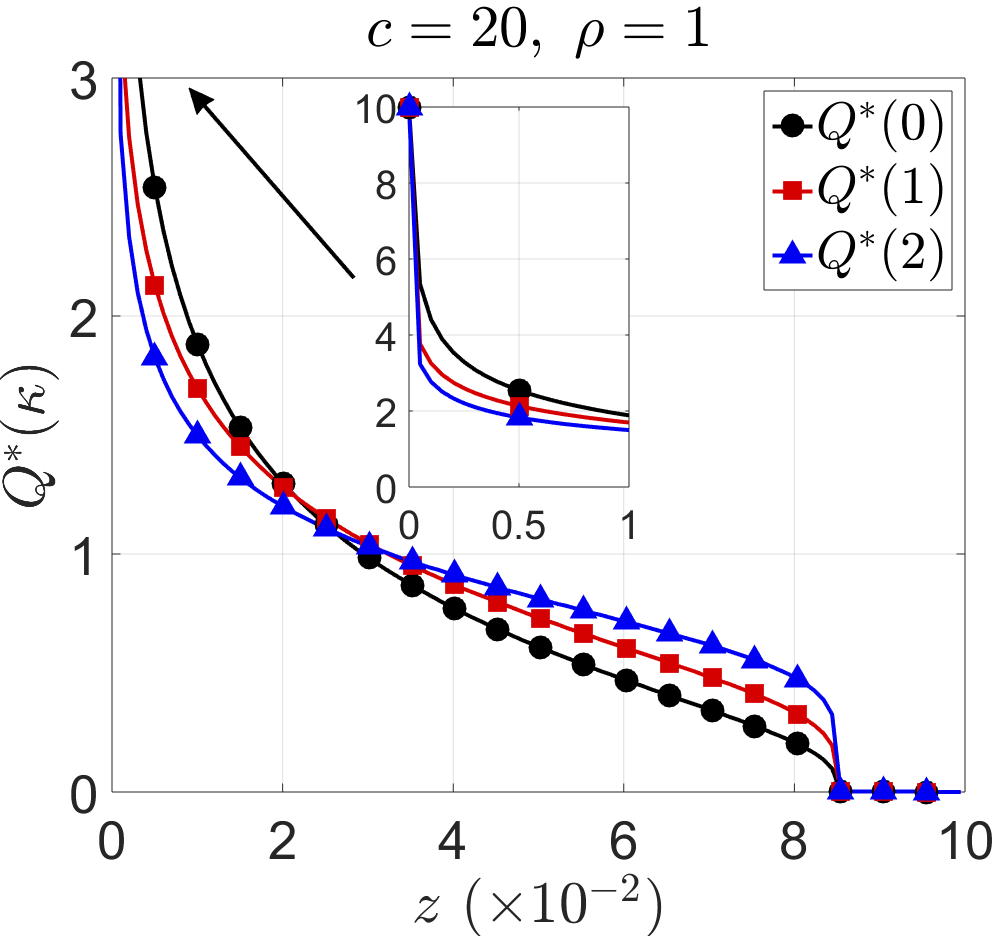}}
	\caption{The optimal data cap $\dcap^*(\mechanism)$ under different data mechanisms $\mechanism\in\{0,1,2\}$.\vspace{-5pt}}
	\label{fig: optimal Q}
\end{figure*}

\begin{figure*}
	\centering
	\setlength{\abovecaptionskip}{2pt}
	\setlength{\belowcaptionskip}{0pt}
	\subfigure[$\pcap^*(\mechanism)$ vs $\QoS$]{\label{fig: QoS_Optimal_PI}\includegraphics[width=0.28\linewidth]{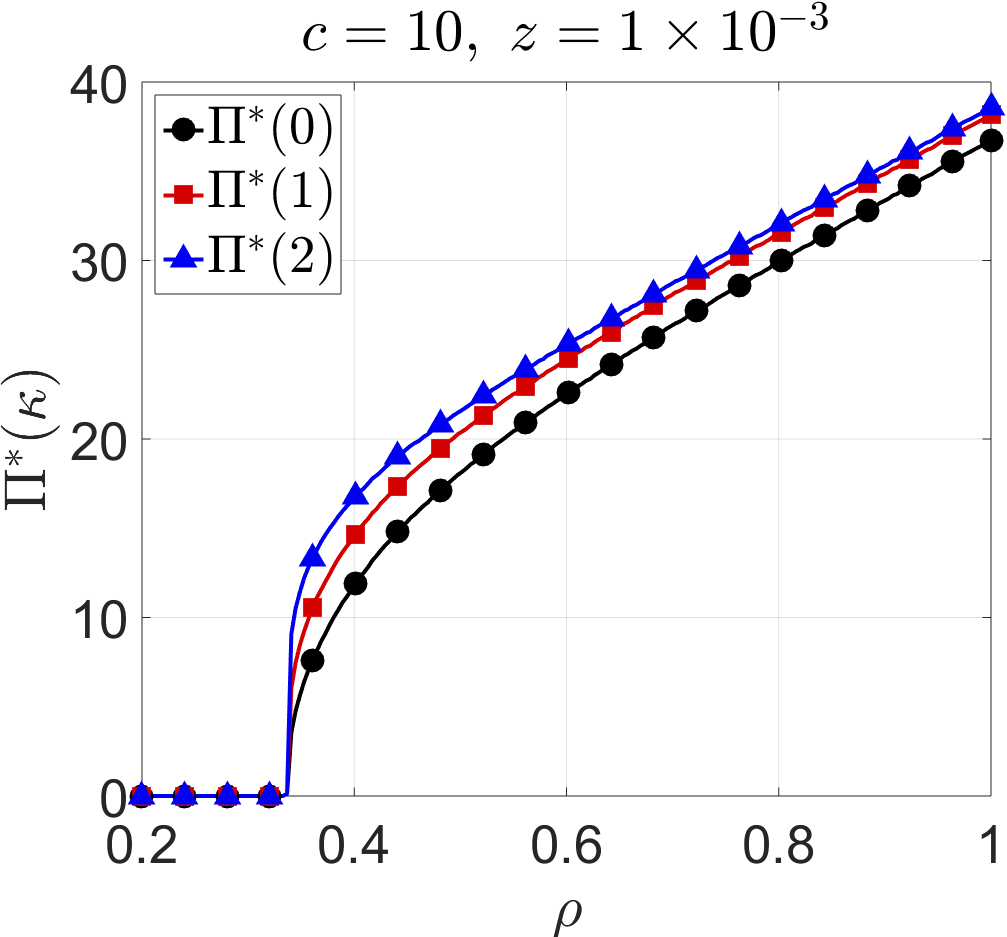}}\qquad
	\subfigure[$\pcap^*(\mechanism)$ vs $c$]{\label{fig: c_Optimal_PI}\includegraphics[width=0.28\linewidth]{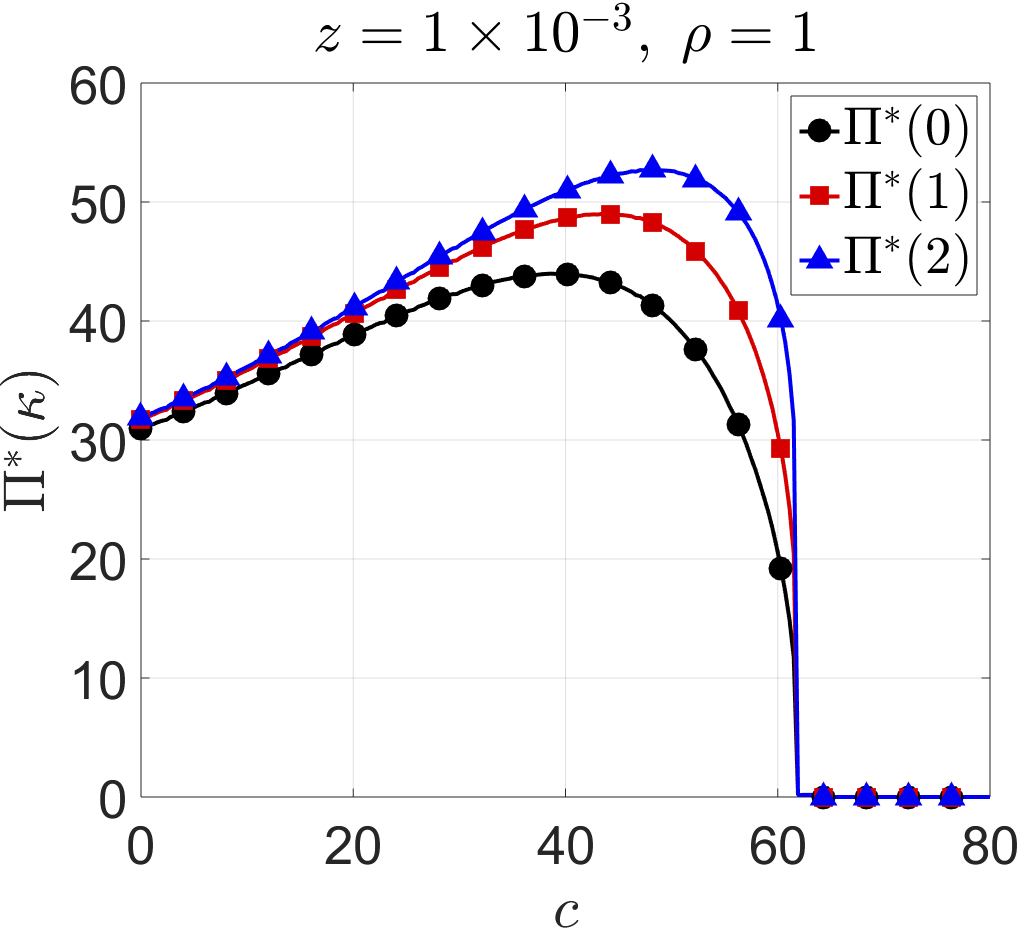}}\qquad
	\subfigure[$\pcap^*(\mechanism)$ vs $z$]{\label{fig: z_Optimal_PI}\includegraphics[width=0.28\linewidth]{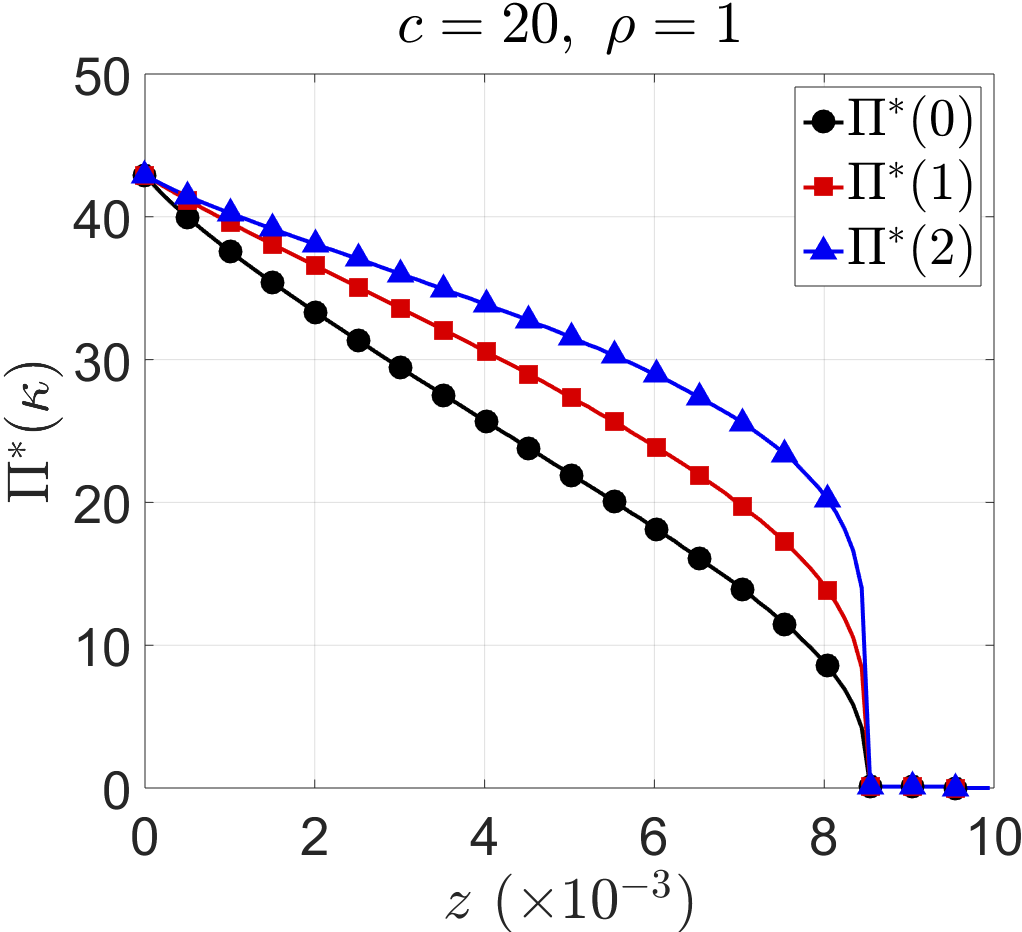}}
	\caption{The optimal subscription fee $\pcap^*(\mechanism)$ under different data mechanisms $\mechanism\in\{0,1,2\}$.\vspace{-5pt}}
	\label{fig: optimal PI}
\end{figure*}

\section{Numerical Results\label{Section: Numerical Results}}
To examine the performance of different data mechanisms, we collected some real data from the telecommunication market in mainland China\footnote{The related data is obtained through a survey, and the questionnaire is available through \url{https://www.wjx.cn/jq/16895923.aspx}.}.
In Section \ref{Subsection: data}, we analyze the distributions of the data valuation and the network substitutability.
In Section \ref{Subsection: simulation}, we simulate the optimal data plan and investigate the effect of the time flexibility on users' payoffs and the MNO's profit.

\subsection{Empirical  Results\label{Subsection: data}}
To estimate the statistic information of users' data valuation $\val$ and network substitutability $\cut$, 
we collected some mobile users' behavioral data from the telecommunication market in mainland China.
The PDFs of these two parameters $\val$ and $\cut$ are shown by the green bars in Fig. \ref{fig: Theta_pdf} and Fig. \ref{fig: Beta_pdf}, respectively. 
We observe that a large proportion of users' data valuations $\val$  falls into the range  between $10$ RMB/GB to $60$ RMB/GB, and most people would like to shrink $70\%\sim100\%$  of their  overage data demand.
Moreover, we also find that the Pearson correlation coefficient between $\val$ and $\cut$ is less than 0.05, which allows us to fit the two distributions independently.


Next we estimate the data valuation $\val$ PDF by assuming a gamma distribution, which satisfies the increasing failure rate (IFR).
The PDF of a gamma distribution in the shape-rate parametrization is
\begin{equation}
h(\val,k,r) =  \frac{r^k\val^{k-1}e^{-r\val}}{\Gamma(k)},
\end{equation}
where $\Gamma(k)$ is a complete gamma function.
We choose the parameters $k=4.5$ and $r=0.11$ by minimizing the least-squares divergence between the estimated and empirical PDF.
In Fig. \ref{fig: Theta_pdf}, the black curve is the estimated PDF.
Visually, it is qualitatively similar to the empirical PDF.
To further investigate the goodness-of-fit statistically, we use the Kolmogorov-Smirnov test on the null hypothesis that the data valuation comes from the gamma distribution, at the significance level of $0.05$ (i.e., if the Kolmogorov-Smirnov test returns a $p$-value less than 0.05, then we need to reject this hypothesis) \cite{massey1951kolmogorov}. 
The test shows a $p$-value of $0.31$, hence the hypothesis that the data valuation follows the gamma distribution is valid.
 

Next we estimate the network substitutability $\cut$ PDF by assuming a truncated normal distribution, since the network substitutability $\cut$ has a concrete upper bound and lower bound, i.e., $0\le\cut\le1$.
The PDF of a normal distribution $N(\mu,\sigma^2)$ truncated between $[a,b]$ is
\begin{equation}
g(\cut;\mu,\sigma,a,b) =  \frac{ \phi\left(\frac{\cut-\mu}{\sigma}\right) }{ \sigma\left[ \Phi\left(\frac{b-\mu}{\sigma}\right) - \Phi\left(\frac{a-\mu}{\sigma}\right) \right] },
\end{equation}
where $\phi(\cdot)$ is the probability density function of the standard normal distribution, and $\Phi(\cdot)$ is its cumulative distribution function.
Similarly, we find the truncated normal distribution $\cut\sim {\rm TN}(0.91,0.22,0,1)$ by minimizing the least-squares divergence between the estimated and empirical PDFs.
The corresponding Kolmogorov-Smirnov test shows a $p$-value of $0.38$, hence the hypothesis of the truncated normal distribution is valid.


\subsection{Performance Evaluation\label{Subsection: simulation}}
Next we will use the fitted market distribution to investigate how the MNO's QoS and marginal costs affect its optimal data plan, and examine the impact of time flexibility on the users' payoff and the MNO's profit.


We set the minimum data unit to 1MB.
Following the data analysis results in \cite{lambrecht2007does},\cite{nevo2016usage}, we assume that users' monthly data demand follows a truncated log-normal distribution with a mean $\dmean=10^3$ on the interval $[0,10^4]$, i.e., the mean value is $\dmean=1$GB and the maximal usage is $D=10$GB.
Fig. \ref{fig: demand distribution} shows the PMF $f(d)$ and the expected monthly overage usage $A_\mechanism(\dcap)$ under the four data mechanisms, which indicates that  $A_0(\dcap)> A_1(\dcap)> A_2(\dcap)= A_3(\dcap)$ for all $\dcap\in(0,\dmax)$.
Since the degree of time flexibility under $\mechanism=2$ and $\mechanism=3$ is equivalent, in the following we will neglect $\mechanism=3$ and only plot the results for $\mechanism=0,1,2$.

\begin{figure*}
	\centering
	\setlength{\abovecaptionskip}{2pt}
	\setlength{\belowcaptionskip}{0pt}
	\subfigure[Impact of $\QoS$]{\label{fig: QoS_Optimal_Gain}\includegraphics[width=0.28\linewidth]{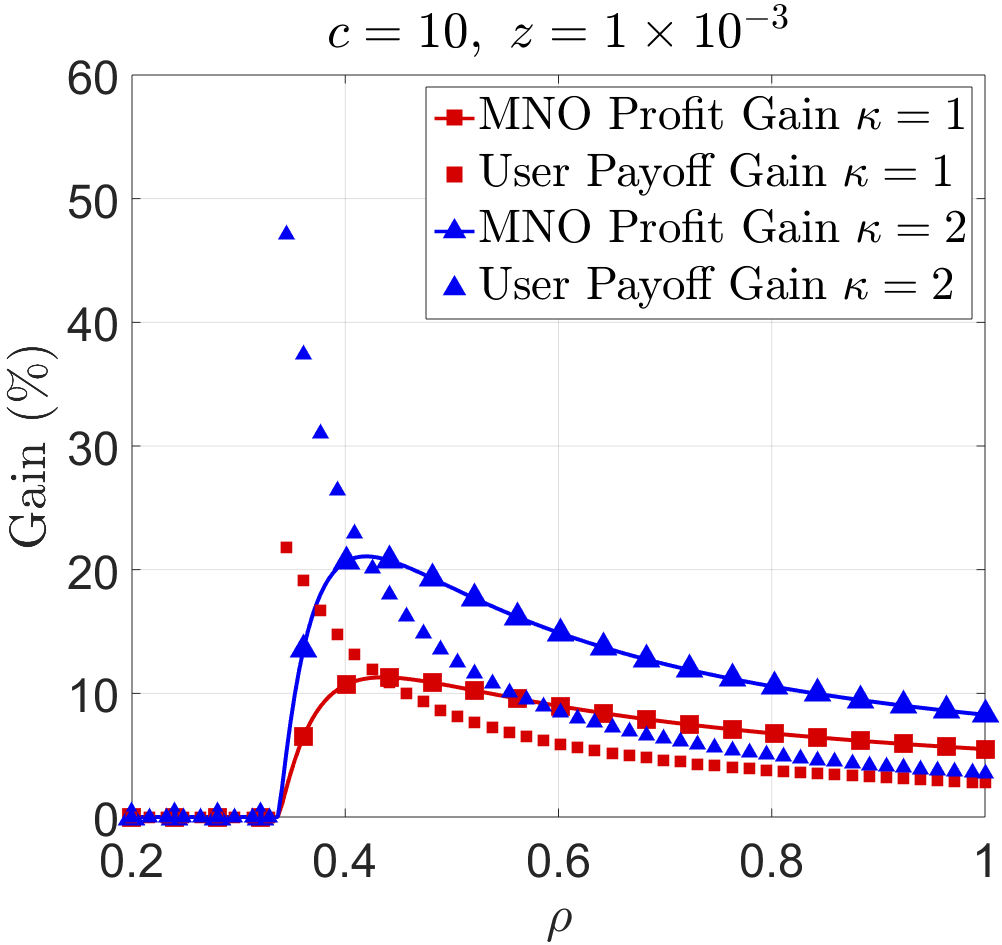}}\qquad
	\subfigure[Impact of $c$]{\label{fig: c_Optimal_Gain}\includegraphics[width=0.28\linewidth]{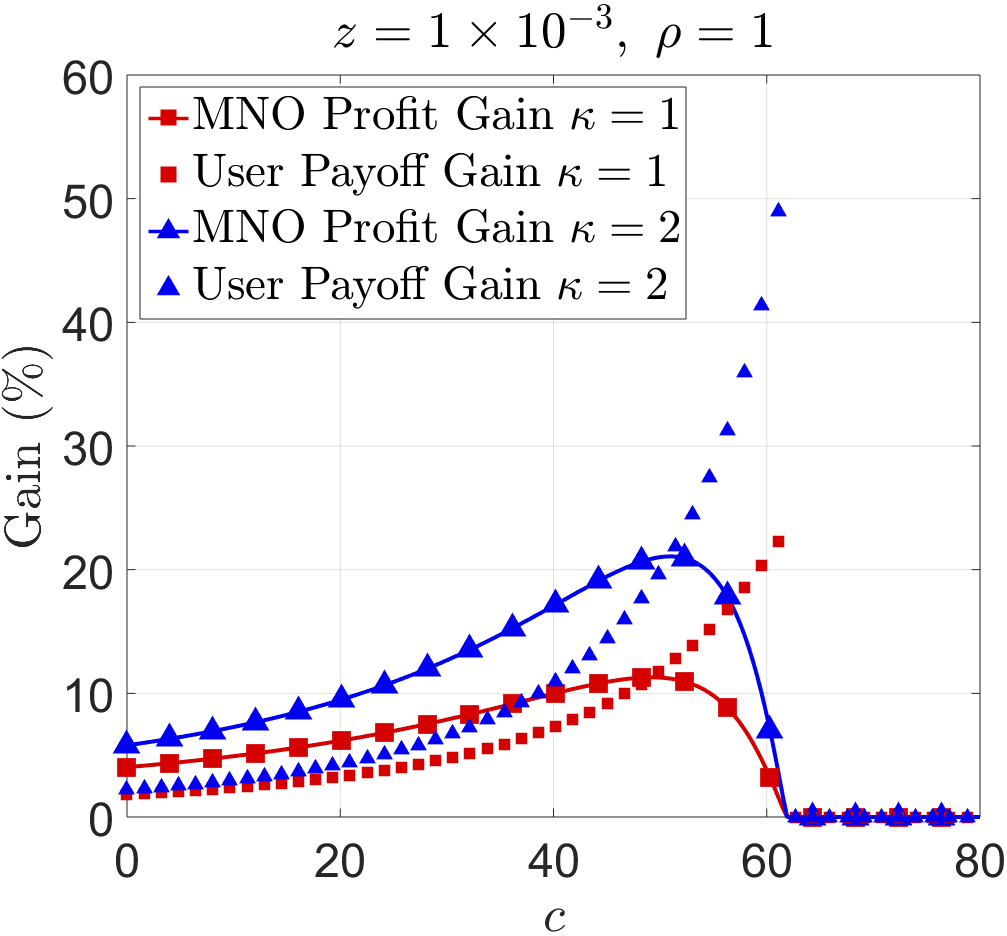}}\qquad
	\subfigure[Impact of $z$]{\label{fig: z_Optimal_Gain}\includegraphics[width=0.28\linewidth]{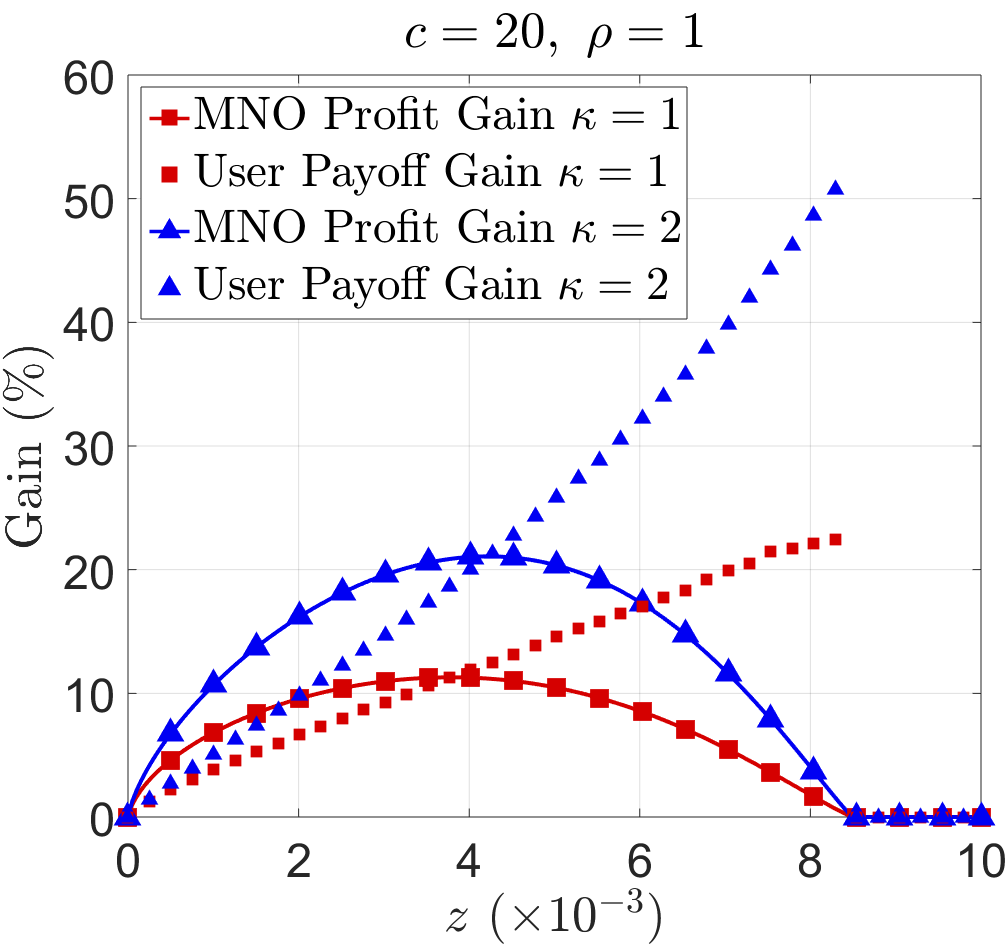}}
	\caption{MNO's profit gain and users' payoff gain with $\mechanism=0$ as the benchmark. \vspace{-10pt}}
	\label{fig: optimal Gain}
\end{figure*}

\subsubsection{Optimal Data Plan \label{Subsubsection: Optimal Data Plan}}
We investigate the impact of QoS $\QoS$, marginal operational cost $c$, and the marginal capacity cost $z$ individually.

In Fig. \ref{fig: optimal Q}, we use three sub-figures to plot the optimal data cap $\dcap^*(\mechanism)$ versus the MNO's QoS $\QoS$, marginal operational cost $c$, and the marginal capacity cost $z$, respectively.
In addition, the three curves in each sub-figure correspond to the three data mechanisms $\mechanism\in\{0,1,2\}$.

Overall, the optimal data cap $\dcap^*(\mechanism)$ increases (from zero) as the MNO becomes stronger, in terms of  
\begin{itemize}
	\item a better QoS $\QoS$, as shown in Fig. \ref{fig: QoS_Optimal_cap},
	\item a smaller operational cost $c$, as shown in Fig. \ref{fig: c_Optimal_cap},
	\item a smaller capacity cost $z$, as shown in Fig. \ref{fig: z_Optimal_cap}.
\end{itemize}

Particularly, the pure usage-based data plan appears when the MNO's QoS $\QoS<0.35$ in Fig. \ref{fig: QoS_Optimal_cap}, marginal operational cost $c>62$ RMB/GB in Fig. \ref{fig: c_Optimal_cap}, or marginal capacity cost $z>8.5\times10^{-3}$ RMB/GB in Fig. \ref{fig: z_Optimal_cap}.
In addition, Fig. \ref{fig: z_Optimal_cap} shows that the flat-rate data plan appears when $z=0$, since the corresponding data cap is the maximal data demand $10$GB.

As we mentioned in Section \ref{Subsubsection: Optimal Data Cap}, a better time flexibility does not necessarily correspond to a smaller data cap.
By comparing the three curves in each sub-figure of Fig. \ref{fig: optimal Q}, we find that a \textit{better} time flexibility would lead to an even \textit{larger} data cap if the MNO is weak in terms of
\begin{itemize}
	\item a poor QoS, e.g., $\QoS=0.4$ in Fig. \ref{fig: QoS_Optimal_cap},
	\item a large marginal operational cost, e.g., $c=50$ RMB/GB in Fig. \ref{fig: c_Optimal_cap},
	\item a large marginal capacity cost, e.g., $z=6\times10^{-3}$ RMB/GB in Fig. \ref{fig: z_Optimal_cap}.
\end{itemize}

The intuitions behind the counter-intuitive result include 
\begin{itemize}
	\item When the MNO is \textit{weak}, it chooses a small data cap, under which the main revenue comes from users' overage payments.
	In this case, offering a better time flexibility can significantly reduce its revenue from users' overage payments.
	Therefore, the MNO will  increase the data cap and the subscription fee to compensate its revenue loss.
	\item When the MNO is \textit{strong}, it chooses a large data cap, under which  its main revenue comes from users' subscription fee.
	In this case, offering a better time flexibility does not reduce its revenue too much, and the MNO will decrease the data cap to further reduce its cost.
\end{itemize}

Next we examine the impact of the time flexibility on the monthly subscription fee. 

Fig. \ref{fig: optimal PI} plots the optimal subscription fee $\pcap^*(\mechanism)$ under different data mechanisms.
By comparing the three curves in each sub-figure, we observe that a higher subscription fee is always associated with  a data mechanism with better time flexibility, even though it may correspond to a smaller or larger data cap as shown in Fig. \ref{fig: optimal Q}.

The above discussions together with Proposition \ref{Proposition: optimal pi}  provide answers to \textbf{\textit{Question \ref{Question: impact}}} mentioned in Section 1: a better time flexibility  corresponds to a smaller data cap $\dcap^*$ if the MNO is strong or a larger data cap if the MNO is weak. 
Meanwhile, a better time flexibility always leads to a higher subscription fee $\pcap^*$.
Finally, it does not affect the optimal per-unit fee $\adfee^*$.

\subsubsection{Users' Payoffs and MNO's Profit\label{Subsubsection Users' Payoffs and MNO's Profit}}
We investigate the performance of different data mechanisms in terms of all users' payoffs and the MNO's profit.
Specifically,  we set the traditional data mechanism $\mechanism=0$  as the benchmark, and plot the performance gain of other schemes comparing to the benchmark.
The three sub-figures in Fig. \ref{fig: optimal Gain} plot the performance gains versus the MNO's QoS $\QoS$, marginal operational cost $c$, and the marginal capacity cost $z$, respectively.
The two \textit{solid curves} in each sub-figure correspond to the MNO's profit gains for $\mechanism\in\{1,2\}$, the other two \textit{dash curves} represent users' payoff gains for $\mechanism\in\{1,2\}$.

Overall, Fig. \ref{fig: optimal Gain} show that the users' payoff gains (i.e., the dash curves) decrease as the MNO's QoS $\QoS$ increases as in Fig. \ref{fig: QoS_Optimal_Gain} or the MNO's marginal costs $c$ and $z$ decrease as in Fig. \ref{fig: c_Optimal_Gain} and Fig. \ref{fig: z_Optimal_Gain}.
In this process, however, the MNO's profit gains (i.e., the solid curves) first increase then decrease  in all three sub-figures.

From the two \textit{solid curves} in each sub-figure of Fig. \ref{fig: optimal Gain}, we find that the MNO's profit gains under $\mechanism=1$ and $\mechanism=2$ are both non-negative, thus the time flexibility can increase the profit of MNO compared with the benchmark $\mechanism=0$.
From the two \textit{dash curves} in each sub-figure of Fig. \ref{fig: optimal Gain}, we find that the users' payoffs gains under $\mechanism=1$ and $\mechanism=2$ are both non-negative, thus the time flexibility increases the users' payoffs as well. 
Moreover, we also note that $\mechanism=2$ always outperforms $\mechanism=1$ in terms of both MNO's profit gain and users' payoffs gain, which indicates that a  better time flexibility leads to a  larger improvement.

Now we know that both the MNO and the users can benefit from the time flexibility, a natural question is who will benefit more?
By comparing the two red curves with squares (or the two blue curves with triangles) in each sub-figure, we find that the MNO benefits more from the time flexibility than users if the MNO is strong, in terms of 
\begin{itemize}
	\item a good QoS, e.g., $\QoS>0.43$ in Fig. \ref{fig: QoS_Optimal_Gain},
	\item a small marginal operational cost, e.g., $c<50$ RMB/GB in Fig. \ref{fig: c_Optimal_Gain},
	\item a small marginal capacity cost, e.g., $z<4\times10^{-4}$ RMB/GB in Fig. \ref{fig: z_Optimal_Gain}.
\end{itemize}
 
Intuitively,  a stronger MNO has a larger  pricing power, thus the strong MNO can benefit more than users from adding time flexibility.
However, a weaker MNO has to leave users more benefits to maintain its market, thus users benefit more than a weaker MNO from time flexibility.

{Furthermore, we also investigate the impact of the variances of the demand distribution, which indicates that the performance gain of the MNO's profit and users' payoff will increase in the variance.
Due to the space limit, please refer to Appendix \ref{Appendix: Variance of Demand Distribution} for more detailed discussions.}

The above discussions answer \textbf{\textit{Question \ref{Question: who benefit}}} in Section 1.
In a monopoly market, both the MNO and users can benefit from a better time flexibility.
Moreover, the MNO benefits more if the MNO offers good services and experiences  small costs.
Otherwise, the users benefit more.

\section{Conclusions and Future Works\label{Section: Conclusions and Future Works}}
In this paper, we study the MNO's optimal three-part tariff plan with time flexibility.
Specifically, we consider four data mechanisms, and formulate the MNO's optimal data plan design problem as a three-stage Stackelberg Game.
Through backward induction, we analytically characterize the users' subscription choices in Stage III, the MNO's optimal data cap and corresponding pricing strategy in Stage II, and the MNO's  optimal data mechanism in Stage I.
Moreover, we conduct a market survey to estimate the distribution of users' data valuation and the network substitutability.
Then we evaluate the performance of different data mechanisms using the real data.

%
%
%

In the future work, we want to collect more empirical data to estimate the MNO's cost and users' data demand distributions, and validate the insights obtained based on the current linear costs model and homogeneous data demand distribution.
Moreover, we will consider a more general competitive market, and analyze the impact of time flexibility on multiple MNOs' competition.


\bibliographystyle{IEEEtran}
\bibliography{ref}

\vspace{-40pt}
\begin{IEEEbiography}[{	\includegraphics[width=1in,height=1.25in,clip,keepaspectratio]{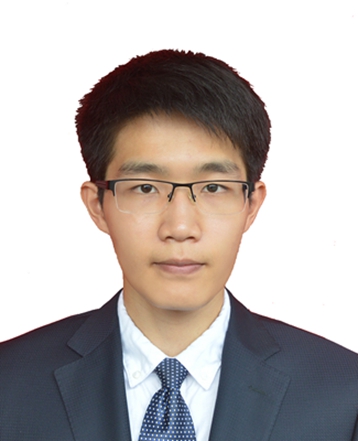}	}]{Zhiyuan Wang}
	received the B.S. degree from Southeast University, Nanjing, China, in 2016.
	He is currently working toward the Ph.D. degree with the Department of Information Engineering, The Chinese University of Hong Kong, Shatin, Hong Kong.  
	His research interests include the field of network economics and game theory, with current emphasis on smart data pricing and fog computing. 
	He is the recipient of the Hong Kong PhD Fellowship.
\end{IEEEbiography}

\vspace{-40pt}
\begin{IEEEbiography}[{	\includegraphics[width=1in,height=1.25in,clip,keepaspectratio]{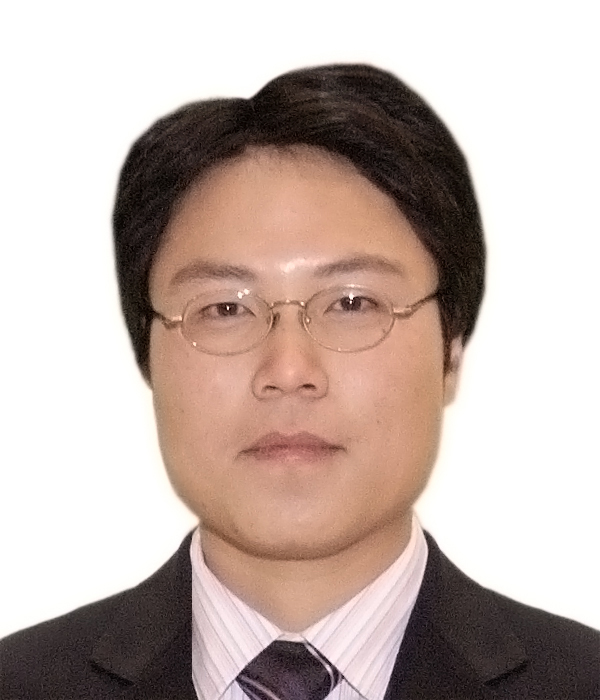}	}]	{Lin Gao}
	(S'08-M'10-SM'16) received the PhD degree in electronic engineering from Shanghai Jiao Tong University in 2010 and worked as a postdoctoral research fellow with the Chinese University of Hong Kong from 2010 to 2015. 
	He is an associate professor in the School of Electronic and Information Engineering, Harbin Institute of Technology, Shenzhen, China. 
	He received the IEEE ComSoc Asia-Pacific Outstanding Young Researcher Award in 2016. His main research interests are in the area of network economics and games, with applications in wireless communications and networking.
	He has co-authored 3 books including the textbook Wireless Network Pricing by Morgan \& Claypool (2013). 
	He is the co-recipient of three Best (Student) Paper Awards from WiOpt 2013, 2014, 2015, and is a Best Paper Award Finalist from IEEE INFOCOM 2016.
\end{IEEEbiography}

\vspace{-40pt}
\begin{IEEEbiography}
	[{	\includegraphics[width=1in,height=1.25in,clip,keepaspectratio]{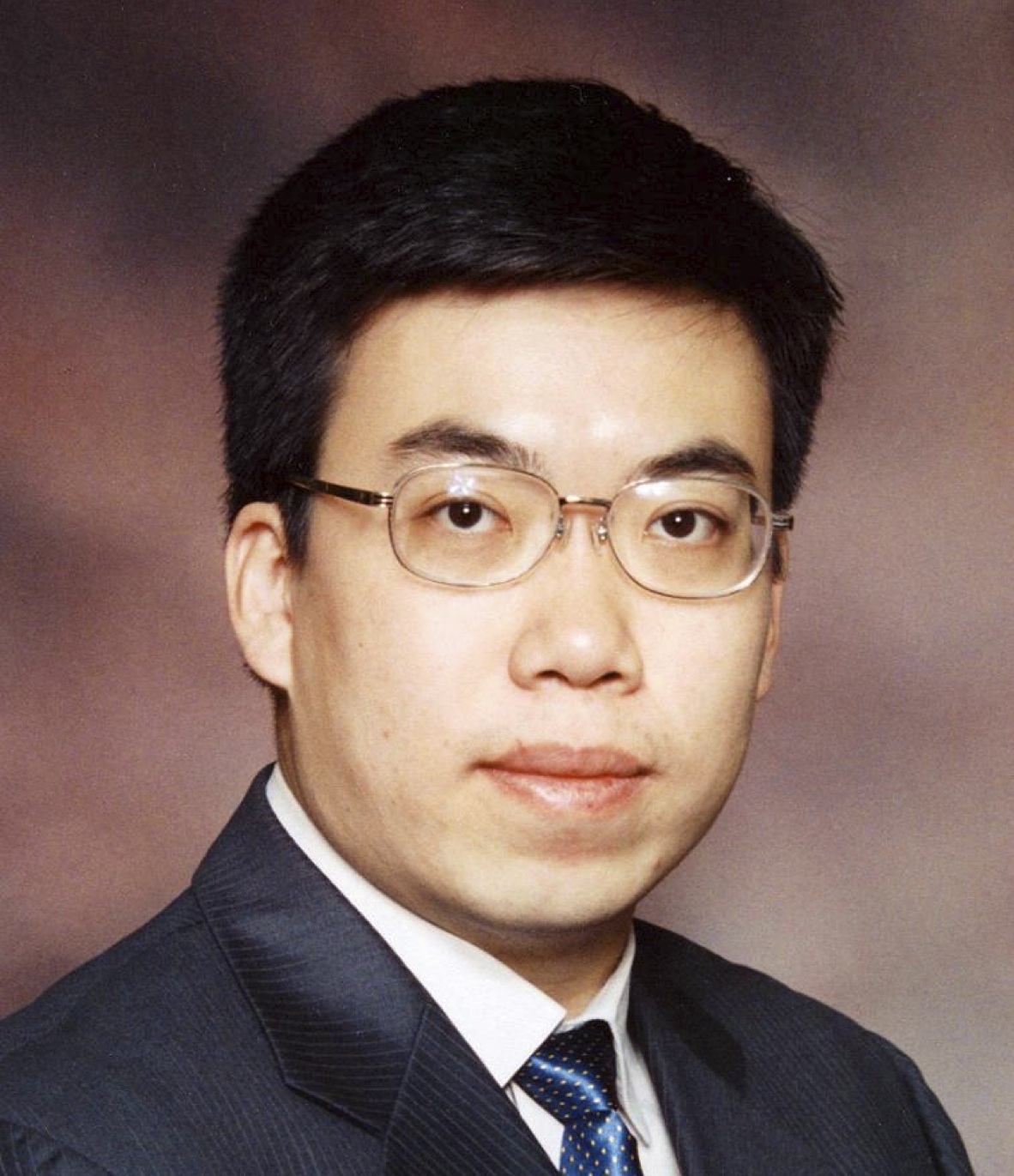}	}]	{Jianwei Huang}
	(F'16) is a Professor and Director of the Network Communications and Economics Lab (ncel.ie.cuhk.edu.hk), in the Department of Information Engineering at the Chinese University of Hong Kong. 
	He received the Ph.D. degree from Northwestern University in 2005, and worked as a Postdoc Research Associate at Princeton University during 2005-2007. 
	He is an IEEE Fellow, a Distinguished Lecturer of IEEE Communications Society, and a Clarivate Analytics Highly Cited Researcher in Computer Science. 
	Jianwei Huang is the co-author of 9 Best Paper Awards, including IEEE Marconi Prize Paper Award in Wireless Communications in 2011. 
	He has co-authored six books, including the textbook on "Wireless Network Pricing." 
	He received the CUHK Young Researcher Award in 2014 and IEEE ComSoc Asia-Pacific Outstanding Young Researcher Award in 2009. 
	He has served as an Associate Editor of IEEE Transactions on Mobile Computing, IEEE/ACM Transactions on Networking, IEEE Transactions on Network Science and Engineering, IEEE Transactions on Wireless Communications, IEEE Journal on Selected Areas in Communications - Cognitive Radio Series, and IEEE Transactions on Cognitive Communications and Networking. 
	He has served as an Editor of Wiley Information and Communication Technology Series, Springer Encyclopedia of Wireless Networks, and Springer Handbook of Cognitive Radio. 
\end{IEEEbiography}

\newpage
\appendices

\section{Proof of Lemma \ref{Lemma: time flexibility comparison}\label{Proof: comparison}}
In order to prove Lemma \ref{Lemma: time flexibility comparison}, in the following three subsections, we prove $A_0(\dcap)\ge A_1(\dcap)$, $A_2(\dcap)=A_3(\dcap)$, and $A_1(\dcap)\ge A_2(\dcap)$, respectively.

\subsection{Proof of $A_0(\dcap)\ge A_1(\dcap)$}
We prove $A_0(\dcap)\ge A_1(\dcap)$ by computing the weighted summations in two steps.
Specifically, $A_0(\dcap)$ and  $A_1(\dcap)$ are given by
\begin{equation}
\begin{aligned}
& A_0(\dcap)=\sum_{d=0}^{\dmax}\left[ d-\dcap \right]^+ f(d),\\
& A_1(\dcap)=\sum_{d=0}^{\dmax}\sum_{\spedata=0}^{\dcap}\left[ d-\dcap-\spedata \right]^+p_1(\spedata) f(d) .
\end{aligned}
\end{equation}

It is obvious that the following inequality is true
\begin{equation}\label{Proof Equ: inequality 1}
\left[ d-\dcap \right]^+ \ge \left[ d-\dcap-\spedata \right]^+ ,\ \forall\ \spedata\in\{0,1,2,...,\dcap\}.
\end{equation}

Now we compute the weighted summation over the weight $p_1(\spedata)$ for  (\ref{Proof Equ: inequality 1}), and we obtain
\begin{equation}
\sum_{\spedata=0}^{\dcap}\left[ d-\dcap \right]^+p_1(\spedata) \ge \sum_{\spedata=0}^{\dcap}\left[ d-\dcap-\spedata \right]^+p_1(\spedata),
\end{equation}
which is equivalent to 
\begin{equation}\label{Proof Equ: inequality 2}
\left[ d-\dcap \right]^+ \ge \sum_{\spedata=0}^{\dcap}\left[ d-\dcap-\spedata \right]^+p_1(\spedata),\ \forall\ d\in\{0,1,...,\dmax\}.
\end{equation}

Then we further compute the weighted summation over the weight $f(d)$ for (\ref{Proof Equ: inequality 2}), and we obtain
\begin{equation}
\sum_{d=0}^{\dmax}\left[ d-\dcap \right]^+f(d) \ge \sum_{d=0}^{\dmax}\sum_{\spedata=0}^{\dcap}\left[ d-\dcap-\spedata \right]^+p_1(\spedata)f(d),
\end{equation}
which implies $A_0(\dcap)\ge A_1(\dcap)$.

\subsection{Proof of $A_2(\dcap)=A_3(\dcap)$}
Second, we prove $A_2(\dcap)=A_3(\dcap)$ through transforming their mathematical expressions.
Specifically, $A_2(\dcap)$ and  $A_3(\dcap)$ are given by
\begin{equation}\label{Equ: Proof A2=A3}
\begin{aligned}
& A_2(\dcap)=\sum_{d=0}^{\dmax}\sum_{\spedata=0}^{\dcap}\left[ d-\dcap-\spedata \right]^+p_2(\spedata) f(d),\\
& A_3(\dcap)=\sum_{d=0}^{\dmax}\sum_{\spedata=-\dcap}^{0}\left[ d-2\dcap-\spedata \right]^+p_3(\spedata) f(d) .
\end{aligned}
\end{equation}

As mentioned in Section \ref{Subsection: Data Mechanisms}, for $\mechanism=3$, the data deficit of the next month $\spedata_3'$ is
\begin{equation}
\spedata_3'=\left\{
\begin{aligned}
& 0,					& \text{if }d\in & [0,\dcap+\spedata_3], \\
& \dcap+\spedata_3-d,	& \text{if }d\in & (\dcap+\spedata_3,2\dcap+\spedata_3), \\
& -\dcap,				& \text{if }d\in & [2\dcap+\spedata_3,\dmax].
\end{aligned}
\right.
\end{equation}

Here we further define $\mu=\spedata_3+\dcap$ and $\mu'=\spedata_3'+\dcap$, thus
\begin{equation}\label{Proof Equ: mu}
\mu'=\left\{
\begin{aligned}
& \dcap,					& \text{if }d\in & [0,\dcap+\spedata_3], \\
& 2\dcap+\spedata_3-d,		& \text{if }d\in & (\dcap+\spedata_3,2\dcap+\spedata_3), \\
& 0,						& \text{if }d\in & [2\dcap+\spedata_3,\dmax].
\end{aligned}
\right.
\end{equation}

Now we substitute $\spedata_3=\mu-\dcap$ into (\ref{Proof Equ: mu}), and get
\begin{equation}
\mu'=\left\{
\begin{aligned}
& \dcap,					& \text{if }d\in & [0,\mu], \\
& \dcap+\mu-d,			& \text{if }d\in & (\mu,\dcap+\mu), \\
& 0,						& \text{if }d\in & [\dcap+\mu,\dmax].
\end{aligned}
\right.
\end{equation}

Therefore, the transition matrix from $\mu$ to $\mu'$ is
\begin{equation}
p_\mu(\mu,\mu')=\left\{
\begin{aligned}
& \textstyle\sum_{d=0}^{\mu}f(d),				& \text{if }\mu' &=\dcap, \\
& f(\dcap+\mu-\mu'),							& \text{if }\mu' &\in(0,\dcap), \\
& \textstyle\sum_{d=\dcap+\mu}^{\dmax}f(d),	& \text{if }\mu' &=0,
\end{aligned}
\right.
\end{equation}
which is the same as the transition matrix of the data mechanism $\mechanism=2$ derived in (16).
Thus they have the same stationary distribution, i.e., $p_\mu(\mu)=p_2(\spedata)$ for any $\mu=\spedata$.

Now we substitute $\spedata_3=\mu-\dcap$ into (\ref{Equ: Proof A2=A3}), and obtain
\begin{equation}
\begin{aligned}
A_3(\dcap)	
&=\sum_{\mu=0}^{\dcap}\sum_{d=0}^{\dmax}[d-\dcap - \mu]^+f(d)p_\mu(\mu)=A_2(\dcap).
\end{aligned}
\end{equation}


\subsection{Proof of $A_1(\dcap)\ge A_2(\dcap)$}
Now we prove $A_1(\dcap)\ge A_2(\dcap)$.
Specifically, $A_1(\dcap)$ and $A_2(\dcap)$ are given by
\begin{equation}\label{Equ: Proof A1>A2 1}
\begin{aligned}
& A_1(\dcap)=\sum_{d=0}^{\dmax}\sum_{\spedata=0}^{\dcap}\left[ d-\dcap-\spedata \right]^+p_1(\spedata) f(d).
\end{aligned}
\end{equation}

\begin{equation}\label{Equ: Proof A1>A2 2}
	\begin{aligned}
		& A_2(\dcap)=\sum_{d=0}^{\dmax}\sum_{\spedata=0}^{\dcap}\left[ d-\dcap-\spedata \right]^+p_2(\spedata) f(d) .
	\end{aligned}
\end{equation}

The only  difference between (\ref{Equ: Proof A1>A2 1}) and (\ref{Equ: Proof A1>A2 1}) lies in the stationary distribution $p_1(\spedata)$ and $p_2(\spedata)$.
Due to the complexity of the Markov chain specified in (16), we are not able to  compute the closed-form expression for $A_1(\dcap)$ and $A_2(\dcap)$.
Nevertheless, we will prove $A_1(\dcap)\ge A_2(\dcap)$ by showing that for an arbitrary realized data demand sequence, the overage data consumption under $\mechanism=1$ is larger than that under $\mechanism=2$.

We consider $T$ months, i.e., $t=1,2,...,T$.
We denote $\bm{d}=\{d^1,d^2,...,d^T\}$ as the realized data demand sequence, where $d^t$ denotes the realized data demand in   month $t$.
Moreover, we denote $\spedata_1^t$ and $\spedata_2^t$ as the subscriber's data surplus under the data mechanism $\mechanism=1$ and $\mechanism=2$ at the beginning of  month $t$.

Therefore, given the realized data usage sequence $\bm{d}$, the subscriber's overage data consumption in month $t$ is
\begin{equation}
	A_\mechanism^t(\dcap,\bm{d})=\left[ d^t - \dcap-\spedata_\mechanism^t \right]^+.
\end{equation}

Next we first introduce Lemma \ref{lemma: T1 T2}.
\begin{lemma}\label{lemma: T1 T2}
	For any realized data demand sequence $\bm{d}$, we have $A_1^t(\dcap,\bm{d}) \ge A_2^t(\dcap,\bm{d})$ and $\spedata_1^t\le\spedata_2^t$ for any $t=1,2,...,T$.
\end{lemma}
\textit{\textbf{Proof of Lemma \ref{lemma: T1 T2}:}}
We prove Lemma \ref{lemma: T1 T2} by induction.

First, if $t=1$, then $\spedata_1^1=\spedata_2^1=0$, thus we know $A_1^1(\dcap,\bm{d})=A_2^1(\dcap,\bm{d})=\left[ d^1 - \dcap \right]^+$.
Hence, Lemma \ref{lemma: T1 T2} is true when $t=1$.

Second, we assume that Lemma \ref{lemma: T1 T2} is true for $t=k$, and  consider the case of $t=k+1$.

We first show that $\spedata_0^{k+1}\le\spedata_1^{k+1}$.
According to we consumption priority, we know that $\spedata_1^{k+1}$ is
\begin{equation}\label{Proof Equ: tau 1}
\spedata_1^{k+1}=\left[ \dcap - d^k \right]^+,
\end{equation}
and $\spedata_2^{k+1}$ is
\begin{equation}\label{Proof Equ: tau 2}
\spedata_2^{k+1}=
\left\{
\begin{aligned}
& \dcap,											& \text{ if }& d^k<\spedata_2^k, \\
& \left[ \dcap+\spedata_2^k - d^k \right]^+,		& \text{ if }& d^k\ge\spedata_2^k.
\end{aligned}
\right.
\end{equation}

Based on (\ref{Proof Equ: tau 1}) and (\ref{Proof Equ: tau 2}), we know that
\begin{itemize}
	\item If $\spedata_2^k=0$, then $\spedata_1^{k+1}=\spedata_2^{k+1}$ for any $d^k$.
	
	\item If $\spedata_2^k>0$, then we have the following relation between $\spedata_1^{k+1}$ and $\spedata_2^{k+1}$,
	\begin{equation}
	\left\{
	\begin{aligned}
	& \spedata_1^{k+1}<\spedata_2^{k+1}, 	&{\rm if}\ & d^k\in(0,\dcap+\spedata_2^k), \\
	& \spedata_1^{k+1}=\spedata_2^{k+1}, 	&{\rm if}\ & d^k\in\{0\}\cup[\dcap+\spedata_2^k,+\infty). \\
	\end{aligned}
	\right.
	\end{equation}
	
\end{itemize}

Therefore, we find that $\spedata_0^{k+1}\le\spedata_1^{k+1}$.
Next we show that $A_1^{k+1}(\dcap,\bm{d}) \ge A_2^{k+1}(\dcap,\bm{d})$.
According to the following formulation
\begin{equation}
\left\{
\begin{aligned}
& A_1^{k+1}(\dcap,\bm{d})=\left[ d^{k+1} - \dcap-\spedata_1^{k+1} \right]^+, \\
& A_2^{k+1}(\dcap,\bm{d})=\left[ d^{k+1} - \dcap-\spedata_2^{k+1} \right]^+,
\end{aligned}
\right.
\end{equation}
we know that
\begin{itemize}
	\item If $\spedata_1^{k+1}=\spedata_2^{k+1}$, then we have $A_1^{k+1}(\dcap,\bm{d})=A_2^{k+1}(\dcap,\bm{d})$ for any realized data demand sequence $\bm{d}$.
	\item If $\spedata_1^{k+1}<\spedata_2^{k+1}$, then  we have the following relation between $A_1^{k+1}(\dcap,\bm{d})$ and $A_2^{k+1}(\dcap,\bm{d})$,
	\begin{equation}
	\left\{
	\begin{aligned}
	& A_1^{k+1}(\dcap,\bm{d})=A_2^{k+1}(\dcap,\bm{d})	&{\rm if}\ & d^{k+1}\le\dcap+\spedata_1^{k+1}, \\
	& A_1^{k+1}(\dcap,\bm{d})>A_2^{k+1}(\dcap,\bm{d})	&{\rm if}\ & d^{k+1}>\dcap+\spedata_1^{k+1}.
	\end{aligned}
	\right.
	\end{equation}
\end{itemize}

Therefore, we obtain $A_1^{k+1}(\dcap,\bm{d}) \ge A_2^{k+1}(\dcap,\bm{d})$, which implies that Lemma \ref{lemma: T1 T2} is true for $t=k+1$.
Hence, Lemma \ref{lemma: T1 T2} is true.  \hfill$\QEDclosed$

According to Lemma \ref{lemma: T1 T2}, we can conclude that 
\begin{equation}
\sum_{t=1}^{T}A_1^{k+1}(\dcap,\bm{d}) \ge \sum_{t=2}^{T}A_1^{k+1}(\dcap,\bm{d}) ,
\end{equation}
which implies that for any realized data demand sequence $\bm{d}$, the overage data consumption under $\mechanism=1$ is larger than that under $\mechanism=2$.
Hence, it also holds when we take the expectation over the stochastic data demand $d$, i.e., $A_1(\dcap)\ge A_2(\dcap)$.
This completes the proof of  Lemma \ref{Lemma: time flexibility comparison}.\hfill$\QEDclosed$

\section{Proof of Theorem \ref{Theorem: Market Share} \label{Proof: Market Share}}
A type-$(\val,\cut)$ subscriber's expected monthly payoff is
\begin{equation} 
\payoffexp(\plan,\val,\cut)
=\QoS\val\left[\dmean-\cut A_\mechanism(\dcap)  \right]  - \adfee(1-\cut) A_\mechanism(\dcap)-\pcap ,
\end{equation}
 by solving $\payoffexp(\plan,\val,\cut)\ge0$, we obtain
\begin{equation}
	\val  \ge  \Val(\plan,\cut)\eq\frac{1}{\QoS}\left[\adfee  + \frac{\adfee\left[\dmean-A_\mechanism(\dcap)\right]-\pcap}{\cut A_\mechanism(\dcap)-\dmean} \right] ,
\end{equation}
thus  the market share of the MNO under the data plan $\plan=\{\dcap,\pcap,\adfee,\mechanism\}$ is
\begin{equation}
	\subscription(\plan)=\{ (\val,\cut): \Val(\plan,\cut)\le \val \le \valmax, 0\le\cut\le1 \}.
\end{equation}

\hfill$\QEDclosed$

\section{Proof of Theorem \ref{Theorem: Optimal Pricing} \label{Proof: Optimal Pricing}}
We prove Theorem \ref{Theorem: Optimal Pricing} by deriving the MNO's profit-maximizing subscription fee and per-unit fee.
Recall that the profit of the MNO is given by
\begin{equation}  
\begin{aligned}
\tilde{\profit}(\plan)	
&= \int_{0}^{1} \int_{\Val(\plan,\cut)}^{\valmax} \big\{ 
\adfee(1-\cut)A_\mechanism(\dcap) +\pcap \\
&\quad- c \cdot \left[ \dmean-\cut A_\mechanism(\dcap) \right] 
\big\} h(\val)g(\cut) d\val d\cut  
-  J(\dcap) .
\end{aligned}
\end{equation}

Since we consider a fixed $\dcap$ in Theorem \ref{Theorem: Optimal Pricing}, then $A_{\mechanism}(\dcap)$ is a constant, and we denote it as $A_{\mechanism}$. 
Accordingly, we obtain
\begin{equation}   \label{Equ: Pricing profit}
\begin{aligned}
\tilde{\profit}(\pcap,\adfee)
&= \int_{0}^{1} \int_{\Val(\plan,\cut)}^{\valmax} \big\{ 
\adfee(1-\cut)A_{\mechanism}  +\pcap \\
& \qquad\qquad\qquad- c \cdot \left[ \dmean-\cut A_{\mechanism}  \right] 
\big\} h(\val)g(\cut) d\val d\cut  \\
&= \int_{0}^{1} \big\{ 
\adfee(1-\cut)A_{\mechanism}  +\pcap  - c \cdot \left[ \dmean-\cut A_{\mechanism}  \right] \big\} \\
&\qquad\qquad \left[ 1-H\left( \Val(\plan,\cut) \right) \right] g(\cut)   d\cut ,
\end{aligned}
\end{equation}
where $\Val(\plan,\cut)$ is given by
\begin{equation}
\begin{aligned}
\Val(\plan,\cut) & =  \frac{\adfee(1-\cut)A_{\mechanism}+\pcap}{\QoS\left(\dmean-\cut A_{\mechanism}\right)} .
\end{aligned}
\end{equation}

Therefore, we can write  $\adfee(1-\cut)A+\pcap$ as 
\begin{equation}\label{Equ: Pricing consumption}
\adfee(1-\cut)A_{\mechanism}+\pcap = \Val(\plan,\cut)\cdot \QoS\left(\dmean-\cut A_{\mechanism}\right).
\end{equation}

When substituting  (\ref{Equ: Pricing consumption}) into (\ref{Equ: Pricing profit}), we can  write the MNO's  profit as follows:
\begin{equation}   
\begin{aligned}
& \tilde{\profit}(\pcap,\adfee) =\int_{0}^{1} 
\QoS\left( \dmean-\cut A_{\mechanism}  \right) \left[ \Val(\plan,\cut)-\frac{c}{\QoS} \right]  \\
&\qquad\qquad\qquad\qquad \cdot\Big[ 1-H\Big( \Val(\plan,\cut) \Big) \Big] g(\cut)   d\cut ,
\end{aligned}
\end{equation}

Note that, under the assumption that $\val$ distribution satisfies the increasing failure rate (i.e., ${h(x)}/\left[{1-H(x)}\right]$ increases in $x$), we have the following inequality
\begin{equation}\label{Proof Equ: IFR example 1}
\left(x-\frac{c}{\QoS}\right)\Big[1-H(x)\Big] \le \frac{\left[1-H\left(x^*\right)\right]^2}{ h\left(x^*\right) },
\end{equation}
where $x^*$ is \textit{uniquely} determined by 
\begin{equation}\label{Proof Equ: IFR example 2}
\frac{ 1-H\left(x^*\right)}{ h\left(x^*\right) } =x^*-\frac{c}{\QoS},
\end{equation}
since $ \left[1-H(x)\right]/h(x) $ decreases in $x$ and $x-c/\QoS$ increases in $x$.

Based on (\ref{Proof Equ: IFR example 1}) and (\ref{Proof Equ: IFR example 2}), we know that when the subscription fee $\pcap^*$ and the per-unit fee $\adfee^*$ satisfy 
\begin{equation}
\left\{
\begin{aligned}
	&   H\left( \frac{\adfee^*}{\QoS} \right) + \frac{\adfee^*-c}{\QoS}\cdot h\left( \frac{\adfee^*}{\QoS} \right) =1, \\ 
	& \pcap^*=\adfee^*\Big[ \dmean-A \Big] ,
\end{aligned}
\right.
\end{equation}
we always have the following inequality
\begin{equation}\label{Proof Equ: integration}
\begin{aligned}
&\QoS\left( \dmean-\cut A  \right) \left[ \Val(\plan,\cut)-\frac{c}{\QoS} \right]  \Bigg[ 1-H\Big( \Val(\plan,\cut) \Big) \Bigg] \\
\le & \QoS\left( \dmean-\cut A  \right)\cdot \frac{\left[1-H\Big(\Val(\dcap,\pcap^*,\adfee^*,\mechanism,\cut)\Big)\right]^2}{ h\Big(\Val(\dcap,\pcap^*,\adfee^*,\mechanism,\cut)\Big) },\ \forall\ \cut\in[0,1].
\end{aligned}
\end{equation}

By computing the integration over $\cut\in[0,1]$ on (\ref{Proof Equ: integration}), we have
\begin{equation}   
\begin{aligned}
& \int_{0}^{1} 
\QoS\left( \dmean-\cut A  \right) \left[ \Val(\plan,\cut)-\frac{c}{\QoS} \right]  \Bigg[ 1-H\Big( \Val(\plan,\cut) \Big) \Bigg] g(\cut)   d\cut  \\
&\le \int_{0}^{1} \QoS\left( \dmean-\cut A  \right)\cdot \frac{\left[1-H\Big(\Val(\dcap,\pcap^*,\adfee^*,\mechanism,\cut)\Big)\right]^2}{ h\Big(\Val(\dcap,\pcap^*,\adfee^*,\mechanism,\cut)\Big) } g(\cut)   d\cut ,
\end{aligned}
\end{equation}
which indicates $\tilde{\profit}(\pcap,\adfee)\le\tilde{\profit}(\pcap^*,\adfee^*)$.
Moreover, $\adfee^*$ is unique according to (\ref{Proof Equ: IFR example 1}) and (\ref{Proof Equ: IFR example 2}).\hfill$\QEDclosed$

\section{Proof of Theorem \ref{Theorem: Optimal Data Cap}\label{Proof: Optimal Data Cap}}
We prove  Theorem \ref{Theorem: Optimal Data Cap} by deriving the MNO's profit-maximizing data cap.
Under the optimal pricing strategy, the MNO's expected total profit is given by
\begin{equation}\label{Proof Equ: profit Q}
\begin{aligned}
& \tilde{\profit}(\dcap,\pcap^*(\dcap,\mechanism),\adfee^*,\mechanism ) \\
=& \frac{ \left[\dmean-\bar{\cut}A_\mechanism(\dcap)\right] \left(\adfee^*-c\right)^2 }{\QoS} h\left(\frac{\adfee^*}{\QoS}\right)-z\cdot\dcap ,
\end{aligned}
\end{equation}
which is concave on the data cap $\dcap$, since $A_\mechanism(\dcap)$ is convex on $\dcap$ and $z\cdot\dcap$ is linear on $\dcap$.
We take the derivative of (\ref{Proof Equ: profit Q}) with respect to the data cap $\dcap$, and obtain the following optimality condition
\begin{equation}
\frac{ -\bar{\cut}A'_\mechanism(\dcap^*)\left(\adfee^*-c\right)^2 }{\QoS} h\left(\frac{\adfee^*}{\QoS}\right)-z =0 ,
\end{equation}
which is equivalent to
\begin{equation}\label{Equ: Proof Optimal Q FOC}
 -A'_\mechanism(\dcap^*)  = \frac{z\cdot\QoS}{ \bar{\cut}\cdot \left(\adfee^*-c\right)^2\cdot h\left(\frac{\adfee^*}{\QoS}\right) }.
\end{equation}

Recall that the optimal per-unit fee $\adfee^*$ satisfies 
\begin{equation}
H\left( \frac{\adfee^*}{\QoS} \right) + \frac{\adfee^*-c}{\QoS}\cdot h\left( \frac{\adfee^*}{\QoS} \right) =1,
\end{equation}
which means that 
\begin{equation}\label{Equ: Proof Optimal Q p^*}
\adfee^*-c = \QoS \cdot \frac{ 1-H\left( \frac{\adfee^*}{\QoS} \right) }{ h\left( \frac{\adfee^*}{\QoS} \right) }.
\end{equation}

We substitute (\ref{Equ: Proof Optimal Q p^*}) into (\ref{Equ: Proof Optimal Q FOC}), then obtain
\begin{equation} 
A_\mechanism'\left( \dcap^* \right) =  \tdr(\QoS,c,z),
\end{equation}
where $\tdr(\QoS,c,z)$ is given by
\begin{equation}
\tdr(\QoS,c,z)=\frac{ z \cdot h\left(\frac{\adfee^*}{\QoS}\right)  }{ \bar{\cut}\QoS \left[ 1-H\left(\frac{\adfee^*}{\QoS}\right) \right]^2 }.
\end{equation}
\hfill$\QEDclosed$

\section{Proof of Lemma \ref{Lemma: Property A(Q)} \label{Proof: Property A(Q)}}
We prove Lemma \ref{Lemma: Property A(Q)} by computing $A_{\mechanism}(0)$, $A_{\mechanism}(\dmax)$, $A'_{\mechanism}(0)$, and $A'_{\mechanism}(\dmax)$, and showing the convexity of $A_{\mechanism}(\dcap)$ on $\dcap$.
Recall that the mathematic expression of $A_\mechanism(\dcap)$ is
\begin{equation} \label{Equ: Proof A(Q)}
\begin{aligned}
A_\mechanism(\dcap)=&\mathbb{E}_{d,\spedata} \big\{\left[ d-\dcap_\mechanism^e(\spedata) \right]^+ \big\} \\
=&  \textstyle\sum\limits_{\spedata}	\sum\limits_{d}	[d-\dcap_\mechanism^e(\spedata)]^+f(d)p_\mechanism(\spedata).
\end{aligned}
\end{equation}

\subsection{Compute $A_{\mechanism}(0)$ and $A_{\mechanism}(\dmax)$}
When substituting $\dcap=0$ and $\dcap=\dmean$ into $A_\mechanism(\dcap)$, respectively, we obtain
\begin{equation}
\begin{aligned}
& A_\mechanism(0)= \textstyle \sum\limits_{d}	[d-0]^+f(d)=\dmean, \\
& A_\mechanism(\dmax)=\textstyle\sum\limits_{\spedata}\sum\limits_{d}0\cdot	 f(d)p_\mechanism(\spedata)=0.
\end{aligned}
\end{equation}

\subsection{Compute $A'_{\mechanism}(0)$, and $A'_{\mechanism}(\dmax)$}
We compute the derivative of $A_\mechanism(\dcap)$ with respect to $\dcap$ as follows:
\begin{equation}\label{Proof Equ: A'}
\begin{aligned}
A_\mechanism'(\dcap)	=& \lim\limits_{\Delta\dcap\rightarrow0^+} \frac{A_\mechanism(\dcap+\Delta\dcap)-A_\mechanism(\dcap)}{\Delta\dcap} .
\end{aligned}
\end{equation}

When substituting $\dcap=0$ into (\ref{Proof Equ: A'}),  we obtain $A'_{\mechanism}(0)$ as following
\begin{equation}\label{Proof Equ: A' 0}
\begin{aligned}
A_\mechanism'(0)	
= \lim\limits_{\Delta\dcap\rightarrow0^+} \frac{A_\mechanism(\Delta\dcap)-\dmean}{\Delta\dcap}.
\end{aligned}
\end{equation}

Now we can compute $A_0'(0)$ by substituting $A_0(\Delta\dcap)$ into (\ref{Proof Equ: A' 0}), as follows
\begin{equation}
\begin{aligned}
A_0'(0)	=& \lim\limits_{\Delta\dcap\rightarrow0^+} \frac{ \sum_{d=0}^{\dmax}\left[d-\Delta\dcap\right]^+f(d)-\dmean}{\Delta\dcap}=-1.
\end{aligned}
\end{equation}


Similarly, we are able to show that $A_\mechanism(0)=-1$ for any $\mechanism\in\{0,1,2,3\}$.

Furthermore, when substituting $\dcap=\dmax$ into (\ref{Proof Equ: A'}), we obtain $A_\mechanism'(\dmax)$ as follows
\begin{equation}\label{Proof Equ: A' D}
\begin{aligned}
A_\mechanism'(\dmax)	
=& \lim\limits_{\Delta\dcap\rightarrow0^-} \frac{A_\mechanism(\dmax+\Delta\dcap)}{\Delta\dcap}
\end{aligned}
\end{equation}

Now we can compute $A_0'(\dmax)$ by substituting $A_0(\dmax+\Delta\dcap)$ into (\ref{Proof Equ: A' D}), as follows
\begin{equation}
\begin{aligned}
A_0'(\dmax)	= \lim\limits_{\Delta\dcap\rightarrow0^-} \frac{ \sum\limits_{d=0}^{\dmax}\left[d-\dmax-\Delta\dcap \right]^+f(d) }{\Delta\dcap} = 0.
\end{aligned}
\end{equation}


Similarly, we are able to show that $A_\mechanism(\dmax)=0$ for any $\mechanism\in\{0,1,2,3\}$.

\subsection{Convexity}
According to (\ref{Equ: Proof A(Q)}), $A_\mechanism(\dcap)$ is the nonnegative weighted summation over $[d-\dcap_\mechanism^e(\spedata)]^+$.
Specifically, $ [d-\dcap_\mechanism^e(\spedata)]^+$ is given by
\begin{equation}
\left[\dcap-\dcap_\mechanism^e(\spedata)\right]^+ =
\left\{
\begin{aligned}
& \left[ d-\dcap \right]^+ , &\text{if }& \mechanism=0,\\
& \left[ d-\dcap-\spedata \right]^+ , &\text{if }& \mechanism\in\{1,2\} ,\\
& \left[ d-2\dcap-\spedata \right]^+ , &\text{if }& \mechanism=3 ,
\end{aligned}
\right.
\end{equation}
which indicates that $ [d-\dcap_\mechanism^e(\spedata)]^+$ is always convex in $\dcap$ for any $\mechanism\in\{0,1,2,3\}$.
Considering the nonnegative weighted summation is a convexity-preserving operation, thus $A_\mechanism(\dcap)$ is convex in $\dcap$.	\hfill$\QEDclosed$

{
\section{Variance of Demand Distribution}\label{Appendix: Variance of Demand Distribution}
We investigate the impact of the demand distribution variance by considering two different values.
Specifically, we consider two log-normal distributions, which have the same mean value of 1GB but different variances (i.e., $\sigma=0.8,0.4$).
In the following, we first show the distributions and the expected overage payments, then compare the performance gains under the two values of variances.

Fig. \ref{fig: Revision variance A} shows the demand distributions and the corresponding expected overage payment.
The horizontal axis in each sub-figure corresponds to the MNO's QoS.
By comparing Fig. \ref{fig: Revision_A_LargeVar} and Fig. \ref{fig: Revision_A_SmallVar}, we note that given the same data mechanism (e.g., $\mechanism=1$ with two red curves in the two sub-figures), a larger demand variance leads to higher expected overage payments.
The impact of time-flexible data mechanism on reducing overage payment is stronger under a larger variance, since the differences among the three curves with markers in Fig. \ref{fig: Revision_A_LargeVar} is larger than that in Fig. \ref{fig: Revision_A_SmallVar}.

Fig. \ref{fig: Revision variance Gain} plots the performance gain of the time-flexible data mechanisms (compare with the traditional data mechanism) under the two values of variance.
We note that a larger variance leads to a higher gain for both MNO's profit gain and users' payoff gain.

Based on the above discussions, we find that the time-flexible data mechanism plays a significant role, especially when the users' demand variance is relatively large.

\begin{figure}
	\centering
	\setlength{\abovecaptionskip}{2pt}
	\setlength{\belowcaptionskip}{0pt}
	\subfigure[$\sigma=0.8$]{\label{fig: Revision_A_LargeVar}\includegraphics[width=0.48\linewidth]{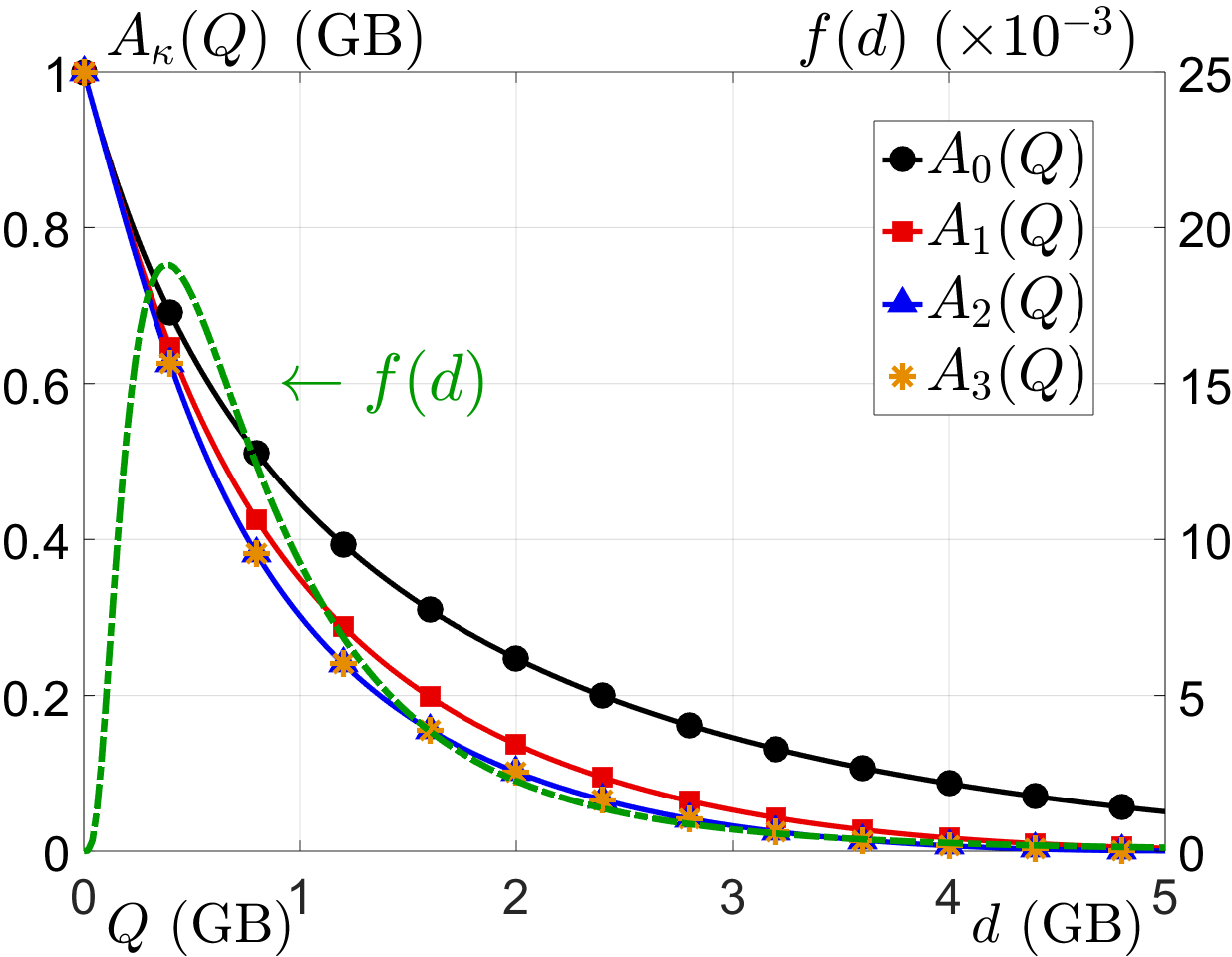}}\quad 
	\subfigure[$\sigma=0.4$]{\label{fig: Revision_A_SmallVar}\includegraphics[width=0.48\linewidth]{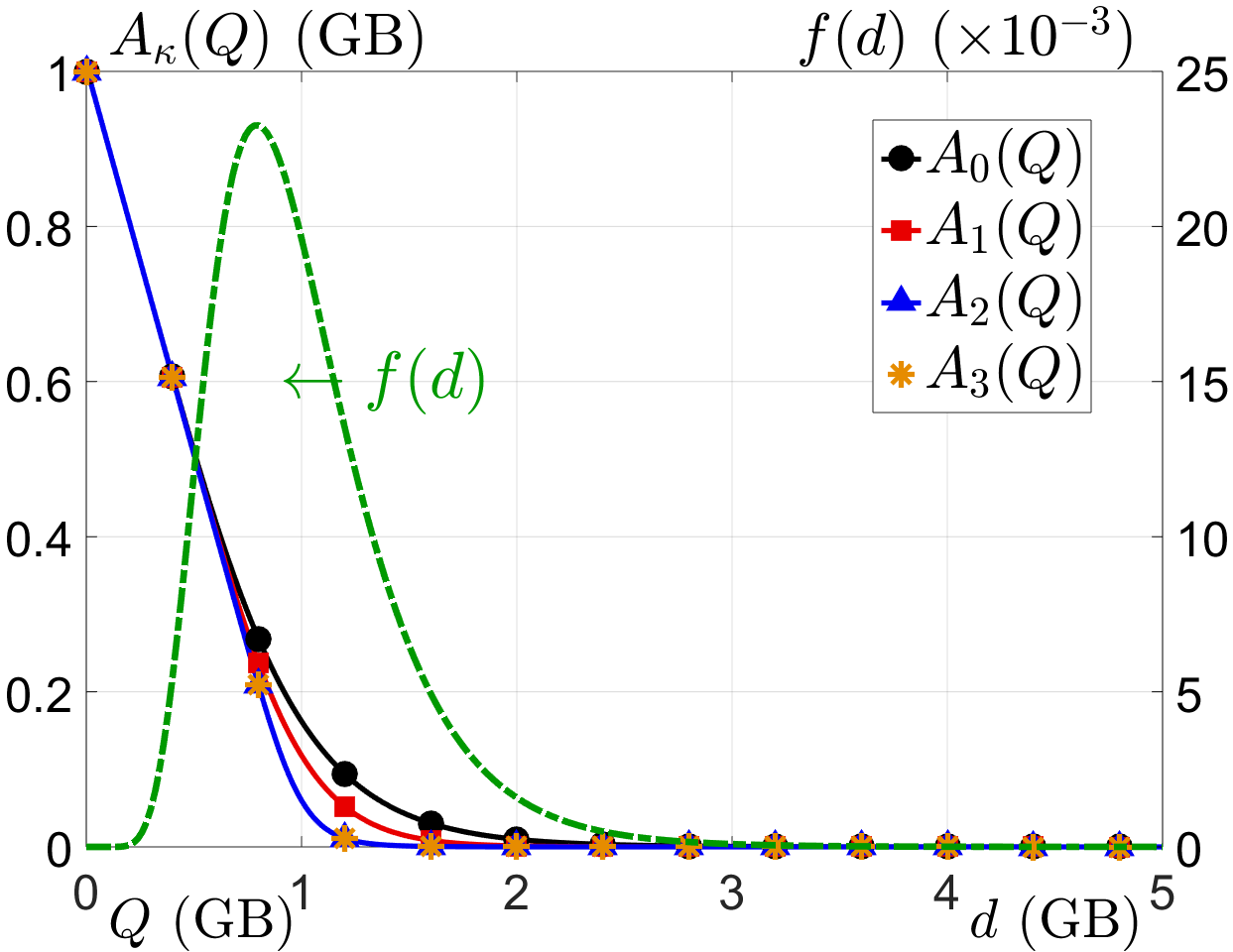}}
	\caption{$A_\mechanism(\dcap)$ vs $\dcap$ and $f(d)$ vs $d$ under different variances.}
	\label{fig: Revision variance A}
\end{figure}



\begin{figure}
	\centering
	\setlength{\abovecaptionskip}{2pt}
	\setlength{\belowcaptionskip}{0pt}
	\subfigure[$\sigma=0.8$]{\label{fig: Revision_LargeVal_QoS_Optimal_Gain}\includegraphics[width=0.48\linewidth]{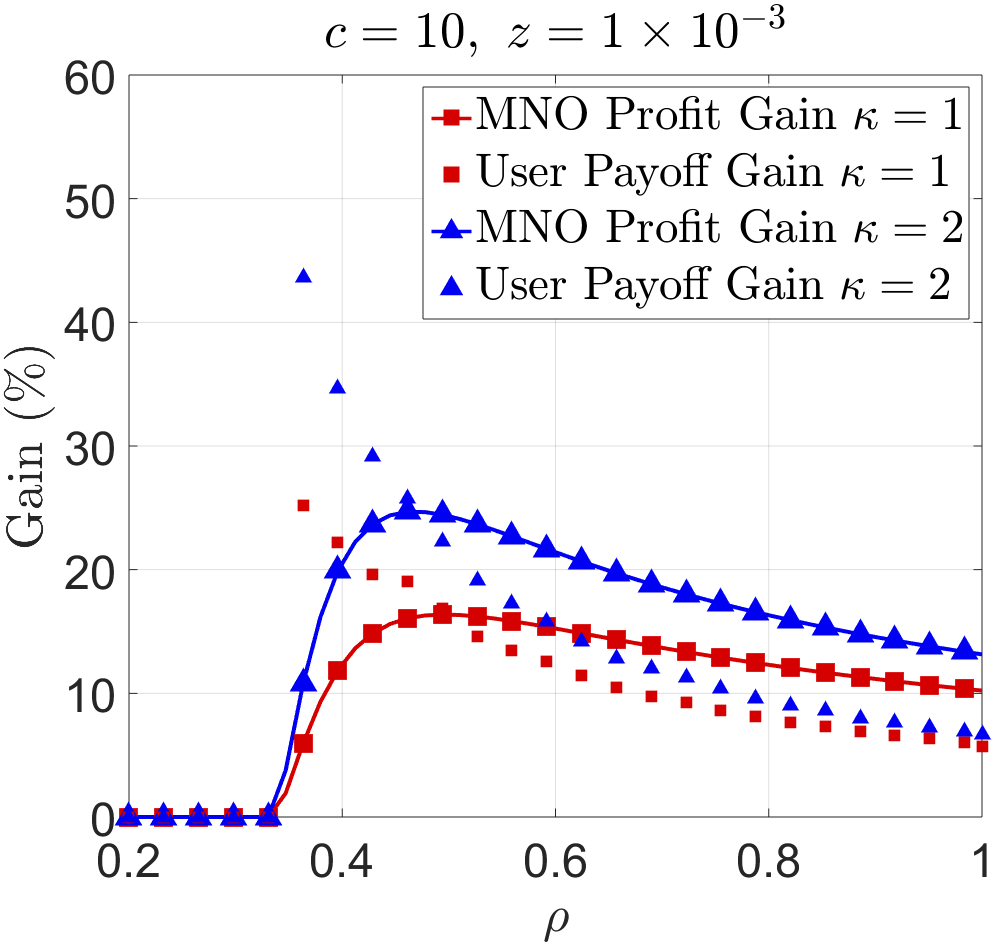}}\quad
	\subfigure[$\sigma=0.4$]{\label{fig: Revision_SmallVal_QoS_Optimal_Gain}\includegraphics[width=0.48\linewidth]{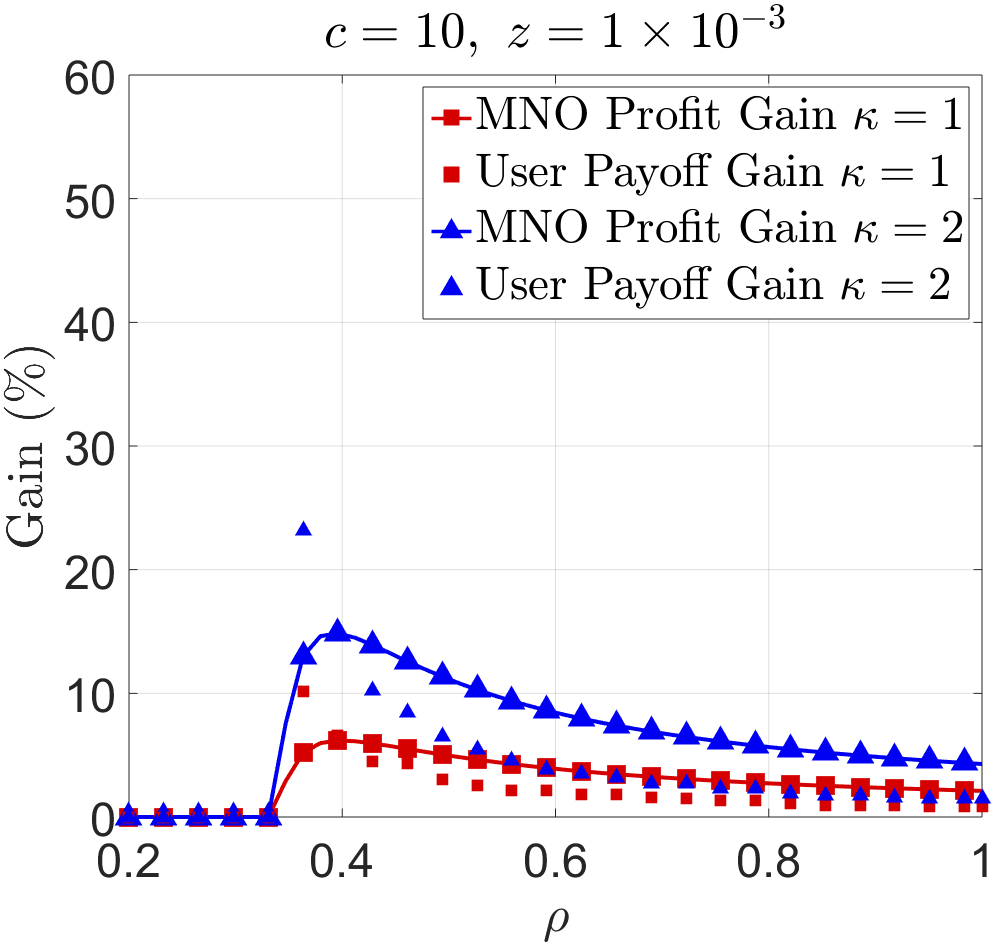}}
	\caption{Performance gain under different variances.}
	\label{fig: Revision variance Gain}
\end{figure}

}

\end{document}